\documentclass[structabstract]{aa}  
\usepackage{graphicx}
\usepackage{natbib}
\bibpunct{(}{)}{;}{a}{}{,}

\def\lya{Ly$\alpha$}

\def\ebv{\ifmmode E_{B-V} \else $E_{B-V}$\fi}

\def\fesc{\ifmmode f_{\rm esc}(\lambda) \else $f_{\rm esc}(\lambda)$\fi}
\def\fescuv{\ifmmode f_{\rm esc}({\rm UV}) \else $f_{\rm esc}(\rm UV)$\fi}
\def\flya{\ifmmode f_{\rm esc}({\rm{Ly}\alpha}) \else $f_{\rm esc}({\rm{Ly}\alpha})$\fi}
\def\wlya{\ifmmode {\rm EW}_{\mathrm{Ly}\alpha}\else ${\rm EW}_{\mathrm{Ly}\alpha}$\fi}

\def\cm2{cm$^{-2}$}
\def\kms{km~s$^{-1}$}

\def\hi{H{\sc i}}

\def\nh{\ifmmode N_{\mathrm{HI}}\else $N_{\mathrm{HI}}$\fi}
\def\vexp{\ifmmode {\rm v}_{\rm exp} \else v$_{\rm exp}$\fi}
\def\nhi{\ifmmode \overline{N_{\mathrm{HI}}}\else $\overline{N_{\mathrm{HI}}}$\fi}
\def\taua{\ifmmode \overline{\tau_{a}}\else $\overline{\tau_{a}}$\fi}
\def\tauc{\ifmmode \overline{\tau_{c}}\else $\overline{\tau_{c}}$\fi}
\def\nratio{\ifmmode n_{\rm IC}/n_{\rm C}\else  $n_{\rm IC}/n_{\rm C}$\fi}
\def\nc{\ifmmode n_{\rm C}\else  $n_{\rm C}$\fi}
\def\nic{\ifmmode n_{\rm IC}\else  $n_{\rm IC}$\fi}
\def\tratio{\ifmmode T_{\rm IC}/T_{\rm C}\else  $T_{\rm IC}/T_{\rm C}$\fi}

\begin{document}
   \title{Lyman $\alpha$ line and continuum radiative transfer in a clumpy interstellar medium}


   \author{F. Duval
          \inst{1}
          \,,
          D. Schaerer
          \inst{2,3}
          \,,
          G. {\"O}stlin
         \inst{1}
         \,,
          P. Laursen
          \inst{4}
}

   \institute{Department of Astronomy, Stockholm University, Oscar Klein Center, AlbaNova, Stockholm SE-106 91, Sweden
              \email{fduva@astro.su.se}
         \and
Observatoire de Gen\`eve, Universit\'e de Gen\`eve, 51 Ch. des Maillettes, 1290 Versoix, Switzerland
         \and 
             CNRS, IRAP, 14 Avenue E. Belin, 31400 Toulouse, France
         \and
             Dark Cosmology Centre, Niels Bohr Institute, University of Copenhagen, Juliane Maries Vej 30, DK-2100 C{\o}penhagen, Denmark.
             }

\authorrunning{}
\titlerunning{\lya\ radiative transfer in a clumpy ISM}
   \date{Received ... ; accepted ...                 }

 
 \abstract
  {}
{Studying the effects of an inhomogeneous interstellar medium
(ISM) on the strength and the shape of the Lyman alpha (\lya) line in
starburst galaxies.}
{Using our 3D Monte Carlo  \lya\ radiation transfer code, we study the radiative transfer of \lya, UV and optical continuum photons in
homogeneous and clumpy shells of neutral hydrogen and dust surrounding a central source.
  Our simulations predict the \lya\ and continuum escape fraction,
the \lya\ equivalent width EW(\lya), and the \lya\ line profile, and their dependence on the geometry of the gas distribution and the main input physical parameters.}
{The ISM clumpiness is found to have a strong impact on the \lya\ line radiative transfer, entailing a strong dependence of the emergent
features of the \lya\ line (escape fraction, EW(\lya)) on the ISM
morphology.
Although a clumpy and dusty ISM appears more transparent to radiation
(both line and continuum) compared to an equivalent homogeneous ISM of equal dust optical depth, we find that the \lya\ photons are, in general, still more attenuated than UV continuum radiation.
As a consequence, the observed equivalent width of the \lya\ line ($EW_{\rm obs}$(\lya)) is
lower than the intrinsic one ($EW_{\rm int}$(\lya)) for nearly all clumpy ISM configurations considered.
There are, however, special conditions under which \lya\ photons escape more easily than the
continuum, resulting in an enhanced  $EW_{\rm obs}$(\lya).
The requirement for this to happen is that the ISM is almost static (galactic outflows $\le$ 200 km s$^{-1}$),
extremely clumpy (with density contrasts $>10^7$ in HI between clumps and the interclump medium), and very dusty (E(B-V) $>$ 0.30).
When these conditions are fulfilled the emergent \lya\ line profile
generally shows no velocity shift and little asymmetry.
Otherwise the \lya\ line profile is very similar to that expected
for homogeneous media.}
{Given the asymmetry and velocity shifts generally observed in star-forming galaxies with \lya\ emission, we therefore conclude that
clumping is unlikely to significantly enhance their relative \lya/UV transmission.}


   \maketitle
%

\section{Introduction}


Being the intrinsically brightest spectral signature of remote young galaxies \citep{patridge67, schaerer03}, and possessing a rest wavelength of 1216 \AA\ (making it accessible for optical/near-IR ground-based telescopes for redshifts  $z \ge 2$), the Lyman alpha (\lya) line has become the most powerful emission-line probe of the distant young universe. The potential of the \lya\ emission line for detection and redshift confirmation of distant galaxies, derivation of star formation rate (SFRs), as well as probe of the ionization state of the intergalactic medium (IGM) \citep{malhotra04, kashikawa06} and the reionisation epoch \citep{fan02, santos04} is enormous, but necessarily relies on a good astrophysical understanding of the processes that regulates the emergent \lya\ emission from a galaxy.



The importance of the \lya\ line in the cosmological context was first proposed by \cite{patridge67} who suggested that young high-$z$ galaxies, undergoing their first star-forming event, should be detectable thanks to their strong \lya\ emission line. 
Unfortunately, the first attempts to detect high-redshift galaxies in \lya\ gave quite meager results.
The observed \lya\ fluxes appeared fainter than those predicted and only few \lya\ emitters (LAEs) had been detected until the late 1990s (cf. Djorgovski \&\ Thompson 1992, Pritchet 1994). This lack of \lya\ emission has nevertheless triggered several studies which have enabled to highlight the high complexity of the resonant \lya\ line radiative transfer in starburst galaxies \citep{meier81, neufeld90, charlot_fall93, kunth98, tenorio99, mashesse03, ostlin09}. 
While the faint measured \lya\ fluxes were originally attributed to the dust attenuation (Pritchet 1994), it has turned out that many physical effects could strongly modify or supress the \lya\ line within galaxies (metallicity, neutral hydrogen kinematics, geometry of the interstellar medium (ISM)). 
It is only during this last decade, and the development of deep and wide surveys, that many \lya-emitting galaxies have been detected \citep{hu98, hu04, cowie98, kudritzki00, rhoads00, taniguchi03, taniguchi05, shimasaku06, gronwall07, nilsson07, guaita10, ouchi03, Ouchi08, ouchi10}.

Because of the factors which contribute to the \lya\ radiative transfer, the \lya\ line features (line profile, equivalent width EW(\lya), offset from other emission/absorption lines) encode much information on the properties of individual galaxies: gas kinematics, 
gas geometry as well as stellar population. 
For instance, the detection of unusually strong \lya\ line in the spectra of high-$z$ galaxies
could indicate the presence of population III stars within them \citep{schaerer03}, whereas the asymmetry of the line profiles would suggest the presence of strong galactic outflows \citep{kunth98}. The derivation of this precious information requires however an accurate interpretation of the \lya\ line features, which implies as well a complete understanding of the \lya\ radiative transfer in the ISM of galaxies. This is one of the aims of this paper. 

Due to the importance of the \lya\ line for cosmology, several studies have attempted  to understand the physical process governing the escape of \lya\ photons from galaxies. Among the parameters which influence the visibility of the \lya\ line, dust content, neutral gas kinematics and the geometry of the neutral gas seem to play  the most important roles. 
Dust was originally invoked to explain the absence or the faint \lya\ emission from galaxies at large redshift (e.g. Meier \&\ Terlevich 1981).
However, \cite{giavalisco96} studied a local sample of star-forming galaxies observed with the IUE space telescope and found no clear correlation between \lya/H$\beta$ or EW(\lya) and the reddening $E(B-V)$. Other studies have also led to a lack of correlation between the dust attenuation and the strength of the \lya\ line, suggesting that other parameters govern the escape of \lya\ photons \citep{kunth94, thuan97a, atek08}. Among them, the role of the neutral gas kinematics was revealed in the 1990s by  \cite{kunth94} and \cite{lequeux95}. 
For 8 local galaxies observed with the Goddard High Resolution Spectrograph (GHRS), \cite{kunth98} found that when \lya\ line appeared in emission, there was a systematic blueshift of low ionisation states (LIS) metal absorption lines with respect to \lya , indicative of outflows in the
neutral medium. Furthermore, the shape of the \lya\ line profiles proved to be asymmetric. Galaxies showing \lya\ in absorption showed 
significantly smaller relative shifts of LIS lines and \lya . This result clearly shows that the \lya\ escape fraction and line shape are strongly 
affected by the kinematical configuration in the ISM. Phenomenologically it is easy to understand that an outflow in the neutral ISM would
promote the escape of  \lya\ photons and create asymmetric line profiles, since the motion Doppler shifts the line out of resonance and more
so for the red side of the line.
Finally, several studies of the resonant \lya\ transfer have emphasized the importance of the ISM clumpiness  on the escape of the \lya\ \citep{neufeld91, giavalisco96}. In particular, \cite{neufeld91} showed that it could be possible to observe an emergent EW(\lya) higher 
than the intrinsic one in a dusty and clumpy ISM. As the clumpiness of the ISM is well established in our galaxy \citep{stutzki90, marscher93}, 
this parameter must therefore be taken into account in the study of the \lya\ radiative transfer.

With the increased number of \lya\ radiative transfer codes developped recently \citep{ahn01, ahn02, cantalupo05,Verhamme06, pierleoni07, laursen09, forero11}, the transfer of \lya\ photons has intensively been investigated in the framework of galaxy simulations. In particular, such simulations allow us to compare the observed \lya\ line properties of individual galaxies, both nearby and distant ones \citep{ahn03,verhamme08,atek09a}.
However, although most studies have treated the \lya\ radiative transfer in either static or expanding media, the main effects of a multiphase ISM on the \lya\ radiative transfer has been the object of few numerical studies \citep{haiman99, richling03, hansen06, laursen12}. 
The aim of this present paper is to carry out a detailed study of both the \lya\ and the UV continuum radiative transfer in a large range of dusty, moving, homogeneous and clumpy ISMs. 
This will allow us to examine in detail the effects of the ISM clumpiness on the features of the \lya\ line (\lya\ escape fraction, EW(\lya), \lya\ line profiles).

One of the main motivations of our study is also to understand the anomalous strong EW(\lya) revealed by several observations of LAEs at high-$z$ \citep{kudritzki00, malhotra02, rhoads03, shimasaku06, kashikawa12}. While normal stellar population models predict a maximum value of $\sim$ 240 \AA\ for the intrinsic EW(\lya) within starburst galaxies (i.e.\ assuming population I/II stars, Charlot \&\ Fall 1993; Schaerer 2003), it is not rare to observe higher EW(\lya) from high-redshift sources.
Several physical possibilities have already been investigated to explain these high EW(\lya), such as the presence of either population III stars or Active Galactic Nuclei (AGNs) in the host galaxies. But none of them prove to be consistent with the observations \citep{dawson04, wang04, gawiser06}. 
Another possibility is that the high EW(\lya) values found are due to the combined effect of IGM absorption (lowering the continuum on the blue side of \lya\ at high $z$) and observational errors biasing the average EW(\lya) to higher values \citep{hayes06}.
The most popular explanation seems, however,  to be the relative boost of \lya\ photons result in a clumpy ISM as originally suggested by \cite{neufeld91}. In this scenario, 
\lya\ and UV continuum photons propagate in a clumpy ISM, where all neutral hydrogen and dust are mixed together in clumps. While \lya\ photons would scatter off of the surface of clumps, having their journey confined to the dustless interclump medium, the UV continuum photons would penetrate into the clumps and would suffer greater extinction. Such a scenario would thus produce larger EW(\lya) than the intrinsic ones, allowing to explain the anomalously high EW(\lya) observed in some high-$z$ galaxies. 
Other studies have also invoked a higher transmission of \lya\ photons than for the UV continuum, to explain observations of some low redshift \lya\ emitters \citep{scarlata09},
to understand the overall SED of LAEs at $z \sim 4$ \citep{finkelstein08, finkelstein09},
and to reproduce the \lya\ and UV luminosity function of distant galaxies \citep{dayal09, forero11}.

In this paper, we investigate the Neufeld scenario further and examine the physical conditions under which a clumpy ISM could produce a boost of
the \lya\ line relative to the continuum.

The remainder of this paper is structured as follows. In Section 2 we outline a description of our numerical model, presenting the features of the clumpy media and our assumptions. In Sections\ 3 and 4 are presented our results. The \lya\ radiative transfer in homogeneous and clumpy media is presented in Section 3, whereas the formation and the features of the emergent \lya\ line profiles are described in Section 4. Section 5 is dedicated to the discussion of these results with, in particular, an application of our study to the Neufeld scenario. Finally, our main conclusions are summarised in Section 6.


\section{Method}

%


\subsection{3D radiation transfer code}
To study the \lya\ line and UV--optical continuum radiation transfer in clumpy geometries, we have used the latest version of the
3D Monte Carlo radiative transfer code MC\lya\ of \cite{Verhamme06} and \cite{schaerer11}. 
To treat the radiation transfer at wavelengths other than \lya, we here also compute the continuum transfer 
at other wavelengths assuming scattering and absorption by dust. 
For the present paper we are interested in three wavelengths, listed in Table \ref{table:dust}:  the \lya\ line ($\lambda$ = 1215.67 \AA) and its neighboring UV continuum, 
and the optical B and V bands.


\subsection{3D geometries, model parameters, and model output}

Both for simplicity, and since spherically symmetric outflows with a homogeneous \hi\ shell are able to reproduce 
a large variety of observed \lya\ line profiles in Lyman break galaxies and \lya\ emitters \citep{verhamme08,schaerer08,dessauges10},
the same geometry is used to study how a clumpy ISM structure alters the \lya\ line and UV continuum. This clumpy geometry is
also chosen since it has been shown to reproduce observable continuum properties of starburst galaxies and the
Calzetti attenuation law \citep{gordon97,witt00,vijh03}. 
Finally, this also allows us to make a detailed study into the continuity of the extensive grid of radiation transfer models by \citet{schaerer11}.


In practice we adopt the following, simple shell geometries (see figure 1):
a static or radially expanding, homogeneous or clumpy shell of \hi\ and dust surrounding the source emitting both 
\lya\ line and continuum photons. Dust and gas (\hi) are assumed to be co-spatial in the shell. 
We assume a point-like central source.

  \begin{figure*}
   \centering
   \includegraphics[width=173mm]{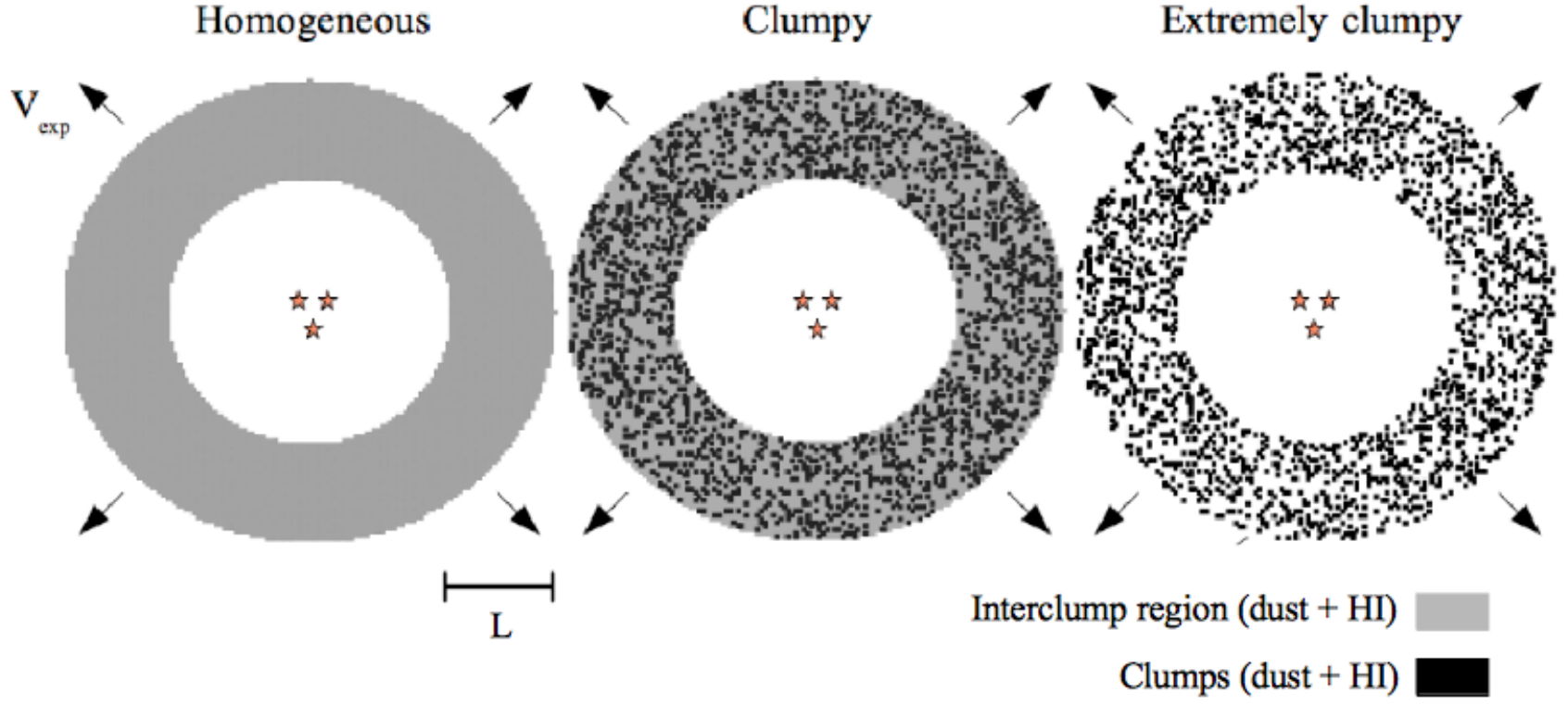}
      \caption{Representation of some 3D homogeneous and clumpy geometries studied in this paper. The star distribution is always localised in the center of the shell, whereas the dust and the \hi\ content are distributed around. The dust and \hi\ distribution can be "homogeneous" (left), "clumpy" (middle) or "extremely clumpy" (right). In a "clumpy" distribution, the clumps and the interclump medium receive, respectively, a high and a low densities of dust and HI. In the case of an "extremely clumpy" distribution, all the dust and the \hi\ content are distributed in clumps.
              }
         \label{fig:geom}
   \end{figure*}
   
\subsubsection{Input parameters}

The four physical and the two geometrical input parameters of our models, listed in Table 1, are the following.
The radial expansion velocity \vexp, the Doppler parameter $b$ of the \hi,
the mean \hi\ column density \nhi, the mean dust absorption optical depth \taua,
the clump volume filling factor FF,  and the  density contrast \nratio\ between the interclump and clumpy medium.
Each parcel of the shell (clump or interclump) exhibits the same radial velocity \vexp.
The Doppler parameter $b = \sqrt{v_{\rm th}^2 + v_{\rm turb}^2}$ reflects the random (thermal +turbulent) 
motions of the \hi. The clumpy (inhomogeneous) medium is defined by the volume filling factor FF of clumps, by their density \nc, and by the density contrast \nratio\
between clumps and interclumps of lower density \nic. The mean \hi\ column density \nhi\ is thus related to the (inter)clump density, FF, and the thickness of the shell $L$ by:
\begin{equation}
  \nhi = ({\rm FF} \nc + (1-{\rm FF}) \nic) L.        
\end{equation}
Similarly one has
\begin{equation}
  \taua = (1-a) \sigma_d  (\frac{m_H}{m_d}) (\frac{M_d}{M_H}) \nhi,
\end{equation}
where $a$ is the dust albedo, $\sigma_d$ the total dust cross section (scattering + absorption),
$m_H$ the proton mass, $m_d$ the dust grain mass,
and ($M_d$/$M_H$) is the dust-to-gas ratio. In the present paper, the dust optical depth \taua\ 
--- the single parameter used to vary the dust content --- 
is derived assuming a dust grain size of 2$\times$$10^{-6}$ cm and a mass $m_d$ = 3$\times$$10^{-17}$ g. 
The total dust optical depth is defined as  
\begin{equation}
\overline{\tau_d}=\frac{\taua}{(1-a)},
\end{equation}
and the dust particle density is 
\begin{equation}
n_d = n_{\rm H} (\frac{m_H}{m_d})  (\frac{M_d}{M_H}), 
\end{equation}
where $n_{\rm H}$ stands for the clump or interclump density. We adopt the SMC dust properties (albedo $a$ and phase function $g$) listed in Table 2. 
These properties, together with the clumpy shell geometry also adopted here, have been shown to reproduce
observable continuum properties of starburst galaxies \citep{gordon97,witt00,vijh03}.
Although detailed model predictions depend to some extent on the
dust properties, the main quantities of interest in this paper
-- the \lya\ and UV continuum escape fractions, and especially
their {\em relative} values -- should not strongly depend
on the exact dust properties. We expect that other poorly known
properties such as the geometry and velocity field, known to
affect sensitively the transfer of \lya\ radiation, are more
important than the detailed dust properties \citep{larsen09}. For these reasons
we have not considered changes of the dust properties, but focus
on the effect of geometry and clumpiness in this paper.

To construct clumpy structures with the desired input parameters in practice, we follow 
a similar approach as \cite{witt00}.
We construct a Cartesian grid of $N=128^3$ cells, within which the shell of thickness $L$
is defined by an inner and outer radius, $R_{\rm min}$ and $R_{\rm max}$.
Assuming a density \nc\ for the high density regions (clumps) 
we then randomly choose a fraction FF of the cells localised in the shell (i.e.\ cells localised at a radius $R$ 
such as $R_{\rm min}$ $\le$ R $\le$ $R_{\rm max}$), which receive a high density \nc. The remaining cells in the 
shell are set to low density \nic. The physical cell size (or equivalently $L$) is then adjusted to 
reproduce the desired mean radial \hi\ column density \nhi, which is computed by drawing random
lines of sight through the shell. Finally the dust content is varied by changing the dust-to-gas ratio ($M_d$/$M_H$), 
yielding different values of the mean dust absorption optical depth \taua.

\begin{table}
\caption{Six input parameters (top) for and derived parameter (bottom) of the homogeneous and clumpy shell models}
\label{table:input}      
\centering                          
\begin{tabular}{lllc c c c c c}        
\hline\hline                 
Parameter & Symbol \\
\hline 
Radial velocity & \vexp  \\ 
\hi\ velocity dispersion & $b$ \\
Mean \hi\ column density & \nhi \\
Mean dust absorption optical depth & \taua \\
Clump volume filling factor & FF \\
Density contrast & \nratio \\
\hline
Covering factor & CF \\
 \hline 
\end{tabular}
\end{table}

\begin{table}[h]
\caption{Dust parameters ($a$ and $g$) taken from \cite{witt00} and adopted for \lya\ line photons, and continuum photons at UV and optical wavelengths (close to the B and V-band). }
\label{table:dust}      
\centering                          
\begin{tabular}{c c c c c c}        
\hline\hline                 
Photons &  & $\lambda$ (\AA) & $\tau_d$/$\tau_V$ & $a$ & $g$\\    
\hline                        
\lya & & 1215.67 & & & \\[-1ex]
 & & & \raisebox{0.5ex}{6.74} & \raisebox{0.5ex}{0.460}&\raisebox{0.5ex}{0.770}\\[-2ex]
UV & & 1235.0 & & & \\
\hline                 
   B-band          &  & 4350.0 & 1.38  & 0.495  & 0.633 \\
   V-band           &  & 5550.0 & 1.00 & 0.490  & 0.607 \\
\hline                                   
\end{tabular}
\end{table}

\begin{table*}
\caption{Range of values of the six input parameters (column 1-6) describing the homogeneous and clumpy shell models, and derived properties
(cols.\ 7, 8). }             
\label{table:pars}      
\centering          
\begin{tabular}{c c c c c c c c c }     
\hline\hline
 $FF$ & \nratio & \vexp [km s$^{-1}$] & $b$ [km s$^{-1}$] & \nhi\ [cm$^{-2}$] & \taua & CF & mass spectrum\\ 
\hline                
0.23 &  [0, 1]  & 0, 50, 100, 200, 250, 300, 400, 600  & 12.8, 20, 40 & [$10^{17}$, $10^{22}$] &  [0, 100] & [0.7, 1] &  $\rho$(m) $\propto$ $m^{-2.04}$\\  
\hline                  
\end{tabular}
\end{table*} 

In Table \ref{table:pars} we summarise the different values that we have explored for the six input parameters describing
our models.
For the present study we have adopted a filling factor $FF= 0.23$, as explained below (Sect.\ 2.2.2). 
The density contrast \nratio\ has been varied from 1 (homogeneous medium) to 0, reflecting the extreme case of 
an empty interclump medium.
Models have been computed for static shells ($\vexp=0$) and expansion velocities up to $\vexp=600$ \kms.
Then, a wide range of parameter space has been considered, as listed in Table \ref{table:pars}.

\subsubsection{Characterisation of clumpy structures}

Given a choice of the clump volume filling factor FF and the thickness of the shell 
(i.e.\ $R_{\rm max}-R_{\rm min}$), two other interesting quantities describing the  inhomogeneous structure can be derived. First, the covering factor CF of the shell corresponding to the fraction of solid angle covered by the clumps as seen from the central source. Models with different covering factors are constructed by varying
$R_{\rm min}$.

Another interesting quantity is the mass spectrum of the clumps. 
As clumps we consider, as \cite{witt96, witt00}, all cells directly connected with each other by at least one face. 
We then determine their mass spectrum, which approximately follows a power law $\rho$(m) $\propto$ $m^{-\alpha}$, where $m$ is the clump mass.
Adopting FF = 0.23, we obtain a power law $\rho(m) \propto m^{-2.04}$ as illustrated in Fig.\ 2. This mass spectrum is consistent with observations of diffuse interstellar clouds showing a power law with $\alpha$ = 2 \citep{dickey89}. The value FF=0.23 in our model is then the most appropriate value if we aim to reproduce the interstellar mass spectrum of nearby galaxies. 
We illustrate in Fig.\ 2 the mass spectrum obtained with FF$=0.23$ (red curve) in a shell geometry defined with $R_{\rm min}$ = 49 and $R_{\rm max}$ = 64 cells. The slope of the mass spectrum does not change noticeably decreasing the covering factor CF (ie.\ decreasing $R_{\rm min}$).

Let's mention that some models with other mass spectra (i.e. other filling factors FF) have been studied, such as $\rho$(m) $\approx$ $m^{-2.70}$ and $\rho$(m) $\approx$ $m^{-3.17}$. However, no notable change is found in any of our results changing only the mass spectrum in clumpy shell structures.

  \begin{figure}
   \centering
   \includegraphics[width=92mm]{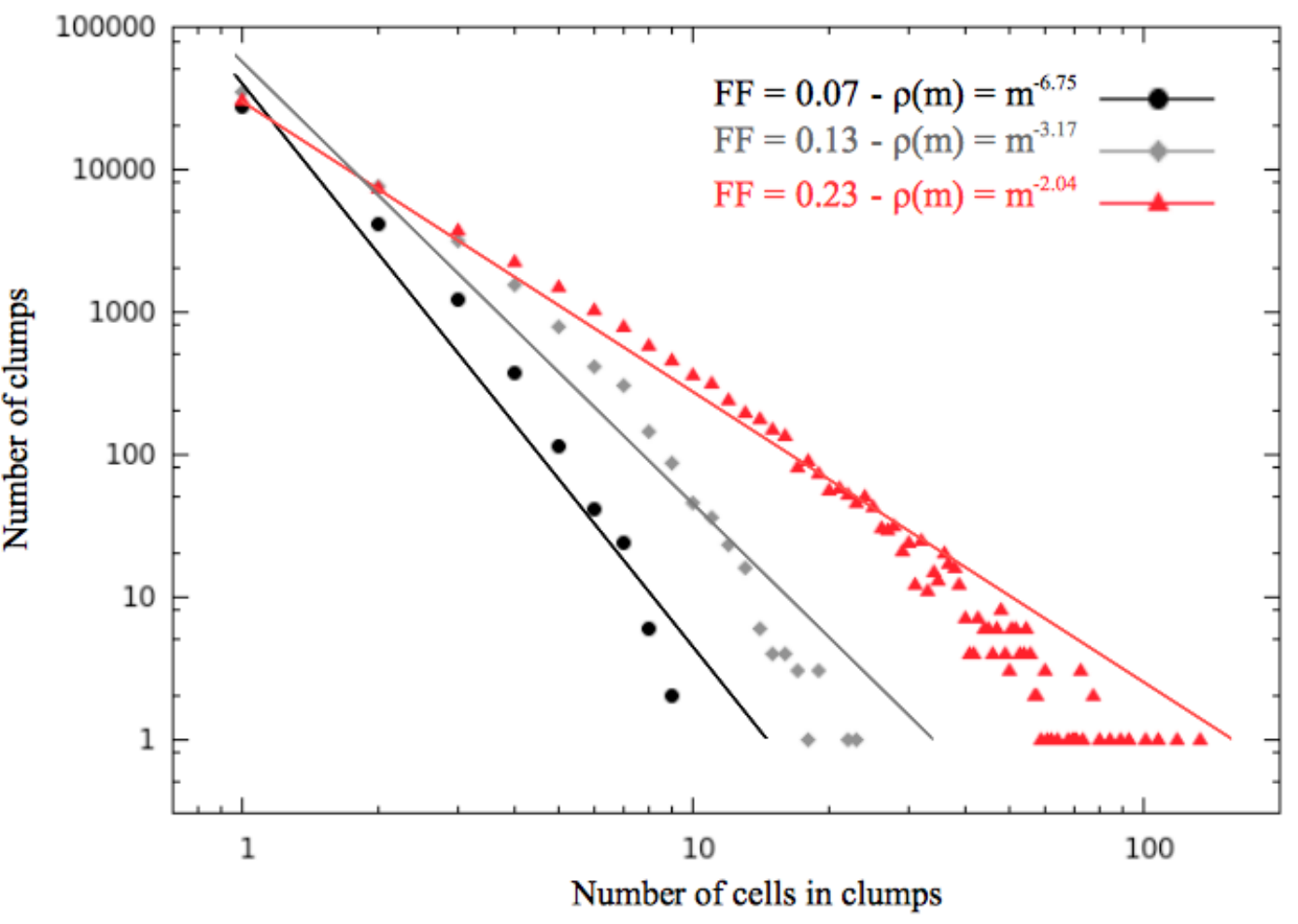}
      \caption{Variation of clump size adopting different filling factors FF in a shell geometry defined with $R_{\rm min}$ = 49 and $R_{\rm max}$ = 64 cells. The filling factors used here are: FF= 0.07 (circles), FF = 0.13 (diamonds) and FF = 0.23 (triangles). The mass spectrum obtained with FF =0.23 ($\rho(m) \propto m^{-2.04}$) is the most consistent with observations of diffuse interstellar clouds. We thus adopt FF = 0.23 throughout this present study.
        }
         \label{FigVibStab}
   \end{figure}

\subsubsection{Input spectra}

In the region close to \lya\ we assume that the spectrum consists of a flat UV continuum 
(i.e.\ constant in number of photons per frequency interval)
plus the \lya\ line, characterised by a Gaussian with an equivalent width EW$_{\mathrm{int}}(\mathrm{Ly}\alpha)$ and 
Full Width at Half Maximum FWHM$_{int}$(\lya). All photons are isotropically emitted from the center of our shell geometries.

Throughout this work, we adopt $FWHM_{int}$(Ly$\alpha$) = 100 km s$^{-1}$, as a typical value for the intrinsic width of the H recombination lines emitted in the ionized gas,
observed in both nearby and distant starburst galaxies. Indeed, based on observations of the velocity dispersion in starburst galaxies, this line width is comparable to the values measured from the velocity dispersion of CO and H$\alpha$ lines in the starburst galaxy 
cB58 \citep{teplitz00,baker04}, the velocity dispersion measured in several starbursts at z $\sim$ 2 by \cite{erb03} and in SMM J2135-0102 at z = 2.32 by \cite{swinbank11}. Furthermore, different values of $EW_{\rm int}$(Ly$\alpha$), specified below if necessary, have been adopted.




\subsubsection{Output parameters}
For the present paper we are interested in the following quantities predicted by our Monte Carlo 
simulations: the average \lya\ escape fraction \flya, the \lya\ line profile and the escape fraction of continuum photons at UV and other wavelengths \fesc.
From these we also derive the observed \lya\ equivalent width $EW_{\rm obs}$(\lya), and the colour excess $E(B-V)$.
All quantities are computed from spatially integrating all the photons escaping our 
spherically symmetric shells. They therefore correspond to average properties
for our homogeneous and clumpy structures. 

In the present study, the \lya\ escape fraction is computed from
\begin{equation}
\flya = \frac{\int_{-\infty}^\infty \fesc \phi(\lambda) d\lambda}{\int_{-\infty}^\infty \phi(\lambda) d\lambda},
\end{equation}
where $f_{\rm esc}(\lambda)$ is the monochromatic escape fraction computed for typically 1000--2000 frequency points around the line
center with a spacing of 20-10 \kms, and $\phi(\lambda)$ describes the input line profile.
The predicted \lya\ line profile can be computed {\em a posteriori} from our simulations
for arbitrary input spectra (line + continuum), as described in Verhamme et al. (2006). \flya\ is slightly dependent on the FWHM of the input line profile, but independent on the value of $EW_{int}$(Ly$\alpha$).

The UV continuum escape fraction \fescuv\ is computed redward of the \lya\ line.
Assuming the \cite{calzetti00} attenuation law we can compute the corresponding
colour excess as  
\begin{equation}
    E(B-V)_{{\rm Calzetti}} = \frac{- 2.5}{k(1235)} \log{(\fescuv)},  
    \label{eq:calz}
    \end{equation} 
with $k(1235) = 11.4$ according to the Calzetti law. 

From the escape fraction of radiation at the optical wavelengths listed in Table \ref{table:dust}
we can also determine the true colour excess
\begin{equation}
  E(B-V)_{{\rm real}} = -2.5 \log{\left(\frac{f_{esc}(V)}{f_{esc}(B)}\right)}.
      \label{eq:real}
\end{equation} 
In practice this is done by calculating the continuum escape fractions at 4350 \AA\ for the $B$ band and at 5500 \AA\ for the $V$ band (Table \ref{table:dust}).

The observed \lya\ equivalent width and the intrinsic one (i.e.\ input value of the source, before radiation transfer)
are related by
\begin{equation}
\label{eq:ew}
  \frac{{\rm EW}_{{\rm obs}}({\rm Ly}\alpha)}{{\rm EW}_{\rm int}({\rm Ly}\alpha)} = \frac{\flya}{\fescuv},
  \end{equation}
where $EW_{int}$(Ly$\alpha$) is the intrinsic \lya\ equivalent width. 

\subsubsection{Validation}
\begin{figure}
   \centering
   \includegraphics[width=91mm]{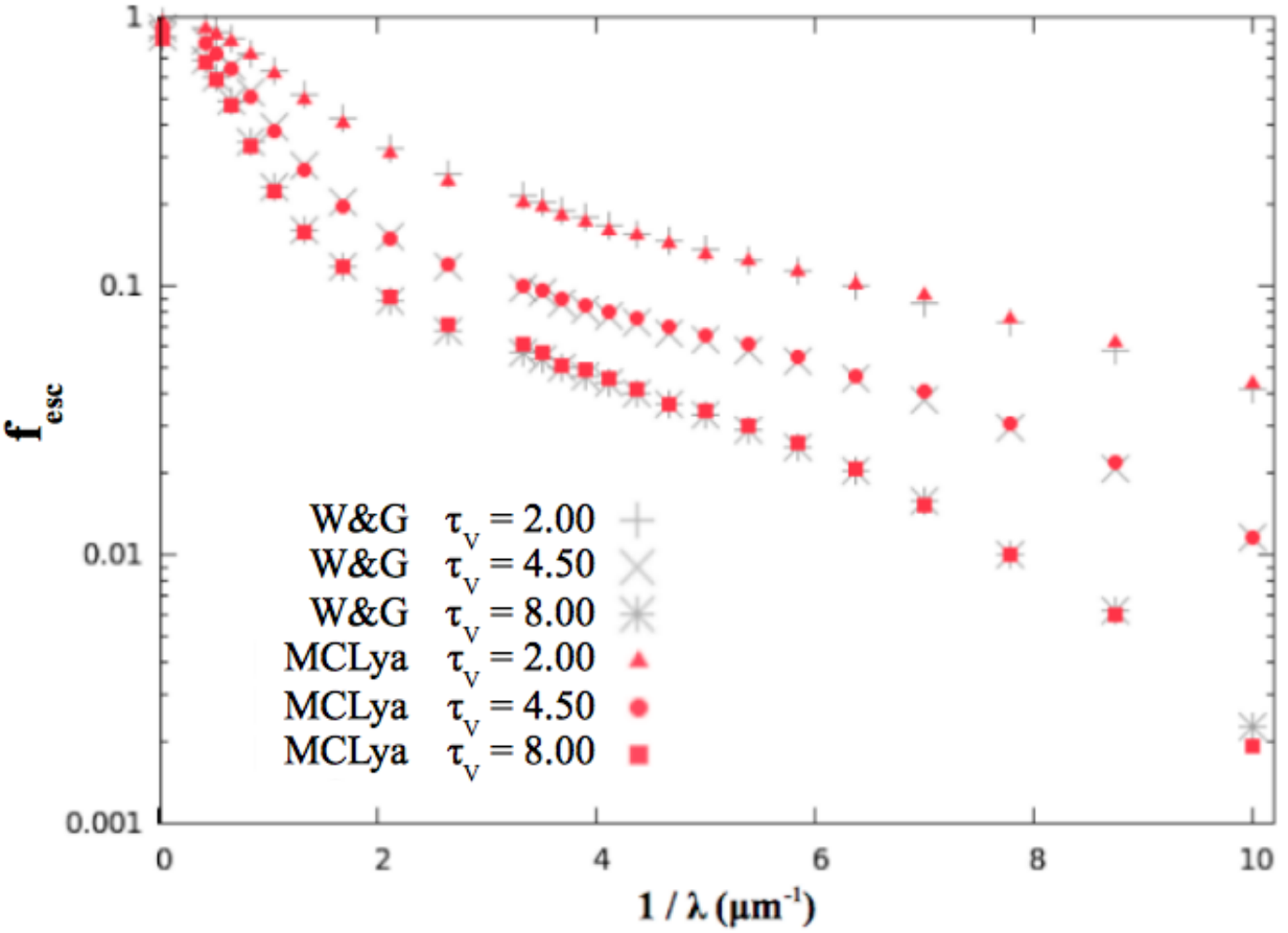}
      \caption{Comparison of the UV-to-optical continuum escape fraction derived from our code with the results of \cite{witt00}. The clumpy shell geometry studied here is built on a cartesian grid of $N = 30^3$ cells and is characterized by the following parameters: FF = 0.15, $\nratio = 0.01$, $R_{\rm min} = 5$ and $R_{\rm max} = 15$. The evolution of the escape fraction at 25 different wavelengths is shown, from $\lambda = 0.1$ $\mu$m to $\lambda = 30$ $\mu$m, assuming three different dust optical depths $\tau_V$ in the shell structure (measured in V band): $\tau_V$ = 2, 4.5 and 8. The escape fractions obtained with MC\lya\ are marked in red, whereas those obtained by \cite{witt00} are marked in grey. We find a very good agreement with \cite{witt00}.
              }
         \label{fig:3}
   \end{figure}

To test the radiation transfer, we have compared our results to \cite{witt00}, whose dust parameters
are adopted in our calculations. We have constructed a clumpy shell model using the same discretisation and input parameters. The derived mass spectrum
is in good agreement with these authors. The resulting continuum escape fractions, shown in Fig.\ \ref{fig:3}, and other results are found in excellent 
agreement with \cite{witt00}, which validates our code. 

\section{The Ly$\alpha$ and UV continuum radiation transfer in homogeneous and clumpy media}

In this section we study the radiative transfer of \lya\ and the UV continuum photons in homogeneous and clumpy shell geometries. 

\subsection{The UV continuum escape fraction}

\begin{figure}
   \centering
   \includegraphics[width=88mm]{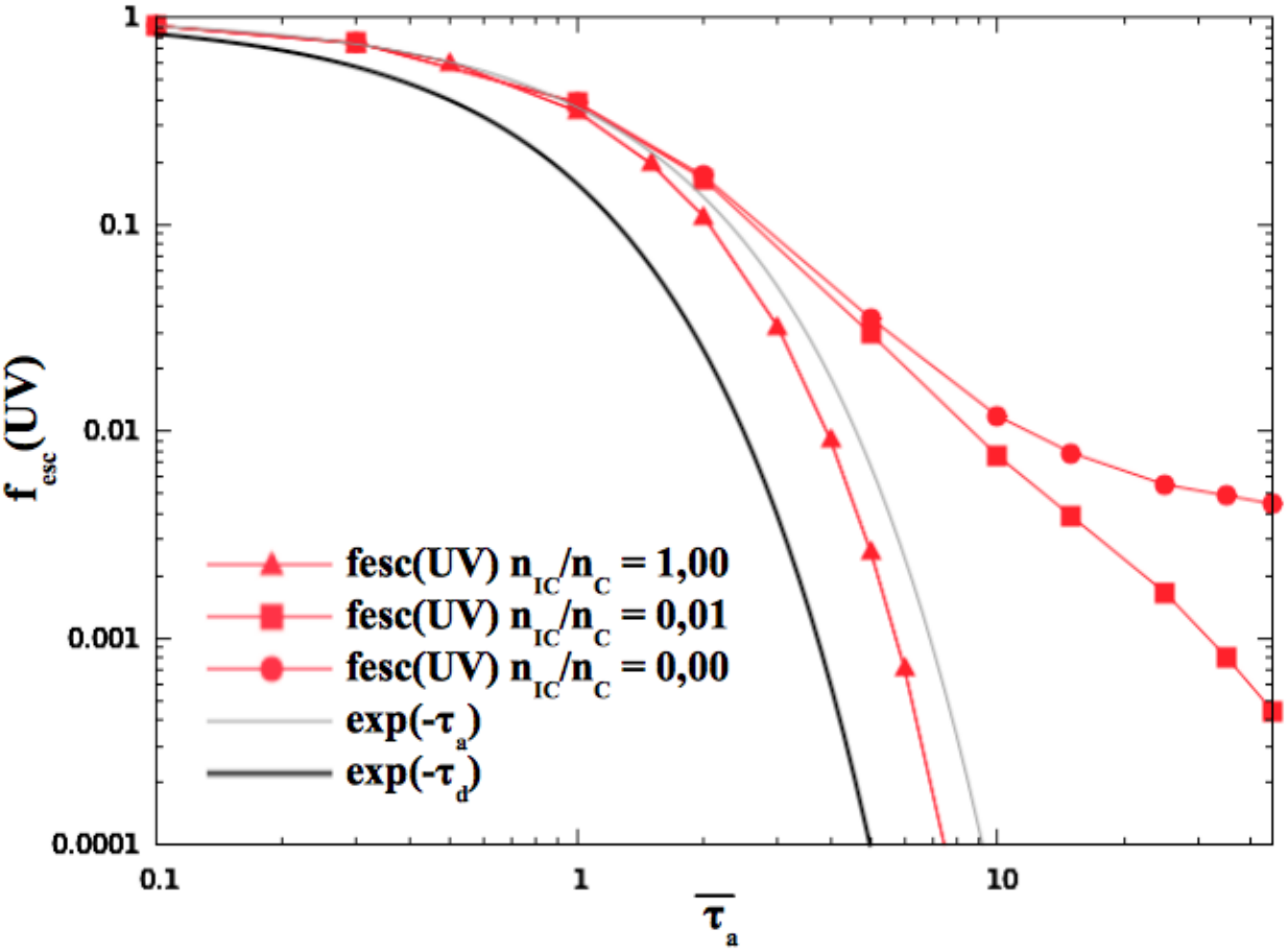}
      \caption{Evolution of the UV escape fraction \fescuv\ as a function of the dust optical depth (\taua) in three different shell geometries: a homogeneous shell ($n_{IC}$/$n_C$ = 1.00), a weakly clumpy shell ($n_{IC}$/$n_C$ = 0.01) and an extremely clumpy shell ($n_{IC}$/$n_C$ = 0.00). The clumpy media are defined with FF = 0.23 and CF = 0.997 ($R_{\rm min}$ = 49 and $R_{\rm max}$ = 64 cells). 
For comparison we also plot the curves for $f_{\rm esc} = \exp(-\taua)$ and $f_{\rm esc} =\exp(-\tau_d)$, which describe respectively the upper and the lower limit of the continuum
escape fraction in homogeneous media.
            }
         \label{fesc_UV}
   \end{figure}

In Figs.\ \ref{fesc_UV} and \ref{Lya_UV_CF}, we examine the evolution of the UV escape fraction \fescuv\ in homogeneous and clumpy media. As shown in both figures, \fescuv\ depends on three main parameters: \\
\\
1) the dust content (\taua) \\
2) the clumpiness of the dust distribution, assumed to trace the \hi\ distribution (\nratio) \\
3) the covering factor (CF) \\

The other parameters describing our shell geometries, the outflows (\vexp), the \hi\ column density (\nhi) and the temperature of the matter ($b$) do not show any effect on the UV escape fraction, as expected.

Figure \ref{fesc_UV} illustrates the effects produced by both the \textit{dust content} (i.e.\ \taua) and the \textit{clumpiness of the dust distribution} (i.e.\ $n_{IC}$/$n_C$) on \fescuv. We adopt here three media with the following conditions: FF = 0.23, CF = 0.997 and $n_{IC}$/$n_C$ = 1.0, 0.01, 0. Qualitatively, we can summarize the effects produced by both \taua\ and $n_{IC}$/$n_C$ on \fescuv\ in the following way: \\  
\\
- \textbf{\taua}: in homogeneous and clumpy media, an increase in the dust optical depth \taua\ always produces a decrease in the UV escape fraction. \\
\\
- \textbf{$n_{IC}$/$n_C$}: a decrease in $n_{IC}$/$n_C$ from 1 to 0 (i.e.\ from a homogeneous to an extremely clumpy dust distribution) always increases the UV escape fraction. A clumpy dust distribution produces indeed higher UV escape fractions compared with an equivalent homogeneous distribution of equal dust content (i.e.\ equal \taua). \\
\\
Clumpy media are thus more transparent to UV continuum radiation, as  previously shown by \cite{boisse90, hobson93, witt96, witt00}. The facts that the dust content concentrates in clumps and that the interclump medium becomes more optically thin, allow UV photons to escape any clumpy media in two different ways \citep{witt96}: 
first, like in homogeneous dusty media, UV photons have to scatter against few dust grains before escaping clumpy media (dust localized in clumps or in between clumps). But, in clumpy media, UV photons take advantage of the weak opacity of the inter-clump medium, which allows them to escape more easily clumpy media than any homogeneous dusty geometry. 
Second, continuum photons can also directly escape clumpy media if several free spaces appear between clumps. However, that can be only possible in extremely clumpy media ($n_{IC}$/$n_C$ $\approx$ 0). In this case, continuum photons are not affected by the weak dust content localized between clumps and can directly escape clumpy media getting through holes which appear between clumps. 

The covering factor CF is thus an important parameter controlling the UV escape fraction in clumpy media. Figure \ref{Lya_UV_CF} shows this dependence in the particular case of extremely clumpy shell geometries ($n_{IC}$/$n_C$ = 0). As expected, the UV escape fraction \fescuv\ decreases when the covering factor increases to unity (i.e.\ all lines-of-sight are covered by one or more clumps from the photon source when CF = 1). Furthermore, in the particular case of extremely clumpy shell geometries ($n_{IC}$/$n_C$ = 0), we can also notice that the covering factor CF provides a general lower limit for \fescuv. As shown in Figs. \ref{fesc_UV} and \ref{Lya_UV_CF}, the UV escape fraction always converges on an asymptote $\fescuv = 1 - {\rm CF}$, corresponding to the direct escape fraction.

Besides this qualitative approach concerning the dependence of \fescuv\ to \taua, $n_{IC}$/$n_C$ and CF, Fig.\ \ref{fesc_UV} illustrates other quantitative results: 
in homogeneous geometries the UV escape fraction decreases very rapidly with the dust optical depth  \taua. If we define \fescuv\ = $e^{-\tau_{\rm eff}}$, the effective optical depth
$\tau_{\rm eff}$ is equal to \taua\ in the absence of scattering. With scattering the effective absorption increases, and one has $\tau_{\rm eff} > \taua$. In clumpy media the situation is different as photons can escape more easily, hence $\tau_{\rm eff} < \taua$.

   The escape fraction of the optical continuum photons evolves in the same way as for the UV photons in homogeneous and clumpy media. Combining the escape fraction of both the B and the V-band in the same media as those studied in Fig.\ \ref{fesc_UV}, we illustrate in Fig.\ \ref{fig:ebv}  the evolution of the derived colour excess E(B-V). This figure can be used to translate the dust optical depth \taua\ of Fig.\ \ref{fesc_UV}  in terms of colour excess E(B-V). 


\subsection{The \lya\ radiative transfer in homogeneous and clumpy media: two regimes appear} 
\label{s:2regimes}

Besides the three main parameters which control the radiative transfer of the UV continuum photons (i.e. \taua, $n_{IC}$/$n_C$ and CF), three other parameters also determine 
the radiative transfer of the resonant scattered \lya\ photons, namely \vexp, \nhi\ and $b$.
We now discuss the influence of these parameters on the UV continuum and on \lya. For simplicity we here assume a constant value of $b$ in all cells (clump or interclump).
Overall we find that we can identify two regimes where the \lya\ propagation is quantitatively different, which we now explain.
The separation between the regimes will be discussed after that (Sect.\ \ref{sect:crit}).



\subsubsection{The ``low contrast" regime: homogeneous and weakly clumpy media}

\begin{figure}
   \centering
   \includegraphics[width=80mm]{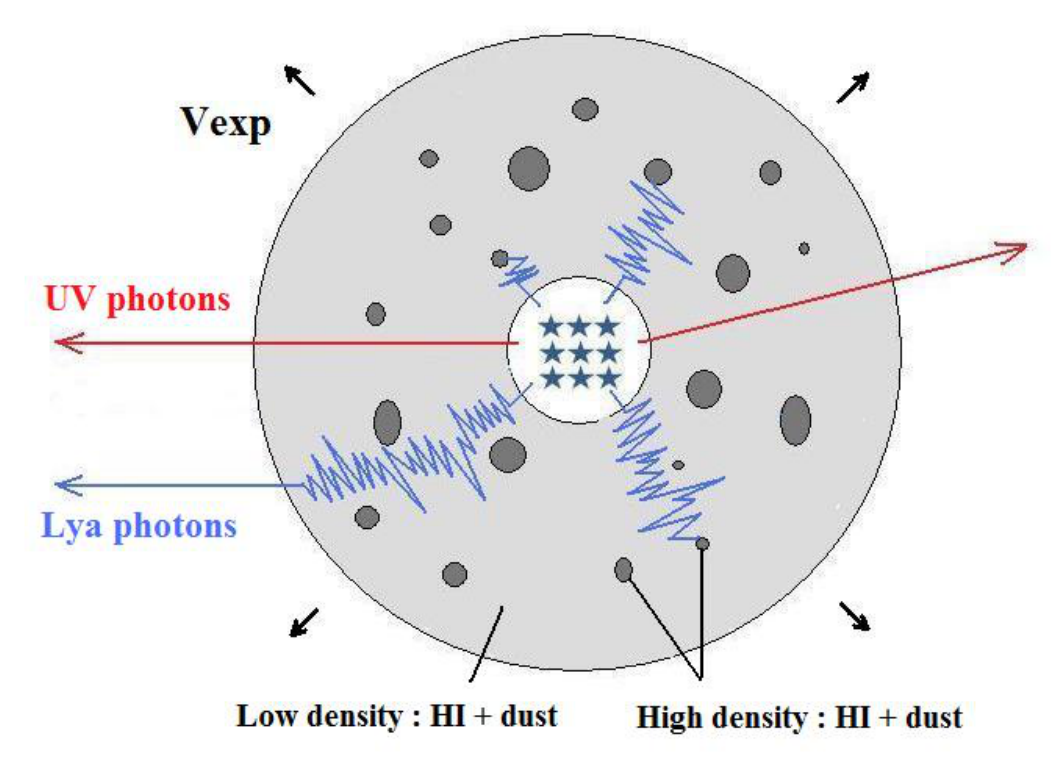}
      \caption{Schematic representation of the way of propagation of UV continuum (red) and \lya\ photons (blue) in the ``low contrast" regime. The medium illustrated here is composed of high density clumps (containing \hi\ + dust), distributed in an interclump medium of low density (HI+dust). In the ``low contrast" regime, the \hi\ content in and in between clumps are relatively high, which renders both regions optically thick for the \lya\ photons. In this way, the \lya\ photons can only escape the medium after undergoing multiple resonant scattering against \hi\ atoms, increasing their probability being absorbed by the dust. The UV photons are not affected by the presence of \hi\ atoms and propagate directly through the medium.
              }
         \label{propagat_low_contrast}
   \end{figure}

\textbf{Propagation of \lya\ photons in the ``low contrast" regime} \\
\\
We show in Fig. \ref{propagat_low_contrast}  the typical way \lya\ photons propagate in the ``low contrast" regime. We deduce this propagation from our numerical simulations, studying the number and the location of each interaction between the \lya\ photons and the HI atoms in clumpy media.
In this regime, we notice that the \lya\ radiative transfer is characterised by a (pseudo-)random walk in the medium, both in and in between clumps. \\
\\
\textbf{\lya\ escape fraction \flya}: \\
\\
  \begin{figure}
   \centering
   \includegraphics[width=90.5mm]{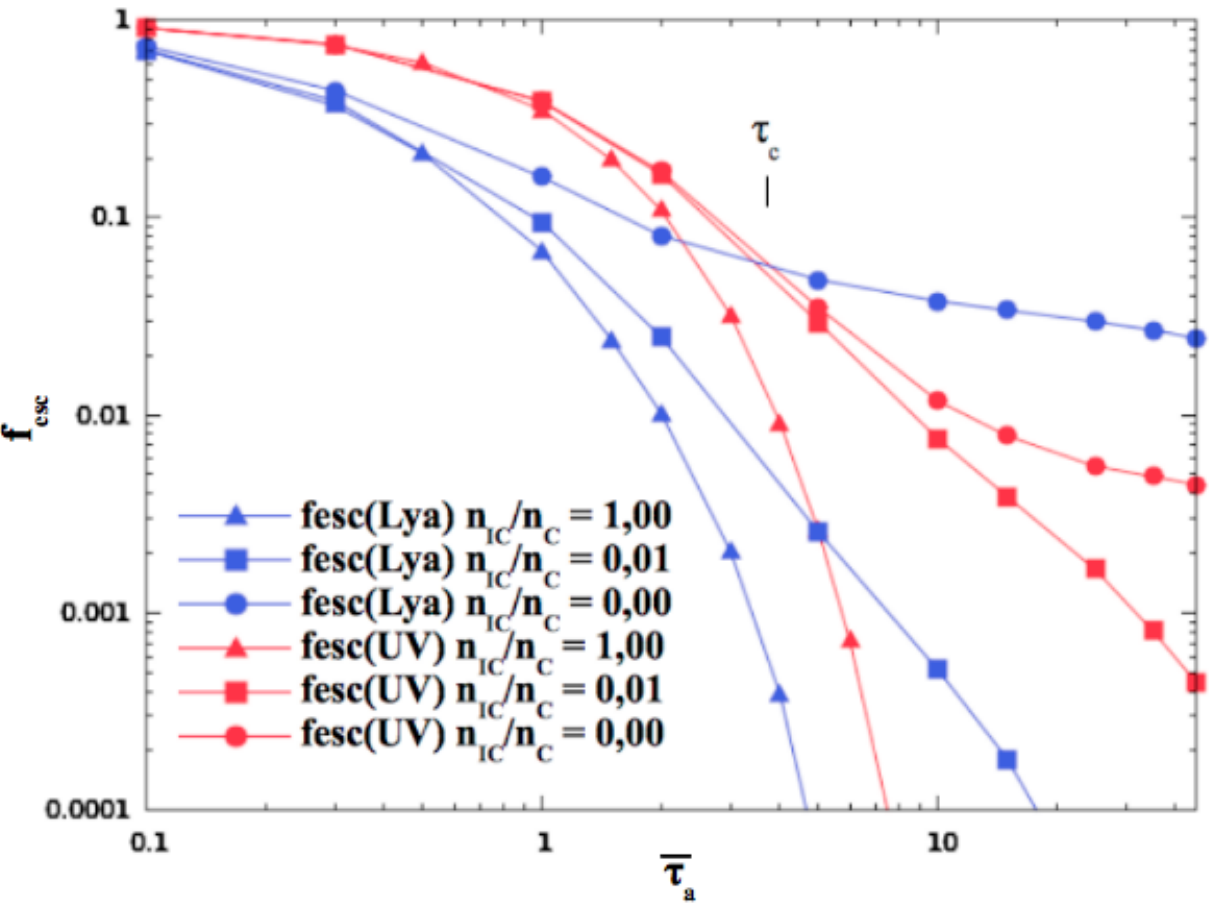}
      \caption{Evolution of the \lya\ (blue lines) and UV continuum (red) escape fraction  as a function of the dust optical depth (\taua) for a homogeneous (triangles), a weakly clumpy (squares), and an extremely clumpy (spheres) medium. In these simulations, we adopt FF = 0.23, CF = 0.997 ($R_{\rm min}$ = 49 and $R_{\rm max}$ = 64 cells), \nhi\ = $10^{19}$ \cm2\ , \vexp\ = 0 km s$^{-1}$ and b= 40 km s$^{-1}$. For comparison, the UV escape fractions (Fig. 4) are overlaid (red lines). For the extremely clumpy medium we note that for \taua\ larger than a certain limit $\tau_c$ (here roughly equal to 3.5), \fescuv\ drops below \flya. 
              }
         \label{Lya_UV_Tau}
   \end{figure}
Figure \ref{Lya_UV_Tau} shows the dependence of \flya\ on the dust content (\taua) of the medium, for different values of clumpiness ($n_{IC}$/$n_C$).
The ``low contrast" regime includes all curves of this figure, except the particular case $n_{IC}$/$n_C$ = 0.00 which belongs to the "high contrast" regime. Note that the quantity \taua\ can be related to the colour excess through Fig. \ref{fig:ebv}.

Qualitatively, we see that \flya\ always decreases with increasing \taua\, as expected. The same is true for increasing $n_{IC}$/$n_C$. However, we see that for a given value of \taua, increasing $n_{IC}$/$n_C$ results in a faster decrease of \flya\ than \fescuv, the reason being the highly increased path length of \lya\ photons due to resonant scattering. Thus, in the ``low contrast" regime \lya\ radiation is more vulnerable to dust than UV continuum radiation.

A change of the covering factor CF has also a noticeable effect on \flya\ in the ``low contrast" regime (CF measuring the proportion of holes which appear between clumps). As the clumps cover an increasing fraction of the sky, it becomes indeed increasingly difficult for the photons to escape, and when CF $\approx$ 1, \flya\ drops drastically.



\begin{figure}
   \centering
   \includegraphics[width=89mm]{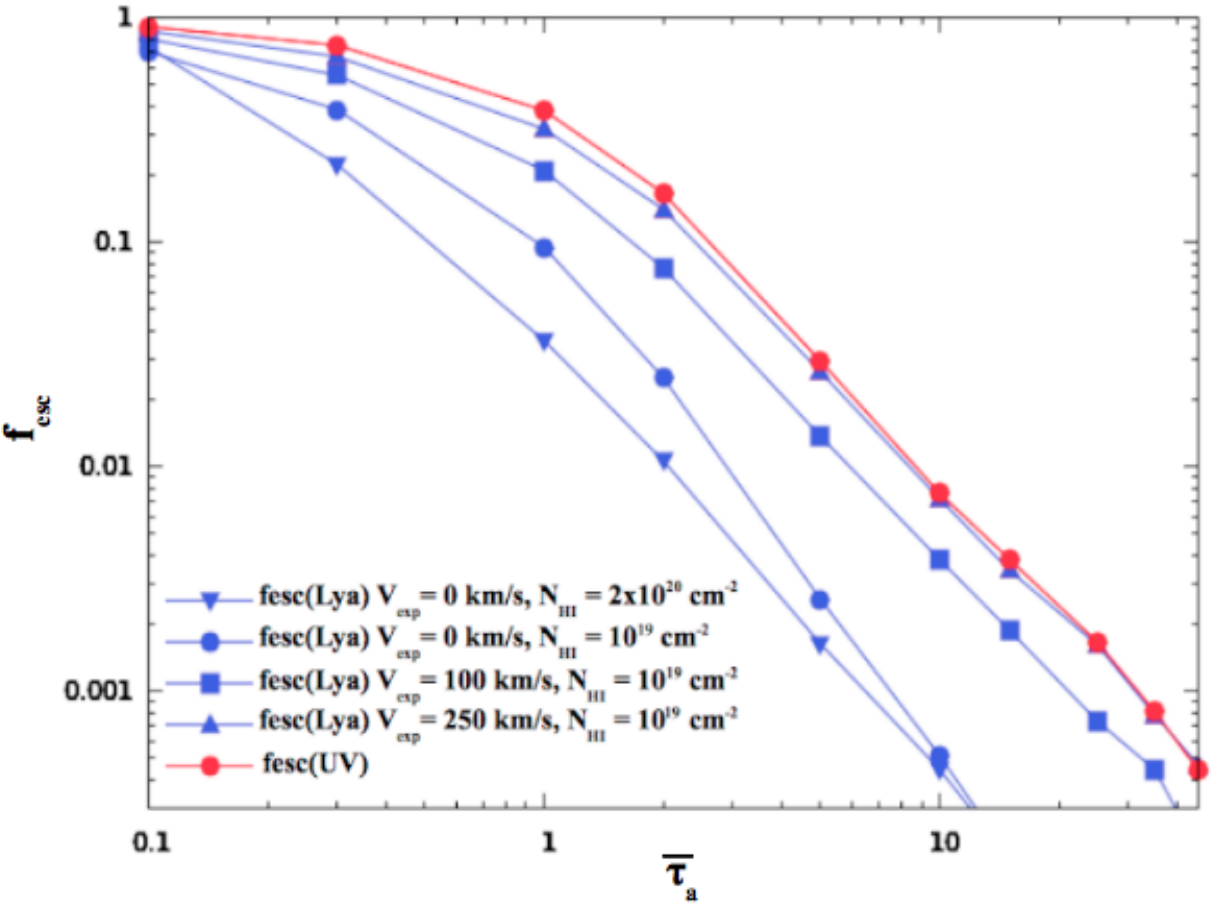}
      \caption{Evolution of the \lya\ escape fraction \flya\ as a function of dust optical depth \taua\, for various values of expansion velocity \vexp\ and mean neutral hydrogen column density \nhi.
The clumpy medium studied here is built with: FF = 0.23, CF = 0.997 ($R_{\rm min}$ = 49 and $R_{\rm max}$ = 64 cells), $n_{ic}$/$n_c$ = 0.01 and b= 40 km s$^{-1}$. The expansion velocities \vexp\ and \hi\ column densities that we adopt are: \vexp\ = 0, 100 and 250 km s$^{-1}$ and \nhi\ = $10^{19}$ \cm2\  and $2$$\times$$10^{20}$ \cm2\ . It is seen that \flya\ increases with increasing \vexp,as well as with decreasing \nhi.
}
         \label{Lya_Vexp_NHI}
   \end{figure}

Finally, the effect of a change of \vexp\ and \nhi\ on \flya\ is shown in Fig. \ref{Lya_Vexp_NHI}. We notice that \flya\ always increases with increasing \vexp\, as well as with \nhi\ decreasing. Since the effect of the expansion velocity is to shift the \lya\ photons away from the line center, they undergo progressively fewer scattering as \vexp\ increases. In fact, for an intrinsic \lya\ line width of $FWHM_{int}$(Ly$\alpha$) (this value is 100 km s$^{-1}$ in Fig. \ref{Lya_Vexp_NHI}),  a galactic outflow showing a velocity \vexp $\ga$ 2 $\times$ $FWHM_{int}$(\lya) is enough to allow \lya\ photons escape as easily as UV continuum photons. As expected, an increase in \nhi\ always leads to a decrease in \flya, since more neutral hydrogen implies more scatterings, and hence an increased total path length before escape, resulting in an increased probability of being absorbed.

Quantitatively, both Figs.\ \ref{Lya_UV_Tau} and \ref{Lya_Vexp_NHI} allow us to generalize the fact that, in the ``low contrast" regime, we always obtain:
   \begin{equation}
   \flya \le \fescuv
   \end{equation}
The strict equality \flya\ = \fescuv\ is only met when \lya\ photons are able to avoid scatterings altogether, i.e. for a sufficiently high expansion velocity, or for a very low \hi\ column density. \\
\\
\textbf{\lya\ equivalent width EW(\lya)}: \\
\\
Combining the definition of the \lya\ equivalent width (Eq.\ 8) and Eq.\ 9, we always have:
   \begin{equation}
   EW_{obs}(Lya) \le EW_{int}(Lya) 
   \end{equation}
In the ``low contrast" regime, the observed \lya\ equivalent width $EW_{\rm obs}(Lya)$ is thus always lower or equal to the intrinsic one $EW_{int}(Lya)$. In other words the \lya\ equivalent width is not ``boosted'' by clumping in this regime.

\subsubsection{The ``high contrast" regime: extremely clumpy shell geometries}

\begin{figure}
   \centering
   \includegraphics[width=80mm]{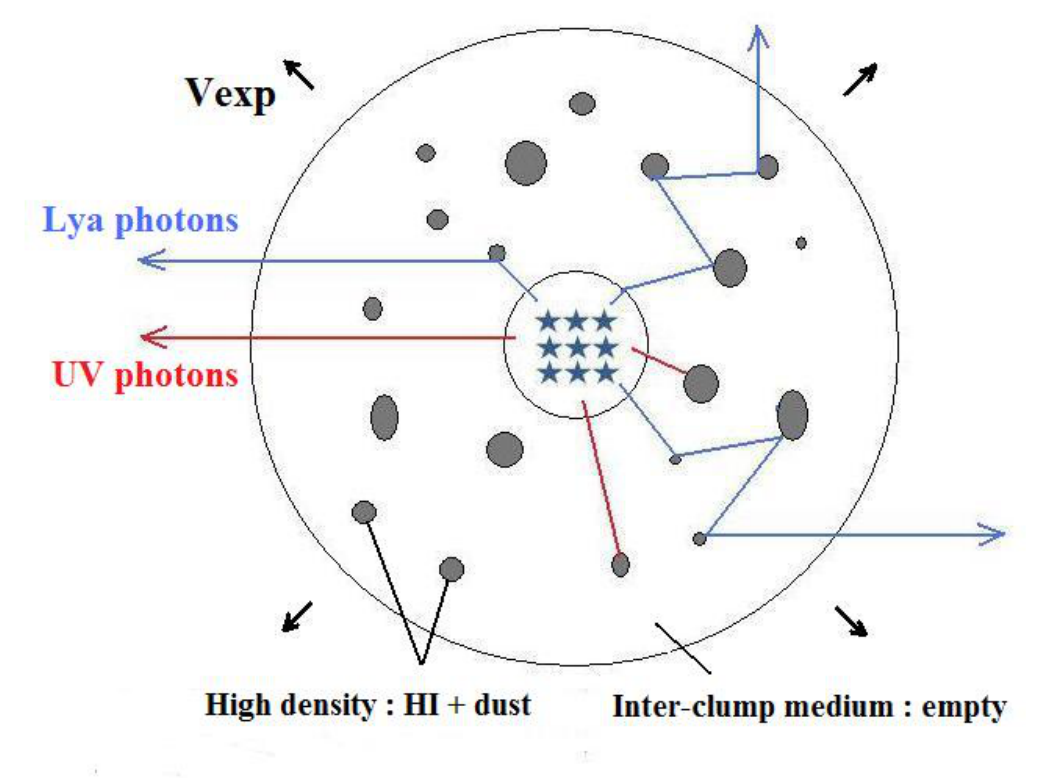}
      \caption{Schematic representation of the way of propagation of the UV continuum (red) and the \lya\ (blue) photons in the ``high contrast" regime. The medium illustrated here is composed by high density clumps (\hi\ + dust), distributed in an empty interclump medium. A fraction of the \lya\ photons scatter on the \hi\ atom comprising the surface of the clumps. The rest of the \lya\ photons, as well as the UV continuum photons, pierce the clumps where they may be absorbed by dust.
              }
         \label{propagat_Lya_high_cont}
   \end{figure}

The ``high contrast" regime of the \lya\ radiative transfer is defined by clumpy media showing a very low ratio \nratio, i.e.\ a high density contrast at least $\nratio\ \la 1.5\times10^{-4}$ for the input parameters considered throughout this study (see Sect.\ 3.3).
To describe the main effects and peculiarities of this regime we here restrict ourselves to the most extreme case with $\nratio=0$. \\
\\
\textbf{Propagation of \lya\ photons in the ``high contrast" regime}: \\
\\ 
The way  \lya\ photons propagate now differs qualitatively from the radiative transfer in the ``low contrast" regime. The details of the way we deduce the propagation of \lya\ photons in the ``high contrast" regime is given in Sect.\ 4, where we study the \lya\ line shape. We sketch in Fig.\ \ref{propagat_Lya_high_cont} the propagation of \lya\ photons in the ``high contrast" regime. The \hi\ content distributed between clumps is now weak enough that scattering between clumps can be neglected, and for a fraction of the \lya\ photons, the radiative transfer is characterised by rebounces on the clumps, as originally suggested by Neufeld (1991). The remaining \lya\ photons propagate in the same way as UV photons, that is penetrating the clumps, being exposed to the dust, or escape the medium freely (if CF $<$ 1).
\\
\\   
 \textbf{\lya\ escape fraction \flya}: \\
\\
Figure \ref{Lya_UV_Tau} shows  the evolution of \flya\ as a function of the dust content (i.e.\ \taua) in the ``high contrast" regime ($n_{IC}$/$n_C$ = 0). Again, note that the quantity \taua\ can be related to the colour excess through Fig.\ \ref{fig:ebv}. From Fig. \ref{Lya_UV_Tau} it is evident that \flya\ always decreases as \taua\ increases. However, the decrease is slower than that of \fescuv, allowing the curve of \flya\ to cross that	of \fescuv\ at a certain  optical depth $\tau_c$ ($\tau_c \approx 3.8$ in Fig. \ref{Lya_UV_Tau}). This is not possible in the ``low contrast" regime, and it is this quantitative difference that defines the threshold between the low and the high contrast regime. 

\begin{figure}
   \centering
   \includegraphics[width=90mm]{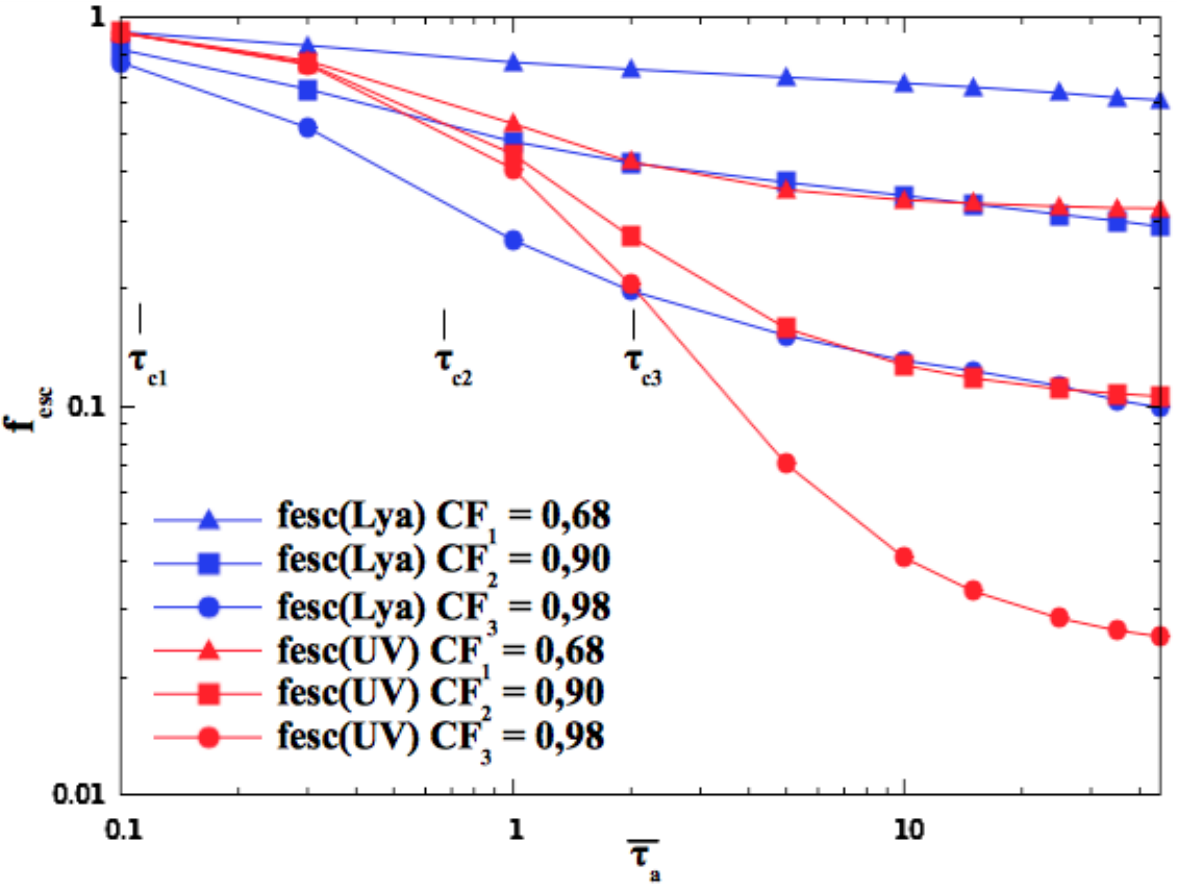}
      \caption{\lya\ (blue lines) and UV (red lines) escape fraction as a function the dust optical depth \taua\ for different values of covering factor CF in the ``high contrast" regime ($n_{IC}$/$n_C$ = 0.00). In this figure, all media have for common parameters: FF = 0.23, $n_{ic}$/$n_c$ = 0.00, \nhi\ = $10^{19}$ \cm2\, \vexp\ = 0 km s$^{-1}$ and $b =$ 40 km s$^{-1}$. 
The values $\tau_{c1}$, $\tau_{c2}$ and $\tau_{c3}$, corresponding to the values of \taua\ where the \lya\ and UV escape fractions of $CF_1$, $CF_2$ and $CF_3$ cross, are marked. While \fescuv\ always converges towards the limit \fescuv\ = 1 - CF in extremely clumpy media, \flya\ shows higher values.
              }
         \label{Lya_UV_CF}
   \end{figure}

Figure \ref{Lya_UV_CF} illustrates the effect of a change of the covering factor CF on \flya\ in the ``high contrast" regime. Again, we see that both \flya\ and \fescuv\ always decreases with increasing CF. However, whereas \fescuv\ approaches asymptotically the value $1-$CF (corresponding to all clumps being fully opaque to the UV so that escape is possible only through direct escape), \flya\ maintains a higher value. Furthermore, we can notice from the Fig.\ 9 that the value of the critical optical depth $\tau_c$ (where \flya\ crosses \fescuv) strongly decreases as CF decreases. 

\begin{figure}
   \centering
   \includegraphics[width=90mm]{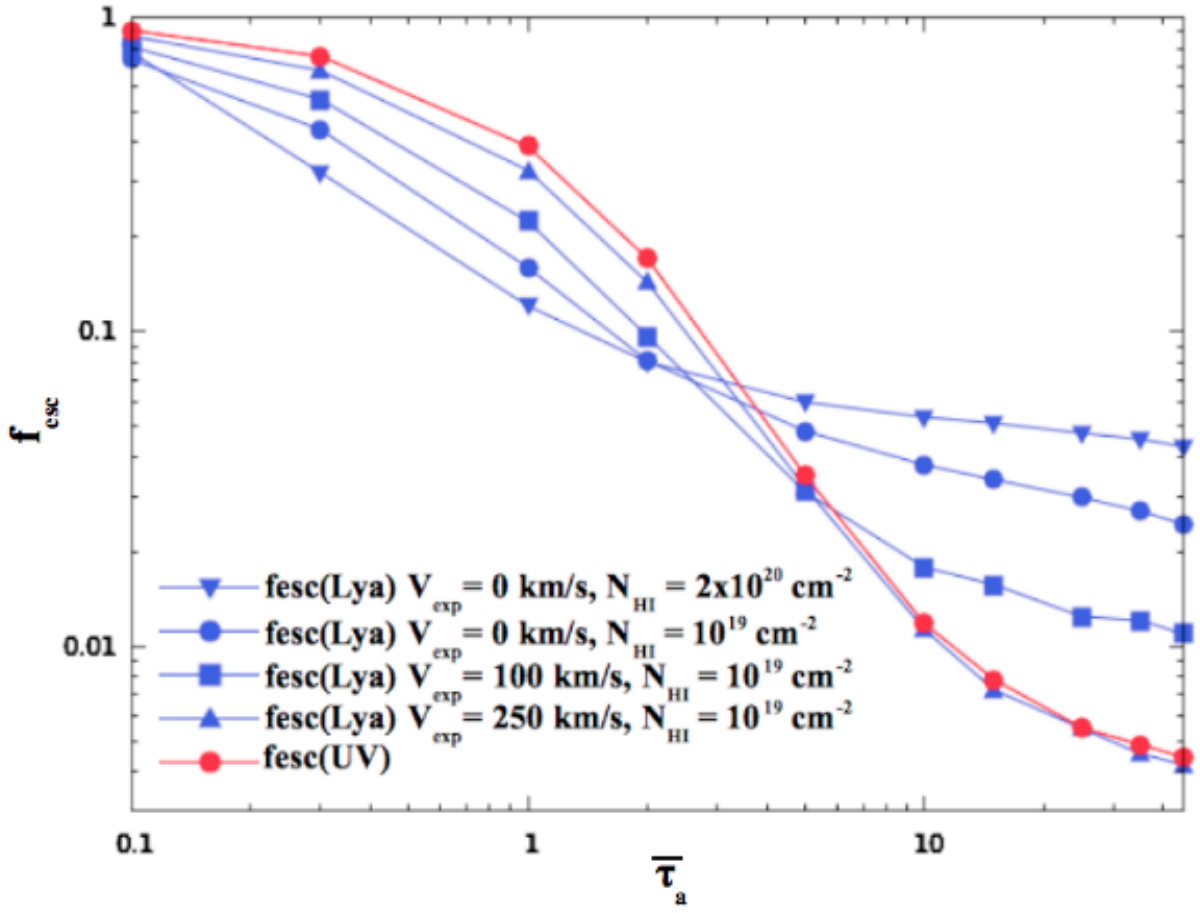}
      \caption{\lya\ (blue lines) and UV (red line) escape fractions as a function of the dust optical depth \taua\ in the ``high contrast" regime, for various values of expansion velocity \vexp\ and hydrogen column density \nhi. The extreme clumpy media have in common: FF = 0.23, CF = 0.997 ($R_{\rm min}$ = 49 and $R_{\rm max}$ = 64 cells), $n_{ic}$/$n_c$ = 0.00 and $b= 40$ km s$^{-1}$. We adopt as well the following expansion velocities \vexp\ and \hi\ column densities \nhi:  \vexp\ = 0, 100 and 250 km s$^{-1}$ and \nhi\ = $10^{19}$ \cm2\ and $2\times10^{20}$ \cm2. 
We indicate the critical dust optical depth $\tau_c$ where the curve of \flya\ for \vexp\ = 0 km s$^{-1}$ cross the curve of \fescuv. }
         \label{Lya_Vexp_high_contrast}
   \end{figure}

The effect of a change of \vexp\ and \nhi\ on \flya\ is shown in Fig. \ref{Lya_Vexp_high_contrast}. We can see that these effects are different and more complex than those of the ``low contrast" regime. More precisely, two different domains appear in the ``high contrast" regime, below and above the critical optical depth $\tau_c$ (where \flya\ crosses \fescuv). Below $\tau_c$, we notice that \flya\ increases with increasing \vexp\, as well as with \nhi\ decreasing. Such behavior is the same than those of the ``low contrast" regime. But above $\tau_c$, the opposite effect is observed, where an increase of \flya\ results from a decrease of \vexp\ and an increase of \nhi. These distinct behaviors can be understood as follows (see Fig. \ref{propagat_Lya_high_cont}):\\
\\
- $\taua \la \tau_c$: In this domain, \lya\ photons are more vulnerable to dust than UV photons. The dust content being relatively low, each clump is optically thin for UV radiation which allows UV photons to escape directly the medium getting through the clumps. However, \lya\ photons have to scatter off of the surface of a high number of clumps before escaping (Fig.\ 8), which increases the probability of being absorbed by the dust. An increase of \vexp\  increases the \lya\ escape fraction \flya. Indeed, increasing the expansion velocity \vexp, all \lya\ photons are Doppler shifted out of resonance, forcing them to pierce the clumps and thus to escape the medium as easily as UV photons. For the same reason, a decrease of \nhi\  increases \flya\ because decreasing the \hi\ density in clumps. It renders therefore clumps more transparent to \lya\ photons.  \\
\\
- $\taua \ga \tau_c$: In this domain, \lya\ photons are less affected by the dust than UV photons. Indeed, while UV photons are now strongly absorbed by the high dust content embedded in clumps, \lya\ photons can avoid interaction with dust scattering off of the surfaces of clumps. 
An increase of \vexp\ results in a decrease in \flya. Increasing the expansion velocity all \lya\ photons are Doppler shifted out of resonance, forcing them to pierce the clumps, where they are strongly absorbed by the dust like UV photons. For the same reason, \flya\ now increases with increasing \nhi\ because increasing the \hi\ content of the clumps the probability of \lya\ photons scattering off the clumps increases, having the journey confined to the dust-free interclump medium. \\
\\
Finally, as in the ``low contrast" regime, we can notice that a velocity $\vexp \ga 2 \times FWHM_{int}$(\lya) is enough to prevent any scattering on the clumps. In this case, \lya\ photons escape the medium in the same way than UV photons. \\
\\
\textbf{\lya\ equivalent width EW(Lya)}: \\
\\
Qualitatively, Figs. \ref{Lya_UV_Tau}, \ref{Lya_UV_CF}, and \ref{Lya_Vexp_high_contrast} reveal that in the ``high contrast" regime \flya\ can be both higher or lower than \fescuv, depending on the actual value of \taua. From the definition of $\tau_{\mathrm{c}}$ (i.e.\ the value of \taua\ where the curves of \flya\ and \fescuv\ cross each other), we have
   \begin{equation}
   \label{eq:11}
    \flya\ \le \fescuv \,\,\, {\rm if} \,\,\, \taua\ \le \tau_{\mathrm{c}}, 
   \end{equation}
and
   \begin{equation}
   \label{eq:12}
   \flya\ \ge \fescuv \,\,\, {\rm if} \,\,\, \taua\ \ge \tau_{\mathrm{c}},
   \end{equation}
where we note that $\tau_c$ is mainly a function of CF.

Combining the definition of the \lya\ equivalent width (Eq. \ref{eq:ew}) with both Eqs.\ \ref{eq:11} and \ref{eq:12}, we obtain:
   \begin{equation}
   EW_{obs}(Lya) \le EW_{int}(Lya)                       \,\,\, {\rm for} \,\,\, \taua \le \tau_{\mathrm{c}},
   \end{equation}
and
   \begin{equation}
   EW_{obs}(Ly\alpha) \ge EW_{int}(Ly\alpha)      \,\,\, {\rm for} \,\,\, \taua \ge \tau_{\mathrm{c}},
      \end{equation}
The fact that \flya\ can exceed \fescuv\ thus allows for an enhancement (``boost") of the equivalent width of the \lya\ line. However, as summarized in Sect.\ 5.2, such an enhancement can only occur under strict physical conditions, concerning the kinematics (i.e. an expansion velocity $\vexp \la 2 \times FWHM_{int}$(\lya)), the clumpiness and the dust content of the clumpy ISM. 


\subsection{The critical ratio $n_{IC}$/$n_C$ separating the two regimes of the \lya\ radiative transfer}
\label{sect:crit}

\begin{figure}
   \centering
   \includegraphics[width=90mm]{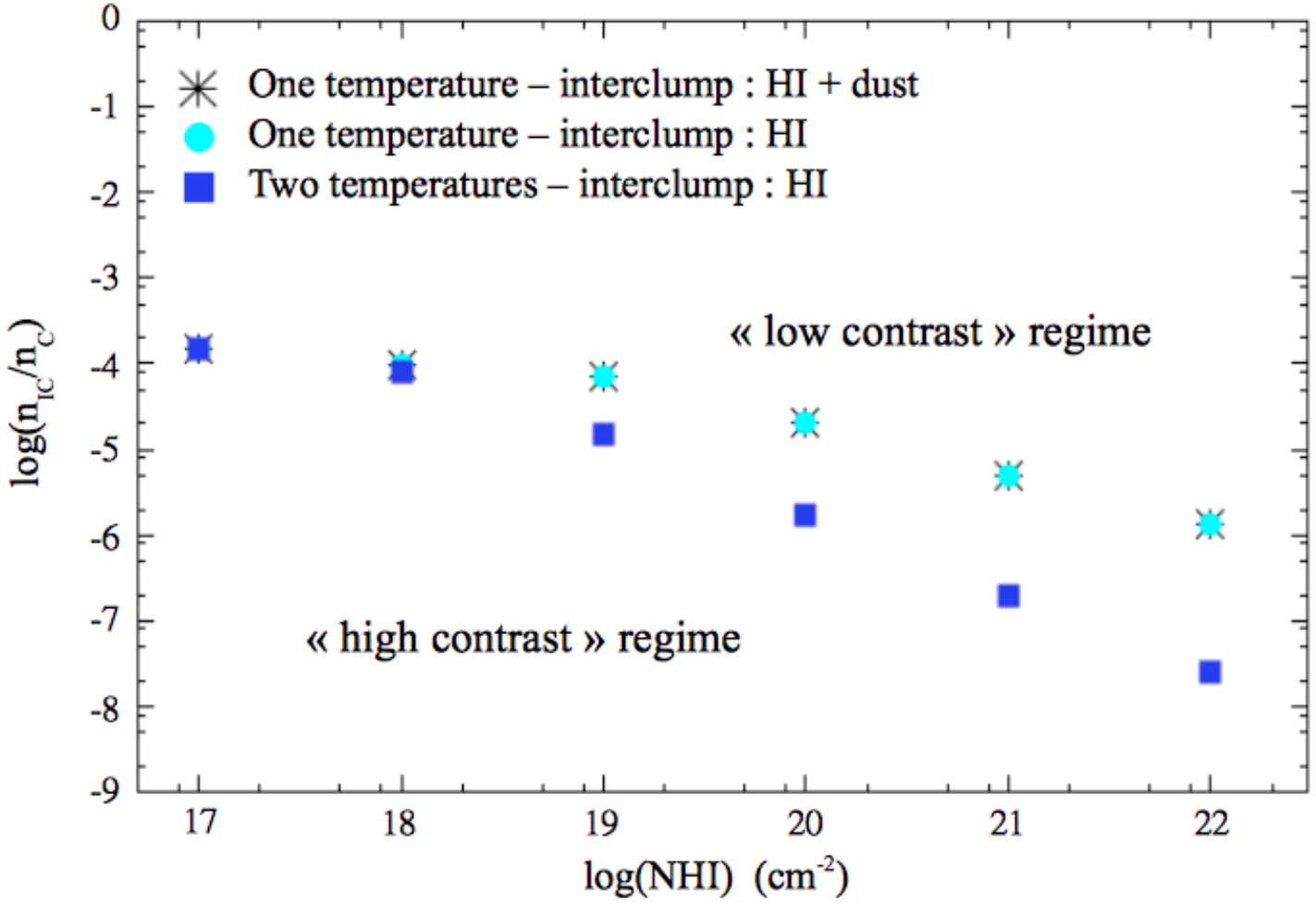}
      \caption{Evolution of the critical ratio $n_{IC}$/$n_C$ separating the  ``low contrast" regime to the ``high contrast" regime of the \lya\ radiative transfer in clumpy shell geometries. For highest ratio $n_{IC}$/$n_C$, all clumpy media belong to the ``low contrast" regime, whereas for lower $n_{IC}$/$n_C$ the ``high contrast" regime is observed. Three curves are represented in this Fig.\ corresponding to three different physical conditions applied to our clumpy media. The star and circle dots have been obtained assuming 1) a single temperature ($b = 40$ \kms) in and in between clumps, but 2) not the same composition between clumps (\hi\ + dust for the star dots and only \hi\ atoms for the circle dots). The square dots assume rather 1) a warmer temperature between clumps than in clumps ($T=10^6$ K between clumps and $b =$ 40 km s$^{-1}$ in clumps) and 2) an inter-clump medium composed only of \hi\ atoms.
              }
         \label{fig:13}
   \end{figure}

We can quantify the distinction between the two regimes of the \lya\ radiative transfer by the low to high density ratio \nratio.
Figure \ref{fig:13} illustrates this limit $n_{IC}$/$n_C$ as a function of the average \hi\ column density \nhi\ in our  clumpy media. This limit is defined in the following way. Above the curves shown in Fig.\ 11, it is impossible to obtain \flya\ $>$ \fescuv\ in models with physical parameters listed in Table 3. Such media belong to the ``low contrast" regime. Conversely, the area localized below the curves corresponds to the ``high contrast" regime where it is possible to observe the inequality \flya\ $>$ \fescuv.

All points shown in Fig.\ \ref{fig:13} have been obtained studying a clumpy medium defined with: FF = 0.23, CF = 0.997, \vexp\ = 0 km s$^{-1}$, \taua\ = 25 and $b =$ 40 km s$^{-1}$. All media built with different parameters show a critical ratio \nratio\ (at the limit between the low and the high contrast regimes) lower than that shown in Fig.\ \ref{fig:13}. Each curves shown in this figure corresponds to three different physical conditions applied in our clumpy media. The star dots are obtained assuming an unique turbulent velocity $b=40$ km s$^{-1}$ in and in between clumps, and an inter-clump medium composed by both \hi\ atoms and dust grains. The circle dots are obtained assuming the same turbulent velocity ($b$) but a dust-free interclump medium. Finally, the squares are obtained applying two different temperatures in clumps and in between clumps ($b =$ 40 km s$^{-1}$ in clumps and $T = 10^6$K between clumps) and assuming an interclump medium only composed by \hi\ atoms. This case is discussed in Sect.\ \ref{s:hotism}.

\begin{figure}
   \centering
   \includegraphics[width=90mm]{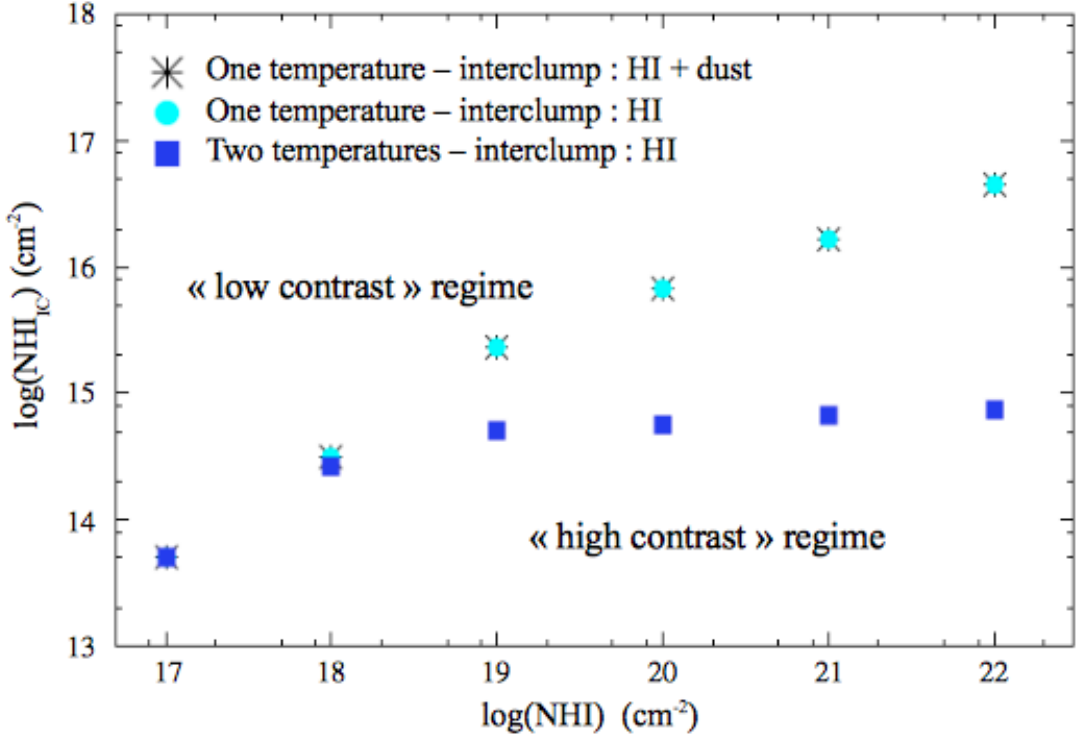}
      \caption{Same as in Fig.\ \ref{fig:13}, but showing the evolution of the \hi\ column density between clumps $N_{\mathrm{HI,IC}}$ (= (1-FF)$\times$$n_{IC}$L, with L as the physical size of the shell), as a function of the total \nhi. When the temperature in and in between clumps are the same (star and the circle dots), the ``high contrast" regime can be observed applying an optically thick interclump medium for \lya\ photons (ie. from $N_{\mathrm{HI,IC}}$ $>$ 6$\times$$10^{13}$ \cm2\ with $b =$ 40 km s$^{-1}$). But, assuming a higher temperature between clumps than those of the clumps (square dots), the ``high contrast" regime disappears as the interclump medium becomes optically thick for \lya\ photons (ie.  from $N_{\mathrm{HI,IC}}$ $>$ 3$\times10^{14}$ \cm2\ for $T = 10^6$K).
              }
         \label{fig:14}
   \end{figure}

From the single temperature case (star and circle dots) we can already mention four main results concerning the border separating the two regimes of the \lya\ radiative transfer: \\
\\
\textbf{The critical ratio $n_{IC}$/$n_C$ separating both regimes is very low}: studying a large range of \hi\ column density \nhi\ [$10^{17}$, $10^{22}$] \cm2, we notice that the limit separating both regimes is reached for very weak ratios $n_{IC}$/$n_C$ ([$1.5\times 10^{-4}$, $1.3\times 10^{-6}$]). That suggests that the ``high contrast" regime can only be found in  galaxies showing the most extremely clumpy ISMs (composed only by cold clouds of neutral hydrogen gas embedded in an extremely ionized interclump medium). The critical ratio has to decrease with increasing \nhi\ to maintain a sufficiently low column density between the clumps.\\
\\
\textbf{The interclump medium can be optically thick for \lya\ photons on the border separating the two regimes}: 
In Fig.\ \ref{fig:14} we show the limit of Fig.\ \ref{fig:13}, but translated in terms of HI column density between clumps $N_{\mathrm{HI,IC}} = (1-{\rm FF}) \times$$n_{IC} L$. 
While the interclump medium is optically thick for \lya\ photons at line center above $N_{\mathrm{HI,IC}}$ = 6 $\times 10^{13}$ \cm2\ (assuming $b =$ 40 km s$^{-1}$ between clumps)\footnote{The optical depth at line center is $\tau_0=3.31 \times 10^{-14} T_4^{-1/2} \nh=3.31 \times 10^{-14} (12.85 {\rm km s^{-1}}/b) \nh$, e.g.\ Verhamme et al.\ (2006).}, we can clearly see that the ``high contrast" regime can extend to somewhat higher interclump column densities, into the optically thick regime.
Nevertheless, this is only possible if the temperatures in and in between clumps are the same (star and circle dots). If the temperatures in and in between the clumps differ (square dots), the interclump medium has to be optically thin in order to observe the ``high contrast" regime (see Sect.\ \ref{s:hotism}). \\
 \\   
\textbf{The limit $n_{IC}$/$n_C$ separating both regimes is not affected by the presence of dust between clumps}: The critical ratio \nratio\ represented by both the star and the circle dots on Fig.\ \ref{fig:13} are the same. Thus, the presence of dust between clumps has no effect on the limit separating both regimes of the \lya\ radiative transfer. 

\subsection{The effects of an inhomogeneous temperature on the \lya\ radiative transfer}
\label{s:hotism}

If we want to render the physics of our clumpy media more realistic, we must assume  different temperatures inside and between the clumps. 
There is ample evidence for such a multi-phase ISM. For example, for the stability of the clumps a pressure equilibrium should intervene between the two phases of our clumpy media, implying a higher temperature in the interclump medium. Also, it is well known that different regions coexist in the real ISM of any galaxies \citep{mckee77}. 
In particular, we can distinguish the warm neutral atomic medium (WNM, where \hi\ atoms are present in the atomic form) to the hot ionized medium (HIM, where \hi\ atoms are in great majority ionized). 
To explore in a simplified manner these effects we have made calculations assuming a temperature of $T = 10^6$ K in the interclump medium (as those measured in the HIM), but a lower temperature   
in all clumps ($b = 40$ km/s, by analogy to the WNM).
%
We then examine the effects on the \lya\ radiative transfer and on the ratio $n_{IC}$/$n_C$ separating the two regimes identified above.

Qualitatively, the \lya\ radiative transfer properties behave in a similar fashion for different interclump temperatures.
However, as shown in Fig.\ \ref{fig:13}, the ratio 
$n_{IC}$/$n_C$ separating both regimes is found to be lower than for the case of constant temperature. In other words, the ``high contrast" 
regime is more limited when the temperature in the interclumps medium is higher than those in clumps.
The reason for this is the following: on one hand the optical depth in the interclump medium decreases (with $T^{-1/2}$), simulating thus 
a medium with even lower density, i.e.\ a higher contrast. On the other hand, the temperature increase leads to a larger frequency
redistribution of the scattered \lya\ photons, which eases their escape due to higher frequency shifts. This effect dominates over the former,
rendering thus the clumps more transparent to \lya\ radiation, where they are strongly absorbed by the dust. This explains why even higher density contrasts are needed to achieve
significant \lya\ ``rebounce" on the clumps, if the interclump medium is hotter than the clumps.

In terms of interclump column densities the limit between the regimes is shown in Fig.\ \ref{fig:14}. In contrast to the 
case of uniform ``cold" temperatures, the limit is now found at quite low column densities of the interclump medium,
corresponding to an optically thin regime. Indeed, such low column densities are needed if one wants to avoid
significant scattering of \lya\ with the corresponding high frequency shifts in a hot interclump medium.

\section{\lya\ line profiles formation in homogeneous and clumpy shells}
\label{s:profiles}
In this section we give an overview of the different emergent \lya\ line profiles produced in the expanding homogeneous and clumpy shell geometries of our model. 

\subsection{\lya\ line profiles from dust free homogeneous and clumpy shell geometries}

First we shall consider the case of a dust free ISM and examine how the \lya\ profiles are modified 
by a clumpy ISM structure. 
The line profiles shown in this section are obtained assuming an intrinsic \lya\ line characterized by $EW_{int}$(Ly$\alpha$) = 80 \AA\ and $FWHM_{int}$ = 100 km s$^{-1}$.

We know that in dust free cases the total \lya\ flux is preserved, and since the continuum is not attenuated (due to the absence
of absorption), the \lya\ equivalent width is thus preserved, in other words the observed EW is identical to the intrinsic one.
This holds obviously both for homogeneous and clumpy structures. The only effect of clumps is to modify the exact frequency
redistribution of photons, i.e.\ the shape of the emergent \lya\ line profile. However, as we will see, the changes to the line profile are only 
relatively small, when the covering factor of the clumps is large.

\begin{figure}
   \centering
   \includegraphics[width=90mm]{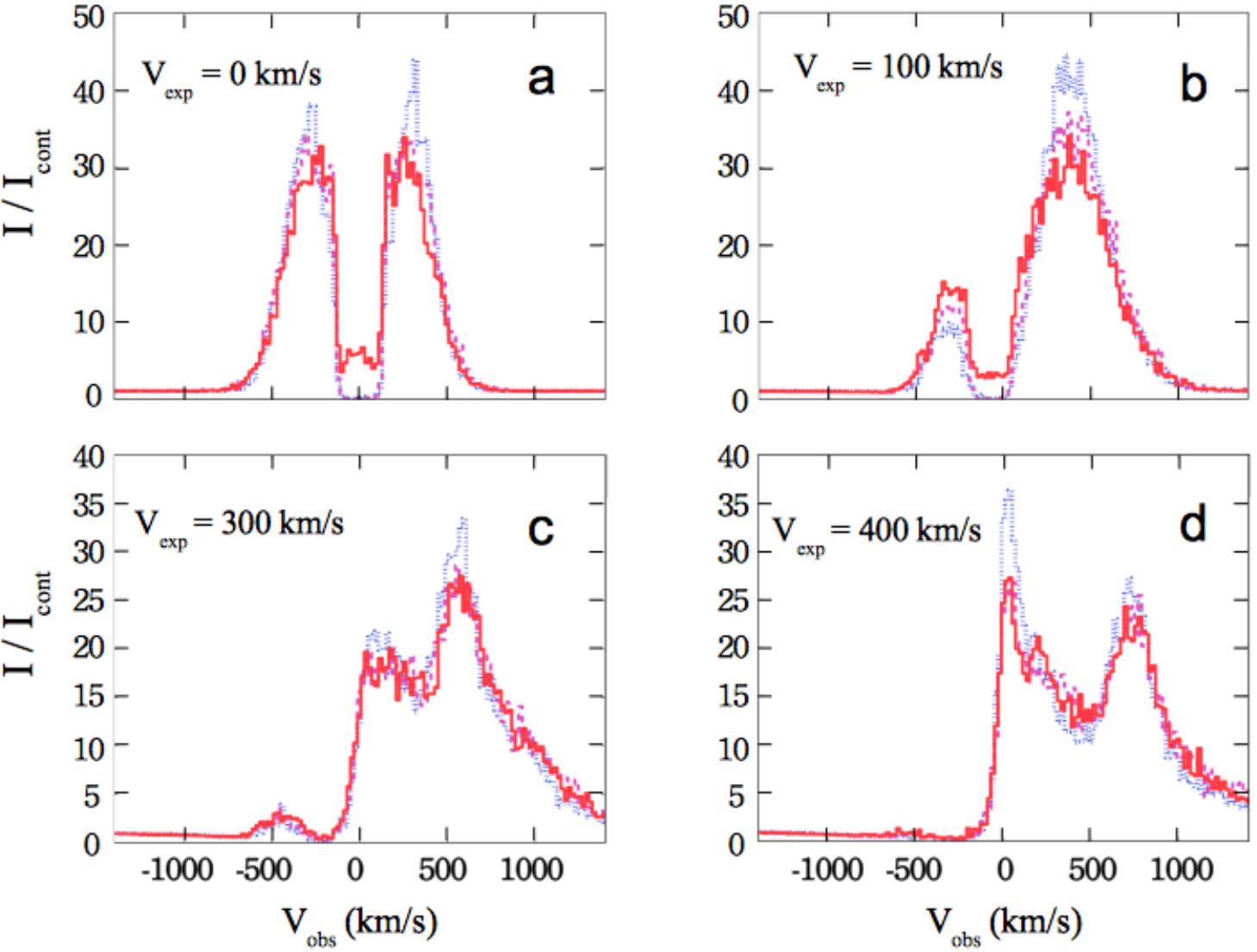}
      \caption{Comparison between the \lya\ line profiles emerging dust free homogeneous shell (dotted blue lines), weakly clumpy shell (dashed magenta lines) and extremely clumpy shells (thick red lines) under different expansion velocities \vexp. All homogeneoues and clumpy shells have for common parameters: \nhi\ = 2$\times$$10^{20}$ /\cm2\ , $b$= 40 km s$^{-1}$, \taua\ = 0 and \vexp\ = 0, 100, 300, 400 km s$^{-1}$. Furthermore, both weakly and extremely clumpy media are built adopting the following parameters: FF = 0.23, CF = 0.997 ($R_{min}$ = 49 and $R_{max}$ = 64 cells) and $n_{IC}$/$n_C$ = 0.01 (weakly clumpy shell) and 0.00 (extremely clumpy shell).     
              }
         \label{Lya_profiles}
   \end{figure}

We first examine the \lya\ line profiles for dust free homogeneous and clumpy structures with {\em low density contrast} (i.e.\ clumpy structures showing an optically thick interclump medium for \lya\ photons). In Fig.\ \ref{Lya_profiles} we study such structures built with the following parameters: $\nhi = 2 \times 10^{20}$ \cm2\ , $b =$ 40 km s$^{-1}$ and \vexp\ = 0, 100, 300, 400 km s$^{-1}$. It clearly appears that the \lya\ line profiles emerging from dust free weakly clumpy shell geometries do not show any noticeable difference compared to homogeneous structures with the same/corresponding properties.

In homogeneous and weakly clumpy media, the mechanisms of formation of the line profiles, as well as their dependance on the parameters \vexp, \nhi\ and $b$, are identical to those explained in detail in \cite{Verhamme06} for homogeneous shell structures. In other words, in the dust free case, weakly clumpy media do not significantly differ from the homogeneous ones in terms of line profiles.

\begin{figure}
   \centering
   \includegraphics[width=90mm]{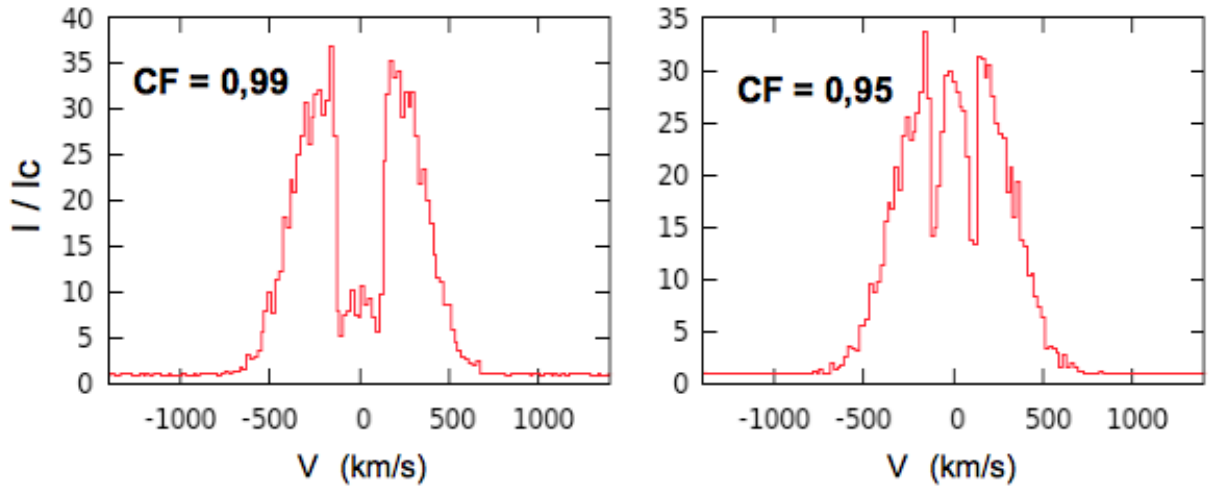}
      \caption{Effect of the covering factor CF on the \lya\ line profiles in extremely clumpy shell geometries. We study here a static extremely clumpy shell geometry defined with the following parameters: FF = 0.23, $n_{IC}$/$n_C$ = 0.00, \nhi\ = 2$\times$$10^{20}$ /\cm2, \vexp\ = 0 km s$^{-1}$, \taua\ = 0 and $b =$ 40 km s$^{-1}$. The only difference between both panels concerns the covering factor CF of the clumpy shell geometries: CF = 0.99 (left panel) and CF = 0.95 (right panel). We can notice that decreasing CF,  the relative intensity of the central peak in the \lya\ line profile is increased. 
              }
         \label{Lya_profiles_CF}
   \end{figure}

Turning now to clumpy media with large density contrasts (i.e.\ clumpy structures showing an optically thin interclump medium for \lya\ photons) we find again very similar line profiles as in the homogeneous case,
as also shown in Fig.\ \ref{Lya_profiles}.
Compared with the line profiles observed from static weakly clumpy and homogeneous media we now see a central peak at line center ($v_{\rm obs}$$\approx$0 km/s). The formation of this central peak is indeed made possible when the interclump medium is optically thin for \lya\ photons. These photons can then propagate in two different ways in the interclump medium: either by escaping through the holes which appear between clumps, or scattering off of the surface of clumps. Both features preserve the intrinsic frequency of the \lya\ photons, which allows to form the central peak seen at center of the \lya\ line ($v_{\rm obs}$$\approx$0). 
The covering factor governs mostly the importance of the central emission, as shown in Fig.\ \ref{Lya_profiles_CF} for a static shell. As expected, the central emission
increases with decreasing the covering factor.

Besides the predicted \lya\ line profiles in dust-free clumpy shell geometries behave in the same way as already shown for homogeneous shells by \cite{Verhamme06},
with the main parameters determining the \lya\ profile being the expansion velocity \vexp, the mean HI column density \nhi, and the Doppler parameter $b$.

  \begin{figure}[htb]
   \centering
   \includegraphics[width=90mm]{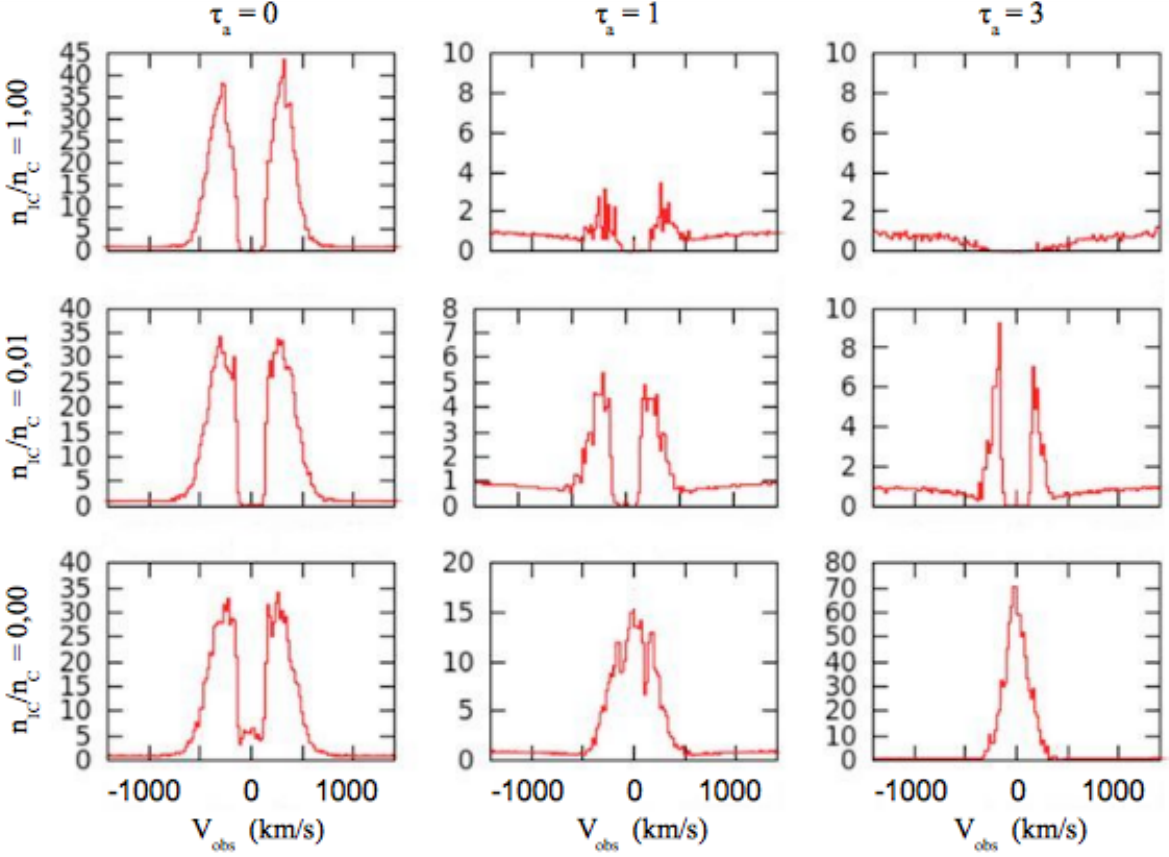}
      \caption{Effects of the dust optical depth \taua\ on the \lya\ line profiles emerging three different static shell geometries: a homogeneous shell ($n_{IC}$/$n_C$ = 1.0 - top line), a weakly clumpy shell ($n_{IC}$/$n_C$ = 0.01 - middle line) and an extremely clumpy shell ($n_{IC}$/$n_C$ = 0.00 - bottom line). Both the weakly and the extremely clumpy shell geometries are those studied, respectively, in figures 14 and 17. We thus adopt here the same physical conditions: FF = 0.23, CF = 0.997, \nhi\ = $2\times10^{20}$ /\cm2\ , \vexp\ = 0 km s$^{-1}$ and $b =$ 40 km s$^{-1}$. Note the variable scales of the different sub panels.
      An enhancement of EW(\lya) (produced by the Neufeld effect) occurs only for
      the physical conditions corresponding to the bottom right panel (\taua\ = 3).
              }
         \label{fig:dust_static}
   \end{figure}

  \begin{figure}[htb]
   \centering
   \includegraphics[width=90mm]{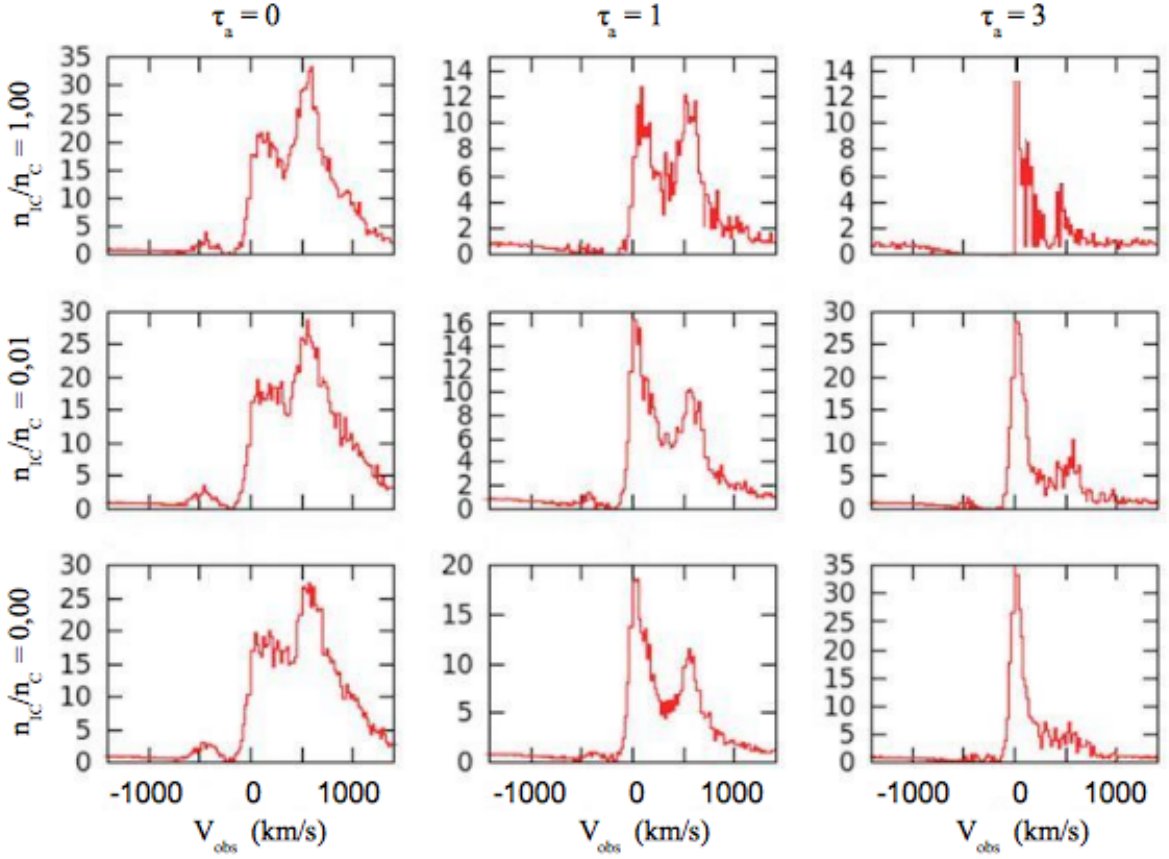}
      \caption{Same as Fig.\ \protect\ref{fig:dust_static} but for an expanding shell with \vexp = 300 \kms. Note that no boost of \lya\ with respect to the continuum is found in any of these models, since the expansion velocity is too high. Discussion in the text.
              }
         \label{fig:dust_vexp}
   \end{figure}

%

%
\subsection{\lya\ line profiles from dusty homogeneous and clumpy shell geometries}

We now examine the main effects produced by dust on the \lya\ line profiles emerging from clumpy shell geometries. 

In Fig.\ \ref{fig:dust_static} we illustrate the evolution of the line profiles predicted for static homogeneous and clumpy shell geometries as a function of the dust optical depth \taua. The top line shows the homogeneous case, the middle the low density contrast, and the bottom line the clumpy medium with a
high density contrast.
Let us first examine the homogeneous and low density contrast cases (\nratio\ = 1.00 and 0.01 respectively).
Increasing \taua\ from 0 to 1, the \lya\ line still appears in emission in both cases. But, a clear decrease in both the width and the intensity of each peak are noticed. For \taua\ = 3, more than 99.8 \% and 98 \% of the \lya\ photons are absorbed by the dust, respectively in the homogeneous and the clumpy media. Therefore, an absorption profile emerges from the homogeneous medium, while faint emission line is predicted from the weakly clumpy medium. Higher dust optical depths \taua\ are needed to
obtain absorption line profiles from weakly clumpy media, typically $\taua\ \ga 35$ in the case of $n_{IC}$/$n_C$ = 0.01.

   For extremely clumpy shell geometries ($n_{IC}$/$n_C$ = 0.00, Fig.\ \ref{fig:dust_static}), the evolution of the \lya\ line profile is different. Increasing \taua\ from 0 to 1, we notice a clear decrease of the width of both lateral peaks, as well as an increase of the relative intensity of the central peak. The photons composing the central peak indeed interact very weakly with the dust, which explains why this peak becomes the dominant one in the line profile as \taua\ increases. 
When \taua\ is further increased from 1 to 3, both lateral peaks are destroyed by dust. The central peak thus becomes the only peak composing the line profile above $\taua \ga 3$. It is interesting to note that we cannot obtain absorption line profiles in extremely clumpy media ($n_{IC}$/$n_C$ = 0.00). Indeed, as the \lya\ photons composing the central peak interact very weakly with dust, they are always able to escape the medium for any dust optical depth \taua, giving thus rise to an emission line.
Comparing the relative escape of the \lya\ and UV continuum photons, we note that the only case in Fig.\ \ref{fig:dust_static} where the \lya\ equivalent width is (slightly) enhanced
is found in the bottom right panel. Indeed, in this case \taua\ is close to critical dust optical depth $\tau_c \approx 3$ for this example of extremely clumpy medium,
where we expect such an enhancement (cf.\ Sect. 3.2.2). 


We now turn to a case with outflows in Fig.\ \ref{fig:dust_vexp}. In this figure, we adopt otherwise identical parameters to those shown in Fig.\ \ref{fig:dust_static}.
Increasing \taua\ in any media (homogeneous or clumpy), we first notice a quick decrease of the intensity of both the dominant red peak (those shifted at $v_{\rm obs}$ = 2$\times$vexp) and the small blue bump. 
This is due to higher number of scatterings these photons undergo, which increases their destruction probability, as already discussed by \citet{Verhamme06}.
For the highest dust content (\taua\ = 3), the \lya\ line escaping homogeneous and clumpy media exhibits an asymmetric profile, but whose the dominant peak is found
at line center ($v_{\rm obs}$=0).
Finally, like in the static case, we notice that the intensity of the line increases as the clumpiness of the medium increases. 
In none of the cases shown here we find a "boost" of the \lya\ equivalent width, since the velocity is too large.

\section{Discussion}

\subsection{Effects of a clumpy ISM on the radiation attenuation}

Given the evolution of the \lya\ and continuum escape fraction in homogeneous and clumpy systems (Sect.\ 3), it is clear that a clumpy medium always produces higher \lya\ and continuum escape fraction compared with an equivalent homogeneous medium of equal dust and hydrogen mass. This main result was demonstrated by several previous studies focused on the transfer of the continuum radiation in clumpy media \citep{boisse90,hobson93, witt96, witt00, varosi99}, but also from other studies focused on the \lya\ line \citep{neufeld91,hansen06}. 

\begin{figure}
   \centering
   \includegraphics[width=90mm]{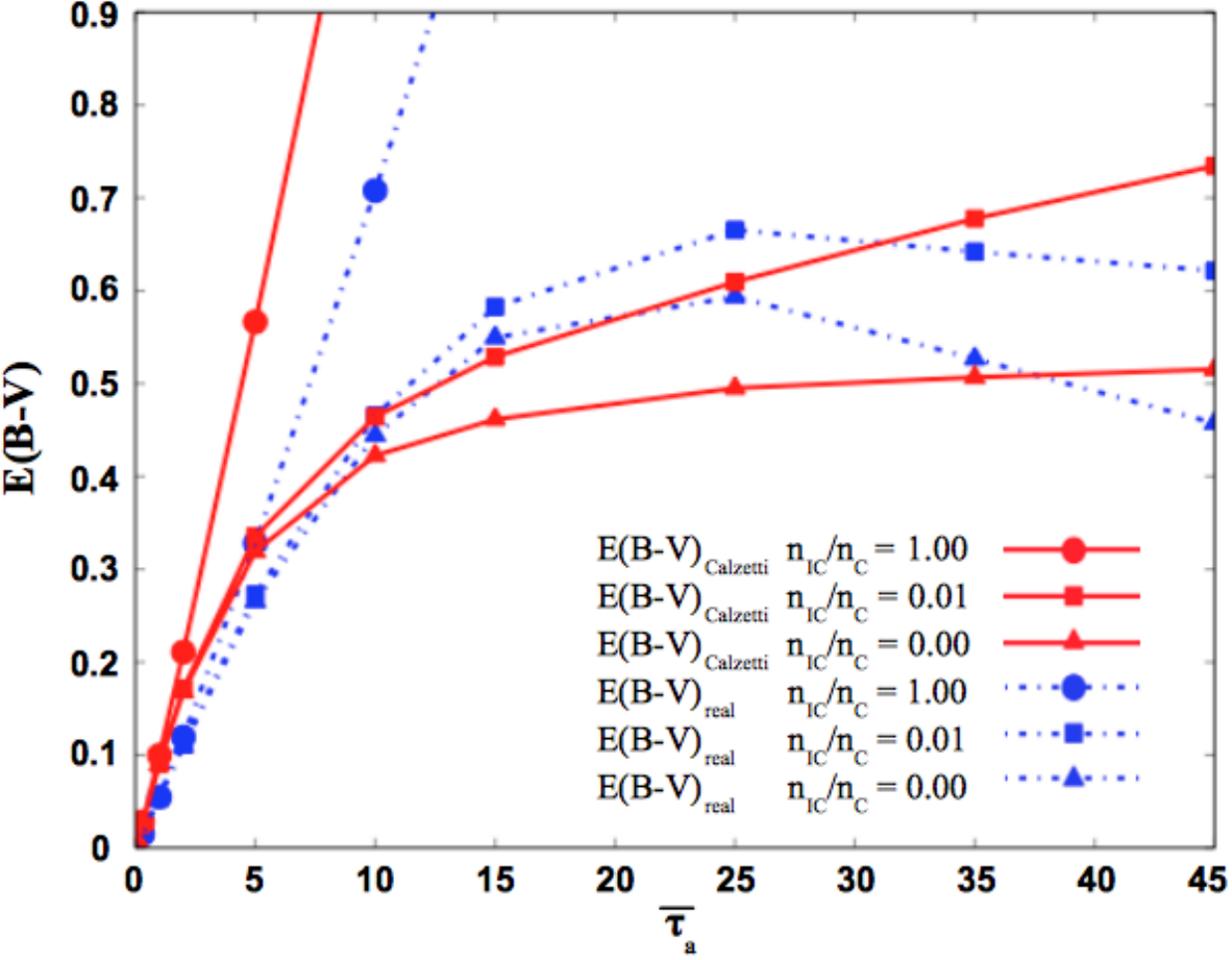}
      \caption{Evolution of the colour excess E(B-V) as a function of \taua\ in the homogeneous and clumpy shell geometries studied in both sections 3 and 4. The shell geometries are defined by the following parameters: FF = 0.23 and CF = 0.997 ($R_{min}$ = 49 and $R_{max}$ = 64 cells), $n_{IC}$/$n_C$ = 1.0, 0.01 and 0. Two different colour excess are shown in this figure. Firstly $E(B-V)_{real}$, which corresponds to the exact value of the colour excess because estimated from the original definition of the colour excess (from the V and B bands). Secondly $E(B-V)_{Calzetti}$ which is estimated from both the attenuation law of  "Calzetti" \citep{calzetti97} and the UV escape fraction. An observer would rather measure $E(B-V)_{Calzetti}$ in practice.
              }
         \label{fig:ebv}
   \end{figure}

The attenuation of the radiation in a galaxy is thus strongly dependent on both the dust content and the dust distribution around the radiation sources. For illustration, we show in Fig.\ \ref{fig:ebv}  
the dependence of the colour excess E(B-V) on both the dust content (\taua) and the clumpiness of the dust distribution in the shell geometries studied throughout this paper. In this figure we compare two different definitions of the colour excess: $E(B-V)_{real}$, which corresponds to the exact colour excess because estimated from the original definition of the colour excess (from the V and B bands), and $E(B-V)_{Calzetti}$ which is estimated from both the Calzetti attenuation law \citep{calzetti00} and the UV escape fraction (see Eqs.\ \ref{eq:calz} and \ref{eq:real}). In practice, the Calzetti attenuation law is usually used to estimate the dust attenuation in starburst galaxies. It is then $E(B-V)_{Calzetti}$ which would be measured by an observer. The Fig. \ref{fig:ebv} allows then us to see in which extend the colour excess $E(B-V)_{Calzetti}$, from the Calzetti law, deviates from the real colour excess $E(B-V)_{real}$ as a function of \taua\ and the clumpiness of the dust distribution.

In a general way, we can notice that the clumpiness of the dust distribution strongly affects both the colour excess $E(B-V)_{real}$ and $E(B-V)_{Calzeti}$ \citep{witt00}. The colour excess decreases as the dust distribution is clumpy and as the dust optical depth decreases in media. Comparing now both definitions of $E(B-V)_{Calzetti}$ and $E(B-V)_{real}$, we can notice that $E(B-V)_{Calzetti}$ does not reproduce very well the real evolution of the colour excess $E(B-V)_{real}$. This deviation between both definitions is explained by a clear evolution of the attenuation law (which measures, at each wavelength, the reduction in the stellar flux from a dusty ISM) as the dust distribution and the dust content change in media. As mentioned in \cite{witt00}, this divergence shows that the use of the same and unique attenuation law in the analysis of a large sample of galaxies (which show different dust geometries and dust content) can become a source of error in the dust attenuation correction for individual galaxies.

\subsection{High \lya\ EWs and the Neufeld model}
\label{s:neufeld}

\subsubsection{Physical conditions needed in the ISM}

\begin{figure}[htb]
   \centering  
   \includegraphics[height=62mm]{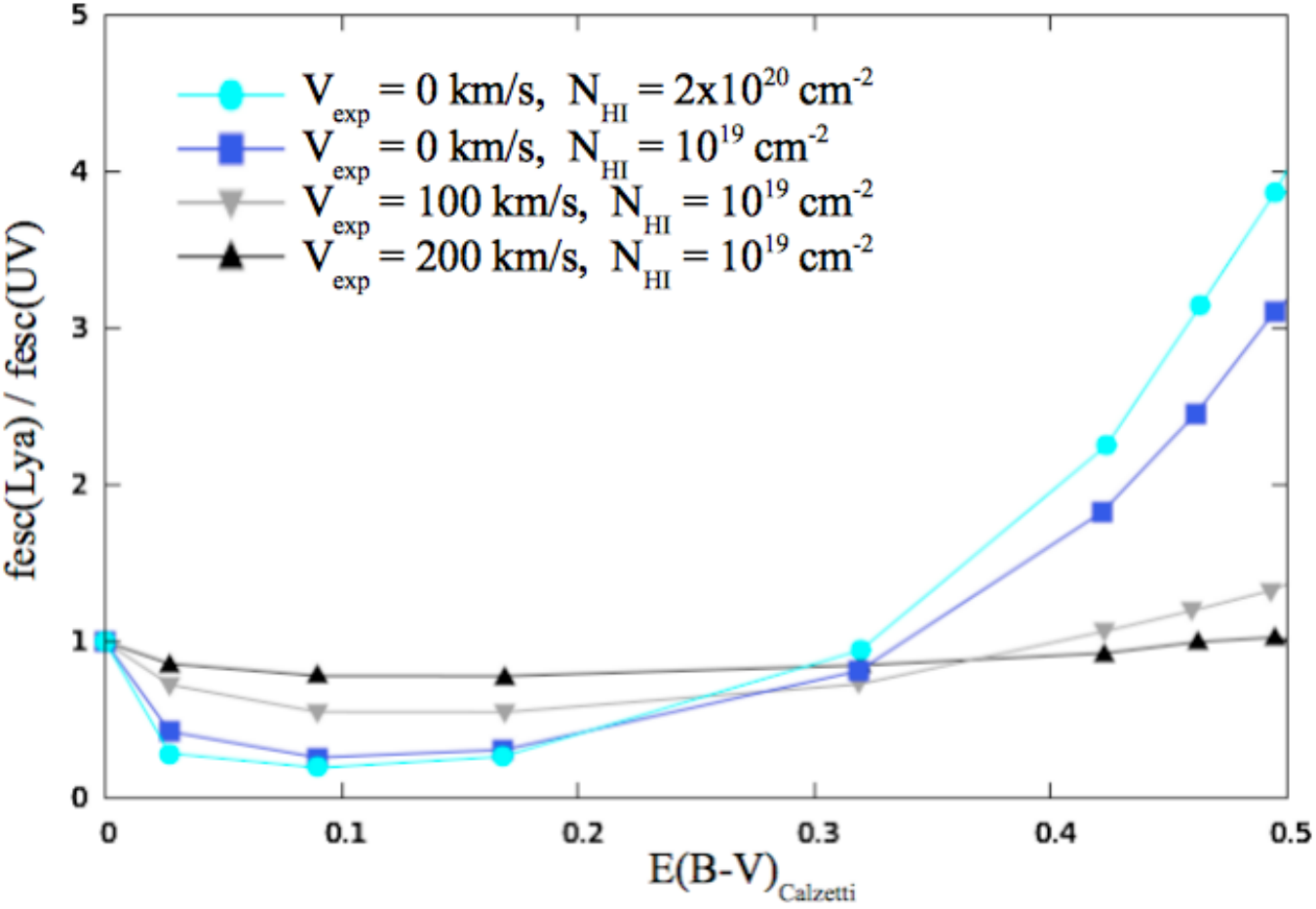} 
      \caption{Evolution of the ratio \flya/\fescuv\ (i.e.\ $EW_{\rm obs}(Lya)$/$EW_{\rm int}(Lya)$) as a function of the colour excess $E(B-V)_{calzetti}$ in four extremely clumpy media. We apply some realistic physical conditions of LAEs to each media: $n_{IC}$/$n_C$ = 0.00, FF = 0.23, CF = 0.997 ($R_{min}$ = 49 and $R_{max}$ = 64 cells), $b =$ 12.85 km s$^{-1}$ (i.e. T = $10^4$ K in clumps), \nhi\ =  $10^{19}$, 2$\times$$10^{20}$ \cm2\  and \vexp\ = 0, 100, 200 km s$^{-1}$. \flya\ is derived assuming an intrinsic line width $FWHM_{int}$ = 100 km s$^{-1}$. From the left to the right,  the models are calculated for \taua\ = 0, 0.3, 1, 2, 5, 10, 15 and 25.}
%
         \label{fig:23}
   \end{figure}

According to our study of the \lya\ transfer in clumpy media, there exists a regime in which the Neufeld scenario works. This regime corresponds to the ``high contrast" regime, as explained in detail in Sect.\ 3.2.2. It is only found in the most extremely clumpy shell geometries of our model, that is composed by clouds of \hi\ and dust embedded in an interclump medium close to be optically thin for \lya\ photons. However, even in this configuration, the Neufeld model only works when the following five main conditions concerning the clumpiness, the kinematic, the dust content and the spatial distribution of the clumps around the stars 
are fulfilled: \\
\\
\textbf{The galaxy outflow has to be relatively slow}: Assuming an intrinsic \lya\ line width $FWHM_{int}$(Lya),  a galactic outflow with an expansion velocity \vexp\ $\la$  2 $\times$ $FWHM_{int}$(Lya) km s$^{-1}$ is needed to be able to enhance EW(\lya) under the Neufeld scenario. In starburst galaxies, the width of the intrinsic \lya\ line is lower than 100 km s$^{-1}$ \citep{teplitz00, baker04, erb03, mclinden11}, which implies an expansion velocity \vexp\ lower than 200 km s$^{-1}$ in the ISM. We illustrate this limit in Fig.\ \ref{fig:23}. This figure shows the evolution of the ratio \flya/\fescuv\ as a function of the dust content (measured here in terms of colours excess $E(B-V)_{calzetti}$) in an extremely clumpy shell geometries. 
We here adopt $b =$ 12.8 km s$^{-1}$, $\nhi\ =  10^{19}$ and $2\times10^{20}$ \cm2\  and $V_{exp}$ = 0, 100, 200 km s$^{-1}$, typical
of values obtained in the analysis of high-$z$ \lya\ line profiles \citep{verhamme08}. 
Finally, let us mention that the curves shown in this figure illustrate the highest enhancements of EW(\lya) we can obtain adopting such physical conditions in a clumpy medium; they reach up to a factor 3--4.
Adopting $FWHM(Lya)_{int}$ = 100 km s$^{-1}$ in Fig.\ \ref{fig:23} we can notice that no significant enhancement of EW(\lya) is obtained for $\vexp \ga 200$ \kms.
Above such expansion velocity, all \lya\ photons are Doppler shifted out of resonance, preventing them to scatter off of the surface of clumps and to escape clumpy ISMs more easily than UV continuum photons. \\
\\   

\begin{figure}
   \centering  
  \includegraphics[width=91mm]{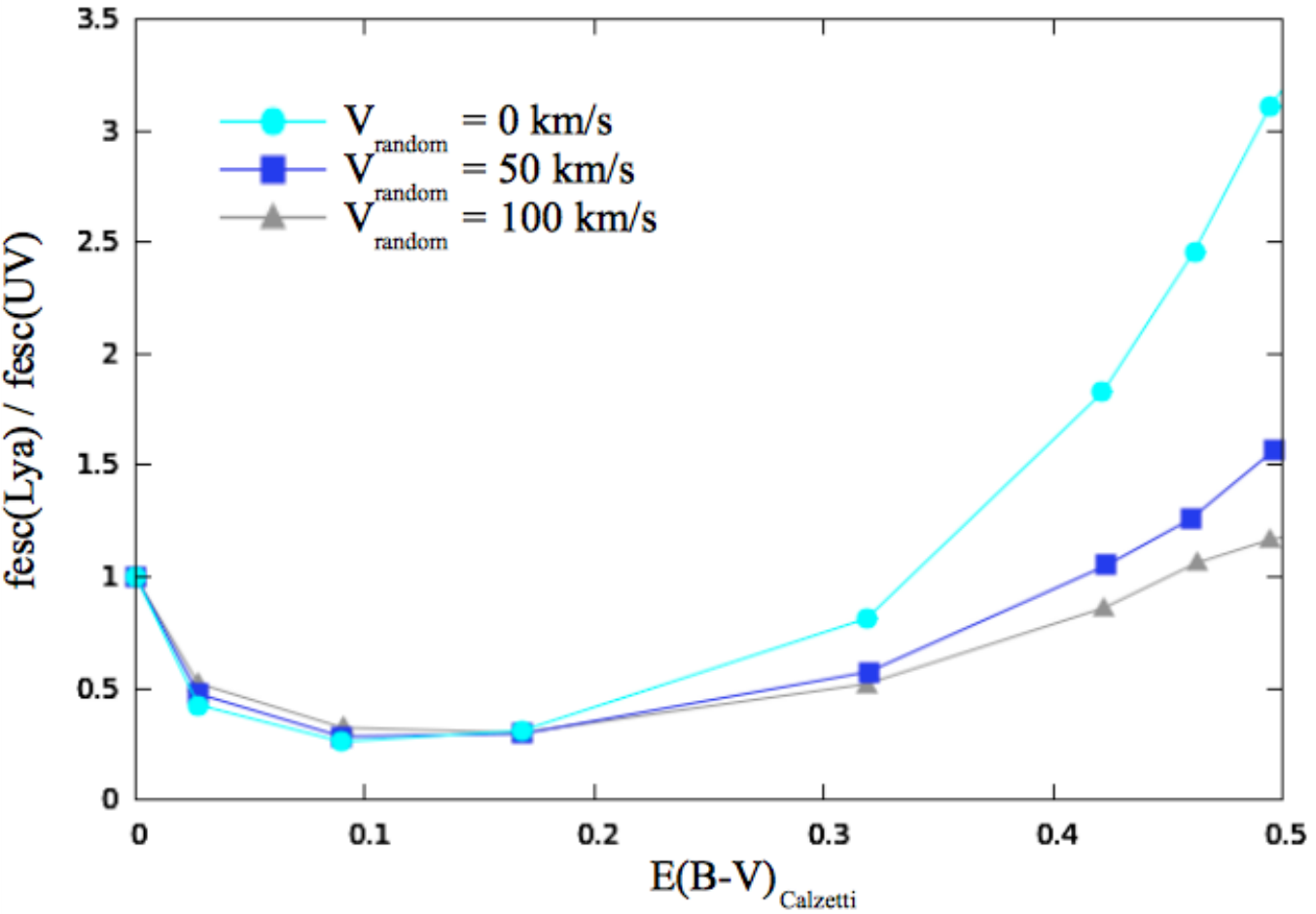}

      \caption{Effect of a nonuniform outflow on the ratio \flya/\fescuv\ in the same clumpy shell geometry studied in Fig. 18 (FF = 0.23, CF = 0.997, $\nhi =10^{19}$ \cm2\ and $b =$ 12.8 km s$^{-1}$). In this figure we assume a radial and random velocity $v_{\rm random}$ for each clump with a maximum velocity of 0 \kms\ (circle), 50 \kms\ (square) and 100 \kms\ (triangle).   
              }
         \label{fig:inho_outflows}
   \end{figure}

\textbf{The galaxy outflow has to be relatively uniform (constant velocity)}: \lya\ photons can scatter on the surface of clumps (Fig. 8) under the condition that each clumps move weakly each other. Should the opposite occur (that is assuming a random component $v_{\rm random}$ in the velocity of each clumps) a strong Doppler shift can occur between clumps which prevents \lya\ photons to scatter against clumps anymore. Finally, such effect strongly decreases the \lya\ escape fraction \flya\ in a clumpy ISM. In Fig.\ \ref{fig:inho_outflows}, we illustrate how the enhancements of EW(\lya) shown in Fig.\ \ref{fig:23} are affected by a nonuniform outflow. In this figure, we assume a radial and random velocity $v_{clump}$ for each clump, such that $v_{\rm clump} = \vexp + v_{\rm random}$, where
$v_{\rm random} = r  v_{\rm max}$ with $r \in [-1,1]$ a random number.
We notice a clear decrease of the ratio \flya/\fescuv\ as $v_{\rm random}$ increases.  \\
\\
\textbf{The \hi\ content between clumps must be extremely small}: The Neufeld model was originally developed assuming an interclump region sufficiently poor in \hi\ atoms,
such that it is completely transparent to \lya\ photons. In our simulations, we have identified the allowed \hi\ content in the interclump medium to allow the Neufeld scenario to work.
There is indeed a certain \hi\  limit above which \lya\ photons cannot freely propagate between clumps, preventing them to escape the medium more easily than UV photons. In all physical conditions, our simulations confirm that the Neufeld model only works if the interclump medium stays optically thin for \lya\ photons, that is if the radial \hi\ column density of the interclump medium ($N_{\mathrm{HI,IC}}$) is lower than 3 $\times$ $10^{14}$ \cm2\ (with a temperature of $T = 10^6$ K between clumps).
For instance, focussing on the clumpy shell geometries studied in Fig.\ \ref{fig:23}, we notice that no enhancement of EW(\lya) is obtained if the radial \hi\ column density between clumps exceeds 1.5 $\times$ $10^{14}$ cm$^{-2}$ (for the curves $\nhi\ = 10^{19}$ cm$^{-2}$) and 2.3 $\times$ $10^{14}$ cm$^{-2}$ (for the curve $\nhi = 2 \times 10^{20}$ cm$^{-2}$). 
In terms of ratio \nratio,  
such limits correspond to a density ratio of $6.90 \times 10^{-6}$ (for the curves $\nhi\ = 10^{19}$ cm$^{-2}$) and 3.45 $\times$ $10^{-7}$ (for the curve $\nhi = 2 \times 10^{20}$ cm$^{-2}$). In reality, lower densities ratios can be observed in a real ISM, if the cold clouds of neutral \hi\ ($T = 10^4$ K, $n_{\mathrm HI}$ = 0.3 cm$^{-3}$) are embedded in a very hot and ionized interclumps medium ($T = 10^6$ K, $n_{\mathrm H}$ $\approx$ 5$\times$$10^{-3} $ $cm^{-3}$ and $x_{\mathrm HI} < 10^{-5.5}$).
In other words, an efficient ``boost'' of \lya\ with respect to the continuum would require such extreme ISM conditions.\\
\\
 \textbf{A high dust content has to be embedded in clumps}: 
As explained in Sect\ 3.2.2 and shown in Fig.\ \ref{fig:23}, no enhancement of EW(\lya) is found below a certain critical dust content (noted $\tau_c$ in Fig.\ \ref{Lya_UV_Tau}). A high dust content is indeed needed in the ISM in order to absorb more efficiently UV continuum photons than \lya\ photons, which thus produces an enhancement of EW(\lya). As shown in Fig. \ref{fig:23}, a colour excess higher than $E(B-V)_{calzetti}$ = 0.32 (ie. $\tau_c$ $\approx$ 5) would be needed in order to enhance EW(\lya) by a factor higher than 3. This dust content limit mainly depends on the covering factor CF of the ISM, where it decreases as CF decreases. \\
\\
\textbf{The distribution of the clumps around the stars}: The spatial distribution of the clumps around the stars plays an important role in the enhancement of EW(\lya). 
A covering factor CF close to unity is needed in order to get a enhancement of EW(\lya) as high as those shown in Fig.\ \ref{fig:23}.
The enhancement of EW(\lya) is indeed maximized when CF = 1, but it strongly decreases as CF decreases around the stars. A low covering Factor CF does not allow to block effectively UV continuum photons than \lya\ photons, thereby decreasing the enhancement of EW(\lya). In particular, under the physical conditions adopted in Fig.\ \ref{fig:23}, we notice that the enhancement of EW(\lya) stays lower than 1.38 for CF $\la 0.68$. \\

Let's mention that, in both figures \ref{fig:23} and \ref{fig:inho_outflows}, the total \lya\ flux is taken into account to derive the EW(\lya) enhancement. However, an observer could measure a lower EW(\lya) in practice
if \lya\ is scattered into an extended low surface brightness region, as recently observed around distant starburst galaxies \citep{steidel11} or in nearby galaxies \citep{ostlin09}.

Concerning the required dust extinction and the high density contrast of neutral hydrogen (i.e. \nratio\ $<$ $10^{-7}$), these conditions seem more characteristic of molecular clouds embedded in a very hot and ionized medium. The Neufeld model could therefore only work in such a galactic environment. It is nevertheless interesting to notice that these necessary conditions to get an enhancement of EW(\lya) in a clumpy ISM are in perfect agreement with those originally predicted by \cite{neufeld91}. In his original paper, Neufeld proposed indeed three suitable conditions for an enhancement of EW(\lya) in a clumpy ISM:
the interclump medium (ICM) must exhibit a very low density with a negligible small absorption and scattering coefficients to \lya\ photons,
the clumpy ISM must show a large covering factor (CF) and a sufficiently small volume filling factor that the ICM can "percolate" (i.e.\ every part of the interclump medium must be connected to every other part), and the probability for \lya\ photons of being reflected by the clumps must be higher than the probability to be absorbed. 
In particular, this last probability should increase when \vexp\ decreases and \nhi\ increases, which is well consistent with our results. 
In addition to the criteria proposed by \cite{neufeld91}, our work highlights the extreme sensitivity of the Neufeld scenario to the kinematic of the clumps (i.e.\ the large scale outflows and the velocity dispersion of the clumps), and its strong dependence on the dust content. Given the ubiquitous evidence for outflows from most star-forming
galaxies and the widespread presence of dust, it is essential to take these effects into account.  
Furthermore, beyond the work of \cite{neufeld91}, our detailed radiation transfer models including the \lya\ line but also transfer of continuum photons at other wavelengths,
also allow us to predict consistently the resulting attenuation (reddening) and the detailed \lya\ line profile, which can directly be compared to observations.

\subsection{Studying the ISM through the \lya\ line profile}
Although the \lya\ line profiles emerging both from homogeneous and clumpy ISMs are quite similar (Fig.\ \ref{Lya_profiles}), we have identified two main effects produced by the ISM clumpiness on the \lya\ line profiles. First, we have seen that an extremely clumpy ISM favours the formation of a peak at the line center of the line profile ($v_{\rm obs}$=0 km s$^{-1}$). Second, since \lya\ escape is facilitated in a clumpy medium, the effect of the dust on the \lya\ line profile is less efficient than in homogeneous media 
(Figs.\ \ref{fig:dust_static} and \ref{fig:dust_vexp}). This can lead to intense \lya\ emission emerging from a very dusty clumpy ISM, whereas an absorption line profiles would emerge from a homogeneous medium with the same dust content.  

This second effect of the ISM clumpiness on the \lya\ line profiles can be a source of uncertainties if we aim to derive some informations on the ISM of distant starburst galaxies (kinematics \vexp, \hi\ column density \nhi, dust content \taua) from \lya\ line fitting \citep{verhamme08, schaerer08}. 
Indeed, depending on the homogeneity/clumpiness of the ISM, 
different derived parameters of the ISM can be obtained studying the same sample of galaxies. Among the ISM parameters it is possible to derive from the fit of 
both the \lya\ line and the UV continuum (\vexp, \nhi, \taua), the dust optical depth (\taua) seems the most uncertain parameter, given the dependence on
the degree of clumping.
On the positive side, the expansion velocity may probably be best determined. In asymmetric \lya\ profiles the frequency of the second bump -- if present -- 
traces quite well $2 \times \vexp$, both in homogeneous or clumpy media, as already pointed out by \citet{Verhamme06}.
However, complications may be that two bumps (i.e. those located at the line center and those  redshifted at $2 \times \vexp$ in Fig.\ \ref{fig:dust_vexp}) are not detectable (e.g.\ due to insufficient spectral resolution), and the second bump (i.e. shifted at $2 \times \vexp$) may not 
always correspond to the peak of the profile. Also, depending on the column density the distinction
of the two peaks may not be very easy \citep{Verhamme06}.

In any case, compared to the typical \lya\ line profiles observed in distant \lya\ emitters and LBGs,
we note that most of them are asymmetric lines with redshifted peaks, which seem difficult to reconcile with
the profiles predicted for very clumpy shell geometries, both static or expanding, since the peak is then
expected to show negligible redshift.
Furthermore, \lya\ absorption lines are usually found among the reddest (presumably more dusty) LBGs,
and the \lya\ equivalent width correlates with reddening
(e.g. Shapley et al. 2003), facts naturally explained by radiation transfer
models with a homogenous ISM \citep{verhamme08, schaerer08}. These findings also argue against a very clumpy, high-contrast ISM, at least for the majority of LBGs.

\subsection{Comparison with Hansen \&\ Oh (2006)}



We now compare our results with the recent numerical study of \lya\ transfer in multiphase and dusty media from \cite{hansen06} (hereafter HO06).
The clumpy media studied in HO06 are only extremely clumpy, that is composed of very optically thick spherical clumps (\hi\ + dust) distributed within an empty interclump medium. From such clumpy media, HO06 deduce analytical formulae fitting the behavior of both the continuum and the \lya\ escape fractions as a function of gas geometry, motion and dust content. In summary, assuming an isotropic scattering by dust grains ($g = 0$) in optically thick spherical clumps, the fitting formula 
is a function of two parameters (Eq.\ 59 in HO06):
\begin{equation}
f_{\rm esc}(\nu) = \frac{1}{\cosh(\sqrt{2 \epsilon_c(\nu) N_0})} 
\label{eq_hansen}
\end{equation} 
where $N_0$ (a geometrical parameter) corresponds to the average number of clumps encountered by photons before escaping the medium in the absence of absorption, and where $\epsilon_c$($\nu$) (a dust parameter) corresponds to the probability of a photon of frequency $\nu$ to be absorbed rather than reflected by a clump. In the clumpy shell geometries of our model, both parameters can be derived in the following way. Firstly we notice that $N_0$ tends to evolve as $N_0$ = 1.1$fc^2$ + 1.42$fc$, where $fc$ corresponds to the mean number of clumps intersected along a random line of sight. The parameter $\epsilon_c$($\nu$) can simply be estimated using the formula derived in HO06 (eq. 27 of their paper):
\begin{equation}
\epsilon_c(\nu) = \frac{2\sqrt{\epsilon(\nu)}}{(1+\sqrt{\epsilon(\nu)})}
\label{eq_hansen}
\end{equation} 
 with $\epsilon(\nu)$ as the absorption probability per interaction (HI or dust) at frequency $\nu$. In particular, $\epsilon(\nu)$ is thus given by $\epsilon(\nu) = 1 - a$ for the UV continuum photons, with $a$ the dust albedo. 



  \begin{figure}
   \centering
   \includegraphics[width=90mm]{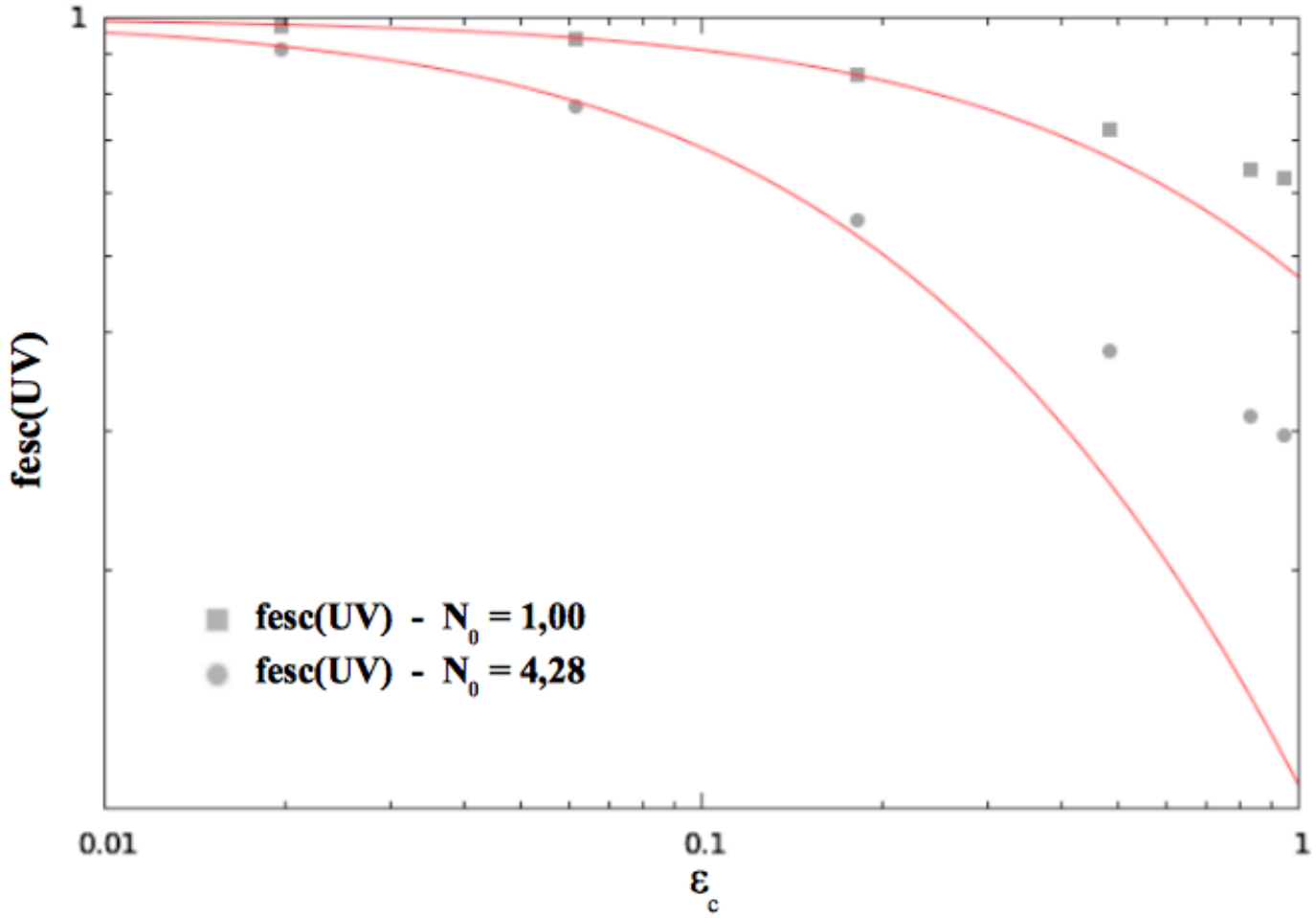}
      \caption{Test of the equation 15 of HO06. We show here the evolution of the UV escape fraction from two different extremely clumpy shell geometries as a function of the absorption parameter $\epsilon_c$. The first medium (square dots) shows $N_0$ = 1.00 and is built with FF = 0.10, CF = 0.38 and \nratio\ = 0. The second medium (circle dots) shows $N_0$ = 4.28 and is built with FF = 0.15, CF = 0.70 and \nratio\ = 0. In both media, each clump is opaque to dust extinction. The clumps are therefore optically thick for UV continuum photons. Both full lines represent respectively the equation 15 for $N_0$ = 1.00 (top line) and $N_0$ = 4.28 (bottom line). 
              }
         \label{comparison_Hansen_Oh}
   \end{figure}

We compare in Fig.\ \ref{comparison_Hansen_Oh} the equation 15 (full lines) to the UV continuum escape fraction derived from the clumpy shell geometries of our model (grey dots). In this figure, two clumpy shell geometries are studied. The first structure, built with FF = 0.10, CF = 0.38 and \nratio\ = 0, shows $N_0 = 1.0$ (top line). The second structure, assuming FF = 0.15, CF = 0.70 and \nratio\ = 0, shows $N_0 = 4.28$ (bottom line). In Fig.\ \ref{comparison_Hansen_Oh}, we just change the values of $\epsilon_c$($\nu$) changing the albedo $a$ of the dust grains, as shown in Eq.\ 16. Although a certain difference appears as $\epsilon_c$ tends towards 1, the continuum escape fraction deduced from our numerical simulations are rather well fitted by the analytical formula of HO06 (Eq.\ 15). In particular, the UV escape fraction follows well the same dependance on $N_0$ and $\epsilon_c$ as predicted by Eq.\ 15 for the low absorption regime ($\epsilon_c$ $<$ 0.4). The difference observed close to $\epsilon_c = 1$ is also observed in HO06 and is explained by a geometrical effect. As suggested by these authors, a better fit can be obtained in this regime rescaling the term $\epsilon_c$($\nu$)$N_0$ (Eq.\ 15) as $\kappa$$\epsilon_c$($\nu$)$N_0$, where $\kappa$ is a unity fitting parameter.  
\\
  \begin{figure*}
   \centering
   \includegraphics[width=90mm]{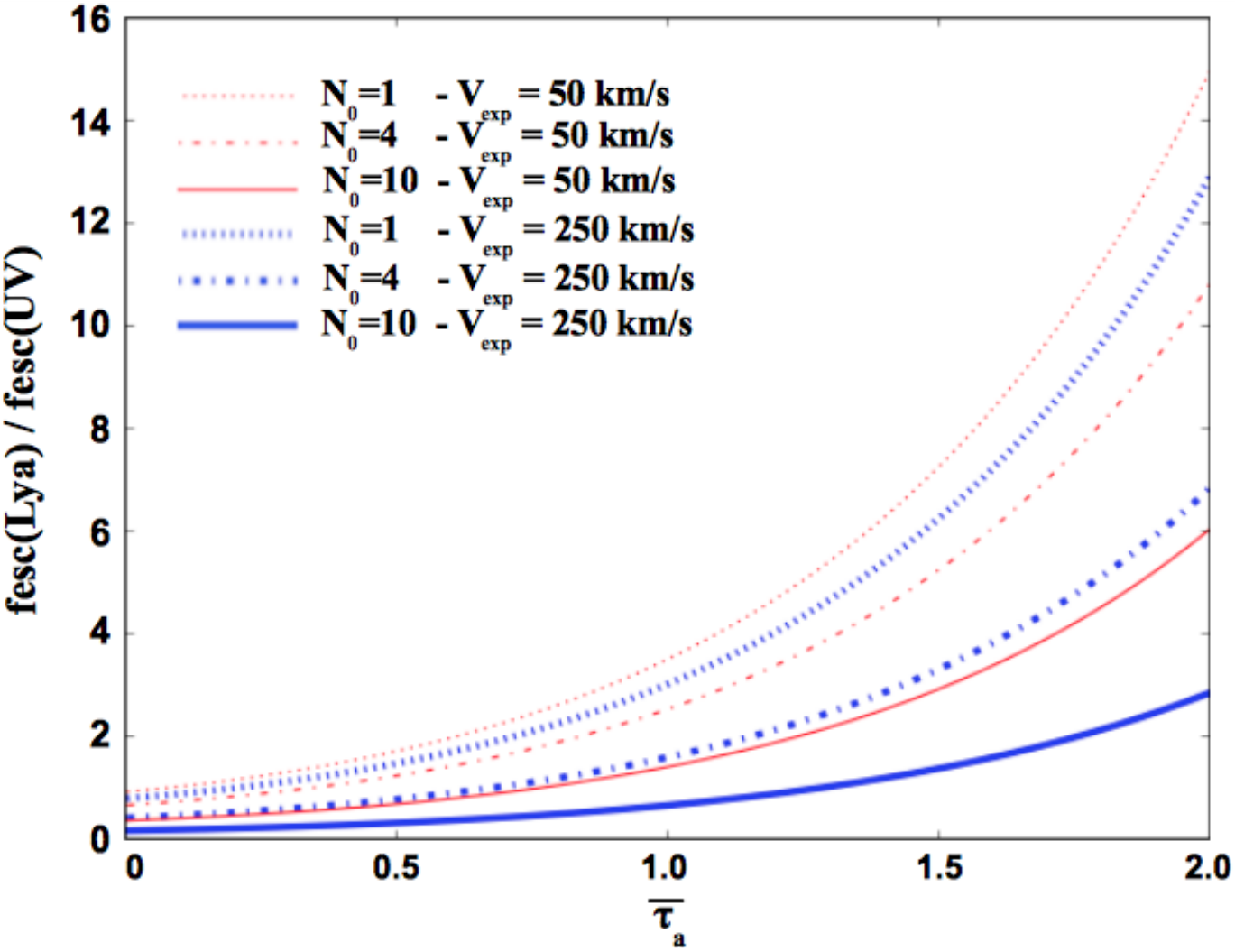}
   \includegraphics[width=90mm]{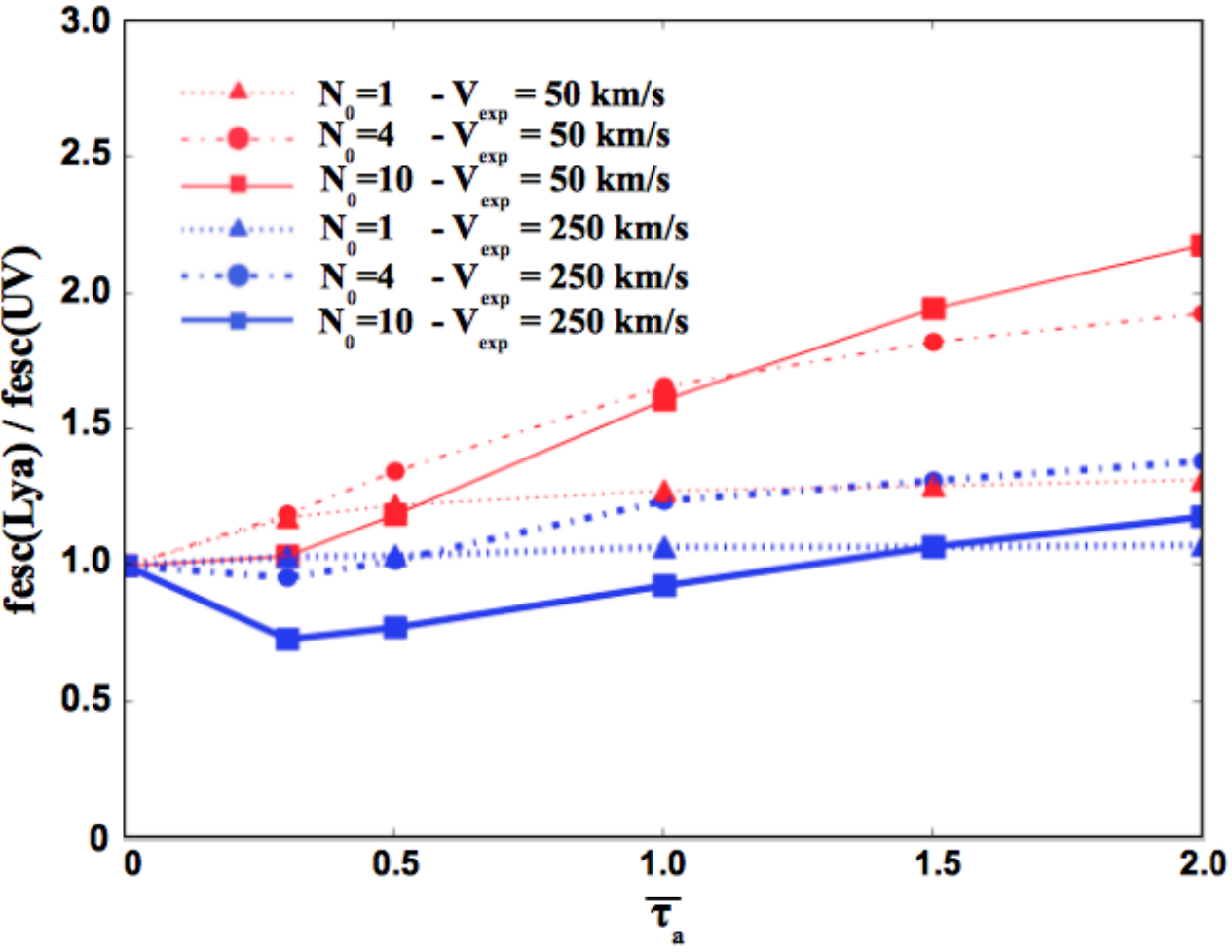}
      \caption{Evolution of the \lya\ line-to-continuum escape fractions ratio, \flya /\fescuv, as a function of the dust optical depth \taua\ in six different extremely clumpy shell geometries. The left panel illustrates the results of HO06 (Fig.\ 18 of their paper), whreas the right panel shows those of our own simulations. Note the different vertical scales. Each clumpy shell geometries are characterised by different values of $N_0$ and are built with different set of parameters. Both structures showing $N_0=1$ (dotted lines) are built adopting FF=0.10, CF=0.38, \nratio=0, T=$10^4$ K, \nhi=$10^{22}$/$cm^{-2}$ and \vexp=50, 250 km s$^{-1}$. Both structures showing $N_0=4$ (dashed-dotted lines) are built adopting FF=0.15, CF=0.70, \nratio=0, T=$10^4$ K, \nhi=$10^{22}$ $cm^{-2}$ and \vexp\ = 50, 250 km s$^{-1}$. Finally, both geometries showing $N_0=10$ (solid lines) are built with FF=0.13, CF=0.90, \nratio=0, T=$10^4$ K, \nhi=$10^{22}$ $cm^{-2}$ and \vexp\ = 50, 250 km s$^{-1}$. In the left panel, each curves are obtained assuming, \fescuv\ given by the equation 17  
and, respectively, \flya\ = (0.94; 0.68; 0.38) for 50 km s$^{-1}$ and (0.81; 0.43; 0.18) for 250 km s$^{-1}$ in each clumpy media ($N_0=1; 4; 10$). Conversely, all curves shown in the right panel are derived from full Monte Carlo simulations. 
Like in HO06, the \lya\ escape fraction is derived in our simulations at the line center (ie. the intrinsic \lya\ line is described as a delta function in all simulations). 
              }
         \label{EW_Hansen_Oh}
   \end{figure*}

Although our numerical simulations reproduce reasonably well all the results and fitting formula of HO06, our conclusions diverge from theirs concerning the Neufeld model. As an application of their numerical study, HO06 give a quantitative estimation of the EW(\lya) enhancements produced by different clumpy and dusty ISMs (Fig.\ 18 of their paper). We reproduce in Fig.\ \ref{EW_Hansen_Oh} (left panel) the results of HO06. These results are based on the following assumptions concerning the clumpy media: 1) the clumps are extremely opaque to \lya\ photons, but 2) each clump is not opaque to dust extinction (i.e. the total dust optical depth for a single clump, taking into account the effect of scattering plus absorption, is $\tau_d$ $<$ 1)\footnote{Nevertheless, the total dust optical depth accross the entire clumpy medium can be greater than unity if many clumps are intersected along the radius of the medium.}, rendering each clump optically thin for UV continuum photons. Given these assumptions, HO06 adopt a constant value of \flya\ (which is derived by an analytical method based on Eq.\ 15), whereas the UV continuum escape fraction is assumed to behave like in homogeneous media. 
In a homogenous medium composed of dust grains with an albedo $\epsilon_d$ $\approx$ 0.5, the UV escape fraction is approximately given by the following equation:
\begin{equation}
\fescuv = \frac{1}{\cosh(\sqrt{4\epsilon_d(\tau_a^2 + \tau_a)})}
\end{equation}  

\begin{figure}
   \centering
   \includegraphics[width=91mm]{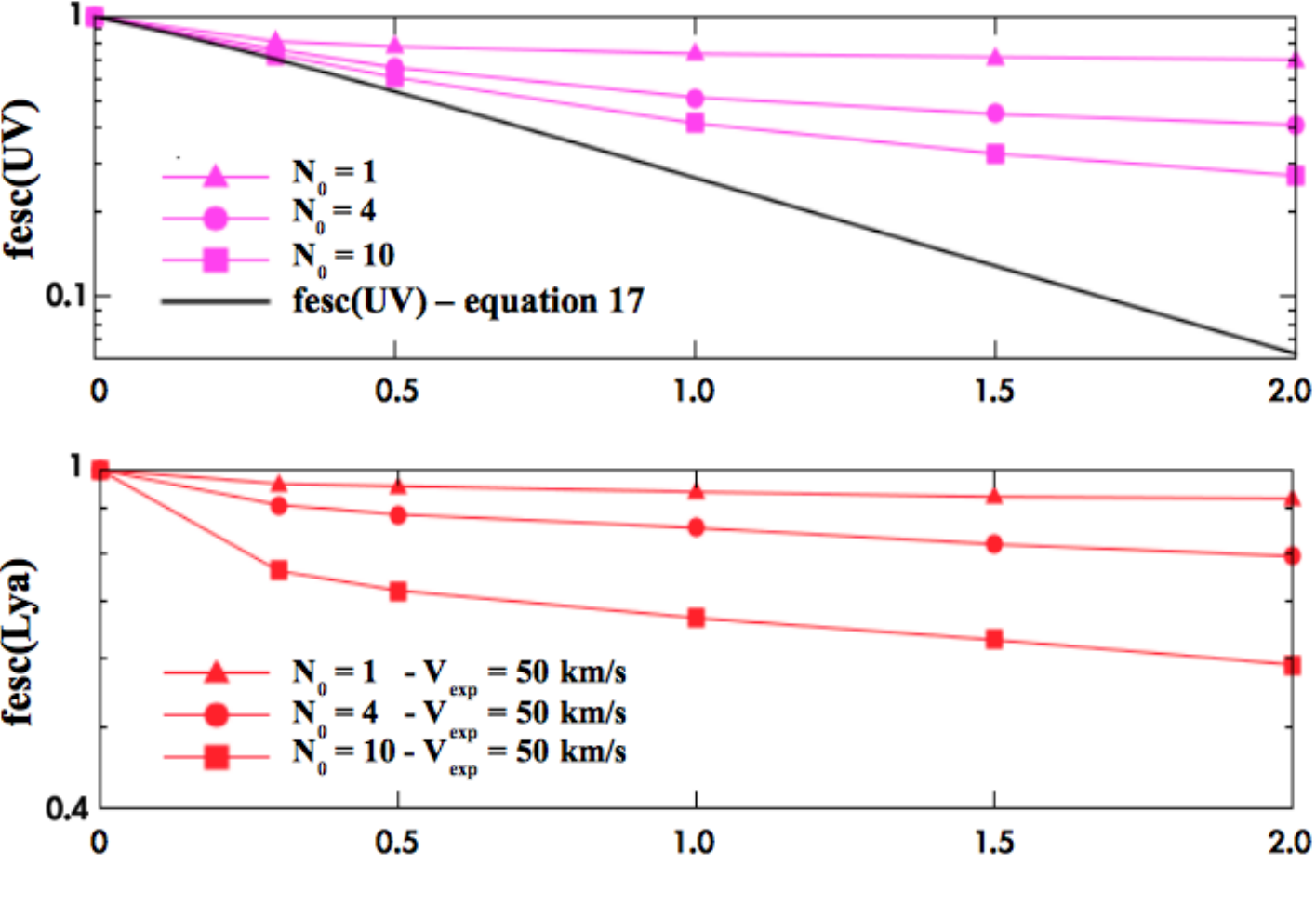}
   \includegraphics[width=91.4mm]{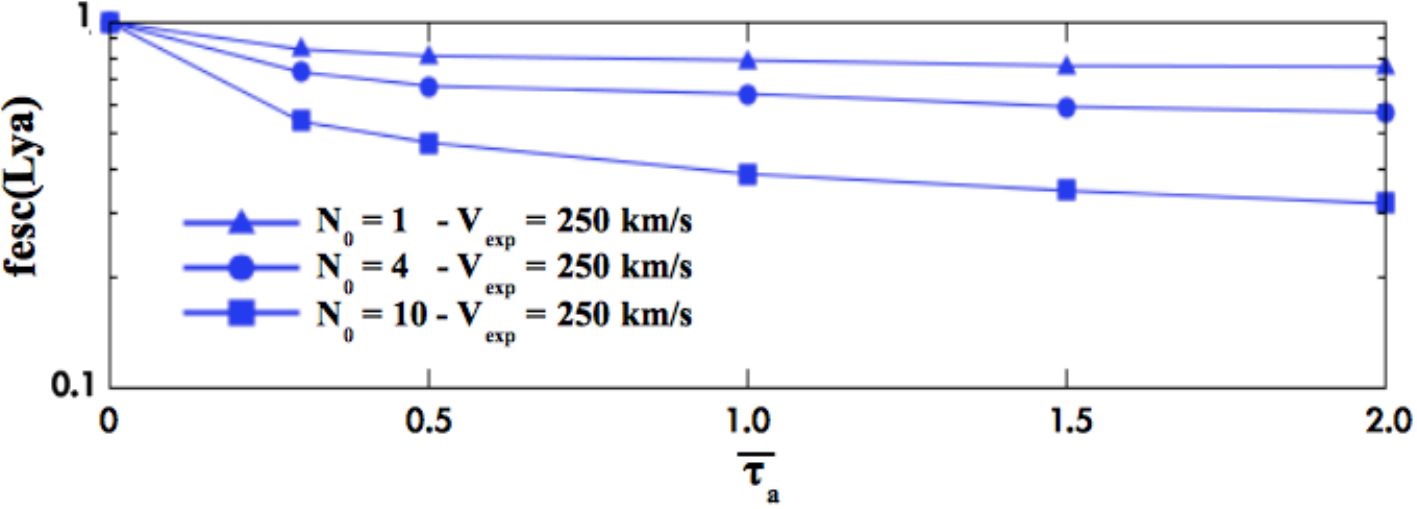}
      \caption{Predicted UV and \lya\ escape fraction as a function of \taua\ for the clumpy media studied in Fig.\ \ref{EW_Hansen_Oh} ($N_0$ = 1 (triangle), $N_0$ = 4 (circle) and $N_0$ = 10 (square)).
            {\em Top}: \fescuv\ obtained from our simulations compared to the equation 17 used by HO06 to carry out their figure shown on the left panel of Fig.\ \ref{EW_Hansen_Oh}. 
            {\em Middle:}  \lya\ escape fraction \flya\ predicted by our models for $\vexp = 50$ \kms.
            {\em Bottom:}  Same as middle panel for \vexp\ = 250 km s$^{-1}$. Like in \cite{hansen06}, the \lya\ escape fraction is derived here at the line center (ie. the intrinsic \lya\ line is a delta function in our simulations)  
           While a clear decrease of the \flya\ with increasing \taua\ is found, HO06 rather assumes a constant value of \flya, indicated in the figure, independently of the dust optical depth \taua. More precisely, HO06 assumes \flya\ = (0.94; 0.68; 0.38) for \vexp\ = 50 km s$^{-1}$ and \flya\ = (0.81; 0.43; 0.18) for \vexp\ = 250 km s$^{-1}$.
              }
         \label{fesc_Hansen_Oh}
   \end{figure}

On the right panel of the Fig.\ \ref{EW_Hansen_Oh} we show the results obtained studying the same media than HO06 (i.e. six clumpy media constructed with three different values of $N_0$ = 1, 4 and 10), but using our own numerical approach. In particular, we derive each curve of the right panel studying the real evolution of \fescuv\ and \flya\ with full Monte Carlo simulations. We can clearly see that we cannot reproduce, neither quantitatively nor qualitatively, the curves of HO06. This strong difference is explained by both assumptions made by HO06 on \fescuv\ (i.e.\ Eq.\ 17) and \flya\ (i.e.\ a constant value under any dust optical depth \taua), which cannot be rigorously met in any clumpy media constructed with low values of $N_0$, as studied in Fig.\ \ref{EW_Hansen_Oh}. We compare in Fig.\ \ref{fesc_Hansen_Oh} the UV and \lya\ escape fractions deduced from our simulations to the assumptions made in HO06. Firstly, while HO06 use the Eq.\ 17 to deduct the evolution of \fescuv\ as a function of \taua\ (i.e. \fescuv\ is therefore assumed to behave like in homogeneous media, which is only correct if the clumpy media are composed of enough \textit{optically thin} clumps distributed in a way they can \textit{intersect all line of sights} around the stars),
this assumption clearly underestimate the correct values of \fescuv\ in the clumpy media studied in Fig.\ \ref{EW_Hansen_Oh}. This discrepancy is explained by the fact that those clumpy media 
are composed of very few clumps which cannot intersect all line of sights around the photon sources\footnote{All clumpy media showing $N_0$ $\le$ 10 have always a covering factor CF $<$ 0.90.}.
Furthermore, this small number of clumps prevent them from staying optically thin to dust extinction in the range of \taua\ [0 : 2]. As a consequence, a large fraction of UV continuum photons can directly escape the clumpy media through several free spaces which appear between clumps, preventing \fescuv\ from behaving like in homogeneous media (i.e.\ Eq.\ 17). Secondly, the constant values of \flya\ assumed by HO06 from their Eq.\ 15 tend to overestimate those obtained from our simulations. This second discrepancy is mainly explained by the fact that \flya\ always increases as the dust optical depth \taua\ decreases in any clumpy media.

In conclusion, the assumptions of HO06 clearly underestimate the correct values of \fescuv\ and overestimate those of \flya. This explains why the enhancements of EW(\lya) obtained in HO06 (left panel in Fig.\ \ref{EW_Hansen_Oh}) are much higher than those deduced from our simulations (right panel). 
Furthermore, the right panel shows the inversion of the curves (in terms of $N_0$) as \taua\ increases. 
The highest enhancement of EW(\lya) are indeed produced by clumpy media containing the highest number of clumps around stars. 


Drawing the parallel with our study of the Neufeld scenario in Sect.\ \ref{s:neufeld}, we can notice that most of the models showing an enhancement of EW(\lya) in Fig.\ 21 (right panel) respect quite well each conditions under which the Neufeld scenario works in a clumpy ISM :
1) the interclump medium is optically thin for \lya\ photons (\nratio\ = 0), 2) the expansion velocities \vexp\ is slow and uniform and 3) the enhancements of EW(\lya) occur above a certain dust optical depth $\tau_c$ equal to (0, 0, 0.3) in each clumpy media ($N_0=1, 4, 10$). Furthermore, as expected when an enhancement of EW(\lya) occurs in a clumpy ISM, the \lya\ line profiles emerging each clumpy geometries studied in Fig.\ 21 are symmetric and peaked at the line center, as expected.

 \subsection{Recent models of \lya\ transfer in clumpy large-scale outflows}
 
 Recently \cite{dijkstra12} have presented radiative transfer calculations of \lya\ photons propagating through clumpy, dusty, large scale 
 outflows using phenomenologically motivated models constrained by absorption line measurement from \cite{steidel11}.
 The calculations of Dijkstra \& Kramer mostly focus on the \lya\ surface brightness distribution. However, since their calculations do not
 follow the behavior of  the UV continuum, it is not possible to infer any information on the presence or absence of an efficient 
 ``Neufeld effect". Furthermore the  \lya\ line profiles predicted by their models are not presented. It is therefore not possible 
 to compare our results with the calculations of \cite{dijkstra12}, and to confront their model to the most direct and sensitive
 observable, the \lya\ line profile itself.
  

\section{Summary and conclusions}

To examine and understand the effects of clumpy ISM structures on the \lya\ line and UV observations of star-forming galaxies
we have carried out detailed radiation transfer calculations using our 3D \lya\ Monte Carlo code MC\lya\ 
\citep{Verhamme06,schaerer11}. Indeed, clumping can in principle significantly alter the transfer of \lya\ in galaxies,
as shown early by \cite{neufeld91}, and has often been invoked to explain strong \lya\ emission or  a higher transmission
for \lya\ photons than for the UV continuum (e.g. Kudritzki et al. 2000, Malhotra \&\ Rhoads 2002, Rhoads et al. 2003, Shimazaku et al. 2006). However, 
only few detailed numerical studies 
of these  effects has so far been undertaken \citep{haiman99, richling03, hansen06, laursen12}, albeit with some simplifying assumptions in most of these works.
Furthermore we wish to identify in which conditions clumping affects the line transfer and how much, and how this is reflected in the emergent \lya\ line profiles.

Our radiation transfer calculations allow us to study simultaneously the dependence of both the \lya\ and the continuum escape fractions, the \lya\ equivalent width EW(\lya), 
and the \lya\ line profiles on the \hi\ content, the dust content, kinematics, gas geometry, and clumping properties.
Since spherically symmetric outflows with a homogeneous \hi\ shell are able to reproduce 
a large variety of observed \lya\ line profiles in Lyman break galaxies and \lya\ emitters \citep{verhamme08, schaerer08, dessauges10},
the same geometry is used to study how a clumpy ISM structure alters the \lya\ line and UV continuum. This clumpy geometry is
also chosen since it has been shown to reproduce observable continuum properties of starburst galaxies and the
Calzetti attenuation law \citep{gordon97,witt00,vijh03}.

Our main results can be summarized as follows: 
\begin{itemize}
\item A clumpy and dusty medium is always more transparent to \lya, UV and optical continuum photons compared to an equivalent homogeneous medium of equal dust content, as already known from earlier studies \citep{boisse90, hobson93, witt96,witt00}.
A clumpy medium thus allows to decrease the global effect of the dust absorption on any radiation.  

\item The UV and optical continuum escape fraction depend on three parameters in homogeneous and clumpy shell geometries: the dust content, the ``clumpiness'' of the dust distribution (described by the density contrast $n_{IC}$/$n_C$), and the covering factor CF (defined here as the fraction of solid angle of the central photons source
covered by the clumps). In a general way, the continuum escape fraction  decreases as the clumpiness of the dust distribution decreases (i.e.\ $n_{IC}$/$n_C$ increases), and as both the dust content and the covering factor increase.   



\item Three additional parameters, i.e.\ in total six, control the \lya\ line transfer: the dust content, the density contrast \nratio, the covering factor, as well as 
the \hi\ column density, the velocity field and the gas temperature.
The \lya\ escape fraction always increases with increasing the clumpiness (i.e.\ $n_{IC}$/$n_C$ decreases), 
and with decreasing dust content and covering factor.
However,  the \hi\ column density and the kinematics of the gas do not affect the \lya\ escape in the same way for homogeneous or clumpy media. That creates two different regimes for the \lya\ radiative transfer in clumpy media. 

\item The first regime (called ``low contrast" regime in this paper) comprises homogeneous and weakly clumpy shell geometries,
corresponding to an interclump density above $\nratio \ga 1.5\times10^{-4}$ for the physical conditions adopted in our model
(such that the interclump medium is also optically thick for \lya). 
In this regime, the \lya\ escape fraction increases with increasing expansion velocity \vexp\ and with decreasing \hi\ column density \nhi,
as for a homogenous ISM. The \lya\ escape fraction is then always less or equal to the UV escape fraction, which implies that the emergent \lya\ equivalent 
width $EW_{\rm obs}$(\lya) is lower than the intrinsic one $EW_{int}$(\lya). 

\item The second regime (called ``high contrast" regime) is found in the most extremely clumpy shell geometries of our model ($\nratio \la 1.5\times10^{-4}$ for the physical conditions adopted in our model). This corresponds to a clumpy medium composed of very dense clumps embedded in an interclump region which is optically thin for \lya\ photons. Two main differences appear compared to the other regime. First, as was originally suggested by Neufeld (1991), it is possible to observe a \lya\ escape fraction which is higher than for the UV continuum. In particular, this is possible above a certain ``critical'' dust optical depth $\tau_c$. 
Second, whereas for $\taua \le \tau_c$ the \lya\ escape fraction behaves as in the ``low contrast" regime, the opposite behavior is found for high enough
dust content ($\taua > \tau_c$), where \flya\ increases with {\em decreasing} \vexp\  and  {\em increasing} \nhi. 

\item Overall we have identified two main effects of the ISM clumpiness on the shape of the \lya\ line profiles.  First, extremely an clumpy ISM favours the formation of a peak at the center of the line profile ($v_{\rm obs}$=0). Second, the intensity of the \lya\ line increases as the clumpiness of the medium increases, as expected.

\item Schematically, the following \lya\ line profile morphologies are predicted from homogeneous and clumpy shell geometries:
``double-peak" profiles with identical/similar peaks symmetric around the source redshift (for static media),  
asymmetric redshifted profiles (from expanding media), 
and absorption line profiles (from very dusty, homogeneous or weakly clumpy media). These types have already been identified
in homogenous models \citep{Verhamme06}. In very clumpy, static and dusty shells, a new category is found:  
``three peaks" profiles similar to the double-peak profiles with an additional third component at line center.

\end{itemize}

As an application of our study, we have examined the conditions under which the Neufeld model \citep{neufeld91} can work in a clumpy ISM,
i.e.\ when an enhancement of the observed EW$_{\rm obs}$(\lya) can be obtained. We find that the following five conditions must be simultaneously fulfilled for the ``Neufeld" effect to be effective.
\begin{itemize}

\item \textbf{The \hi\ content must be very low between clumps},
typically the \hi\ column density in the interclump region must be $\la 3 \times 10^{14}$ \cm2\ to remain optically thin for \lya\ photons. 
Otherwise, \lya\ photons scatter strongly against \hi\ atoms localized between clumps and cannot escape the clumpy medium more easily than UV photons.

\item \textbf{The galactic outflow has to be slow}
with outflow velocities of the order of $\vexp\ \la 2 \times\ FWHM_{int}$(\lya) km s$^{-1}$.
Otherwise \lya\ photons are too redshifted from the clumps and cannot scatter anymore on the surface of clumps, as suggested in the Neufeld model. 

\item \textbf{The galactic outflow has to be as uniform and constant as possible in velocity} for efficient interactions of \lya\ photons with dust.

\item  \textbf{A high dust content must be embedded in clumps} to absorbe as much as possible the UV continuum, which increases the \lya\ equivalent width.
For the physical conditions and clumpy shell geometries adopted here, we find that an enhancement of EW(\lya) by a factor 3--4 can only occur for a 
colour excess E(B-V) $\ga$ 0.3. 

\item \textbf{A large covering factor} is needed in order to get a noticeable enhancement of EW(\lya). 



\end{itemize}

The above conditions are in agreement with the general findings of \citet{neufeld91}.
However, our results differ from those of \cite{hansen06}, who make some simplifying assumptions, which are not consistent with
more rigorous radiation transfer calculations. In our study, the Neufeld model does not work as easily as suggested in \cite{hansen06}.

Given our results it seems quite unlikely/difficult to find conditions in a clumpy, spherically symmetric ISM, where \lya\ photons can escape more easily than the 
nearby UV continuum, i.e.\ where the phenomenon suggested by Neufeld (1991) can be at play. Furthermore, when these conditions are fulfilled we generally find that the
emergent \lya\ line profile shows emission at line center and little asymmetry. Such profiles do, however, not represent the profiles typically observed in
high-redshift galaxies, which are known to be redshifted in the galaxy rest-fame and asymmetric.
Other arguments against the Neufeld effect being effective may be if the sites of \lya\ emission are relatively close to or within cold, dusty environments such as 
molecular clouds, which could absorb more efficiently \lya\ photons than the UV continuum \citep{laursen12, verhamme12}. 

The simulations from this paper and the success of homogenous, spherically expanding models in reproducing the large variety of observed \lya\ line profiles and velocity shifts between photospheric, low ionization absorption, and the \lya\ line \citep{verhamme08, schaerer08} seem to indicate that effects to due inhomogeneities in the ISM and deviations from spherical geometry are not dominant.
Why simple geometries work so well may appear somewhat puzzling, and certainly remains worth understanding more in depth.
More detailed observations and sophisticated radiation transfer models within multi-phase ISM, high-resolution simulations may help
to shed more light on this question and to examine how well the physical parameters derived from \lya\ line profile fits using simple spherically expanding  
model represent reality.

\begin{acknowledgements}
We thank Anne Verhamme for interesting comments on an earlier version of this paper.
Simulations were done on the {\tt regor} PC cluster at the Geneva
Observatory co-funded by grants to Georges Meynet, Daniel Pfenniger, and DS.
The work of DS was supported by the Swiss National Science Foundation. Peter Laursen acknowledges support from the ERC-StG grant EGGS-278202.

\end{acknowledgements}

\bibliographystyle{aa}
\bibliography{article}

\begin{thebibliography}{83}
\expandafter\ifx\csname natexlab\endcsname\relax\def\natexlab#1{#1}\fi

\bibitem[{{Ahn} {et~al.}(2001){Ahn}, {Lee}, \& {Lee}}]{ahn01}
{Ahn}, S.-H., {Lee}, H.-W., \& {Lee}, H.~M. 2001, \aj, 554, 604

\bibitem[{{Ahn} {et~al.}(2002){Ahn}, {Lee}, \& {Lee}}]{ahn02}
{Ahn}, S.-H., {Lee}, H.-W., \& {Lee}, H.~M. 2002, \aj, 567, 920

\bibitem[{{Ahn} {et~al.}(2003){Ahn}, {Lee}, \& {Lee}}]{ahn03}
{Ahn}, S.-H., {Lee}, H.-W., \& {Lee}, H.~M. 2003, \mnras, 340, 863

\bibitem[{{Atek} {et~al.}(2008){Atek}, {Kunth}, {Hayes}, {{\"O}stlin}, \&
  {Mas-Hesse}}]{atek08}
{Atek}, H., {Kunth}, D., {Hayes}, M., {{\"O}stlin}, G., \& {Mas-Hesse}, J.~M.
  2008, \aap, 488, 491

\bibitem[{{Atek} {et~al.}(2009){Atek}, {Schaerer}, \& {Kunth}}]{atek09a}
{Atek}, H., {Schaerer}, D., \& {Kunth}, D. 2009, \aap, 502, 791

\bibitem[{{Baker} {et~al.}(2004){Baker}, {Tacconi}, {Genzel}, {Lehnert}, \&
  {Lutz}}]{baker04}
{Baker}, A.~J., {Tacconi}, L.~J., {Genzel}, R., {Lehnert}, M.~D., \& {Lutz}, D.
  2004, \apj, 604, 125

\bibitem[{{Boisse}(1990)}]{boisse90}
{Boisse}, P. 1990, \aap, 228, 483

\bibitem[{{Calzetti}(1997)}]{calzetti97}
{Calzetti}, D. 1997, \apj, 113, 162

\bibitem[{{Calzetti} {et~al.}(2000){Calzetti}, {Armus}, {Bohlin}, {Kinney},
  {Koornneef}, \& {Storchi-Bergmann}}]{calzetti00}
{Calzetti}, D., {Armus}, L., {Bohlin}, R.~C., {et~al.} 2000, \apj, 533, 682

\bibitem[{{Cantalupo} {et~al.}(2005){Cantalupo}, {Porciani}, {Lilly}, \&
  {Miniati}}]{cantalupo05}
{Cantalupo}, S., {Porciani}, C., {Lilly}, S.~J., \& {Miniati}, F. 2005, \aj,
  628, 61

\bibitem[{{Charlot} \& {Fall}(1993)}]{charlot_fall93}
{Charlot}, S. \& {Fall}, S.~M. 1993, \apj, 415, 580

\bibitem[{{Cowie} \& {Hu}(1998)}]{cowie98}
{Cowie}, L.~L. \& {Hu}, E.~M. 1998, \aj, 115, 1319

\bibitem[{{Dawson} {et~al.}(2004){Dawson}, {Rhoads}, {Malhotra}, {Stern},
  {Dey}, {Spinrad}, {Jannuzi}, {Wang}, \& {Landes}}]{dawson04}
{Dawson}, S., {Rhoads}, J.~E., {Malhotra}, S., {et~al.} 2004, \aj, 617, 707

\bibitem[{{Dayal} {et~al.}(2009){Dayal}, {Ferrara}, {Saro}, {Salvaterra},
  {Borgani}, \& {Tornatore}}]{dayal09}
{Dayal}, P., {Ferrara}, A., {Saro}, A., {et~al.} 2009, \mnras, 400, 2000

\bibitem[{{Dessauges-Zavadsky} {et~al.}(2010){Dessauges-Zavadsky}, {D'Odorico},
  {Schaerer}, {Modigliani}, {Tapken}, \& {Vernet}}]{dessauges10}
{Dessauges-Zavadsky}, M., {D'Odorico}, S., {Schaerer}, D., {et~al.} 2010, \aap,
  510, A26

\bibitem[{{Dickey} \& {Garwood}(1989)}]{dickey89}
{Dickey}, J.~M. \& {Garwood}, R.~W. 1989, \apj, 341, 201

\bibitem[{{Dijkstra} \& {Kramer}(2012)}]{dijkstra12}
{Dijkstra}, M. \& {Kramer}, R.~H. 2012, \mnras, 386, 1

\bibitem[{{Erb} {et~al.}(2003){Erb}, {Shapley}, {Steidel}, {Pettini},
  {Adelberger}, {Hunt}, {Moorwood}, \& {Cuby}}]{erb03}
{Erb}, D.~K., {Shapley}, A.~E., {Steidel}, C.~C., {et~al.} 2003, \apj, 591, 101

\bibitem[{{Fan} {et~al.}(2002){Fan}, {Narayanan}, {Strauss}, {White}, {Becker},
  {Pentericci}, \& {Rix}}]{fan02}
{Fan}, X., {Narayanan}, V.~K., {Strauss}, M.~A., {et~al.} 2002, \aj, 123, 1247

\bibitem[{{Finkelstein} {et~al.}(2009){Finkelstein}, {Rhoads}, {Malhotra}, \&
  {Grogin}}]{finkelstein09}
{Finkelstein}, S.~L., {Rhoads}, J.~E., {Malhotra}, S., \& {Grogin}, N. 2009,
  \apj, 691, 465

\bibitem[{{Finkelstein} {et~al.}(2008){Finkelstein}, {Rhoads}, {Malhotra},
  {Grogin}, \& {Wang}}]{finkelstein08}
{Finkelstein}, S.~L., {Rhoads}, J.~E., {Malhotra}, S., {Grogin}, N., \& {Wang},
  J. 2008, \apj, 678, 655

\bibitem[{{Forero-Romero} {et~al.}(2011){Forero-Romero}, {Yepes}, {Gottlšber},
  {Knollmann}, {Cuesta}, \& {Prada}}]{forero11}
{Forero-Romero}, J.~E., {Yepes}, G., {Gottlšber}, S., {et~al.} 2011, \mnras,
  415, 3666

\bibitem[{{Gawiser} {et~al.}(2006){Gawiser}, {van Dokkum}, {Gronwall},
  {Ciardullo}, {Blanc}, {Castander}, {Feldmeier}, {Francke}, {Franx},
  {Haberzettl}, {Herrera}, {Hickey}, {Infante}, {Lira}, {Maza}, {Quadri},
  {Richardson}, {Schawinski}, {Schirmer}, {Taylor}, {Treister}, {Urry}, \&
  {Virani}}]{gawiser06}
{Gawiser}, E., {van Dokkum}, P.~G., {Gronwall}, C., {et~al.} 2006, \aj, 642,
  L13

\bibitem[{{Giavalisco} {et~al.}(1996){Giavalisco}, {Koratkar}, \&
  {Calzetti}}]{giavalisco96}
{Giavalisco}, M., {Koratkar}, A., \& {Calzetti}, D. 1996, \apj, 466, 831

\bibitem[{{Gordon} {et~al.}(1997){Gordon}, {Calzetti}, \& {Witt}}]{gordon97}
{Gordon}, K.~D., {Calzetti}, D., \& {Witt}, A.~N. 1997, \apj, 487, 625

\bibitem[{{Gronwall} {et~al.}(2007){Gronwall}, {Ciardullo}, {Hickey},
  {Gawiser}, {Feldmeier}, {van Dokkum}, {Urry}, {Herrera}, {Lehmer}, {Infante},
  {Orsi}, {Marchesini}, {Blanc}, {Francke}, {Lira}, \& {Treister}}]{gronwall07}
{Gronwall}, C., {Ciardullo}, R., {Hickey}, T., {et~al.} 2007, \apj, 667, 79

\bibitem[{{Guaita} {et~al.}(2010){Guaita}, {Gawiser}, {Padilla}, {Francke},
  {Bond}, C., {Ciardullo}, {Feldmeier}, {Sinawa}, {Blanc}, \&
  {Virani}}]{guaita10}
{Guaita}, L., {Gawiser}, E., {Padilla}, N., {et~al.} 2010, \apj, 714, 255

\bibitem[{{Haiman} \& {Spaans}(1999)}]{haiman99}
{Haiman}, Z. \& {Spaans}, M. 1999, \apj, 518, 138

\bibitem[{{Hansen} \& {Oh}(2006)}]{hansen06}
{Hansen}, M. \& {Oh}, S.~P. 2006, New Astronomy Review, 367, 979

\bibitem[{{Hayes} \& {{\"O}stlin}(2006)}]{hayes06}
{Hayes}, M. \& {{\"O}stlin}, G. 2006, \aap, 460, 681

\bibitem[{{Hobson} \& {Scheuer}(1993)}]{hobson93}
{Hobson}, M.~P. \& {Scheuer}, P.~A.~G. 1993, \mnras, 264, 145

\bibitem[{{Hu} {et~al.}(2004){Hu}, {Cowie}, {Capak}, {McMahon}, \&
  {Komiyama}}]{hu04}
{Hu}, E.~M., {Cowie}, L.~L., {Capak}, P., {McMahon}, R. G.and~{Hayashino}, T.,
  \& {Komiyama}, Y. 2004, \apj, 127, 563

\bibitem[{{Hu} {et~al.}(1998){Hu}, {Cowie}, \& {McMahon}}]{hu98}
{Hu}, E.~M., {Cowie}, L.~L., \& {McMahon}, R.~G. 1998, \apjl, 502, L99+

\bibitem[{{Kashikawa} {et~al.}(2012){Kashikawa}, {Nagao}, {Toshikawa},
  {Ishizaki}, {Egami}, Hayashi, \& {Ly}}]{kashikawa12}
{Kashikawa}, N., {Nagao}, T., {Toshikawa}, J., {et~al.} 2012, arXiv:1210.4933

\bibitem[{{Kashikawa} {et~al.}(2006){Kashikawa}, {Shimasaku}, {Malkan}, {Doi},
  {Matsuda}, {Ouchi}, {Taniguchi}, {Ly}, {Nagao}, {Iye}, {Motohara},
  {Murayama}, {Murozono}, {Nariai}, {Ohta}, {Okamura}, {Sasaki}, {Shioya}, \&
  {Umemura}}]{kashikawa06}
{Kashikawa}, N., {Shimasaku}, K., {Malkan}, M.~A., {et~al.} 2006, \apj, 648, 7

\bibitem[{{Kudritzki} {et~al.}(2000){Kudritzki}, {M{\'e}ndez}, {Feldmeier},
  {Ciardullo}, {Jacoby}, {Freeman}, {Arnaboldi}, {Capaccioli}, {Gerhard}, \&
  {Ford}}]{kudritzki00}
{Kudritzki}, R.-P., {M{\'e}ndez}, R.~H., {Feldmeier}, J.~J., {et~al.} 2000,
  \apj, 536, 19

\bibitem[{{Kunth} {et~al.}(1994){Kunth}, {Lequeux}, {Sargent}, \&
  {Viallefond}}]{kunth94}
{Kunth}, D., {Lequeux}, J., {Sargent}, W.~L.~W., \& {Viallefond}, F. 1994,
  \aap, 282, 709

\bibitem[{{Kunth} {et~al.}(1998){Kunth}, {Mas-Hesse}, {Terlevich}, {Terlevich},
  {Lequeux}, \& {Fall}}]{kunth98}
{Kunth}, D., {Mas-Hesse}, J.~M., {Terlevich}, E., {et~al.} 1998, \aap, 334, 11

\bibitem[{{Laursen} {et~al.}(2012){Laursen}, {Duval}, \&
  {\"Ostlin}}]{laursen12}
{Laursen}, P., {Duval}, F., \& {\"Ostlin}, G. 2012, arXiv:1211.2833

\bibitem[{{Laursen} {et~al.}(2009{\natexlab{a}}){Laursen}, {Razoumov}, \&
  {Sommer-Larsen}}]{laursen09}
{Laursen}, P., {Razoumov}, A.~O., \& {Sommer-Larsen}, J. 2009{\natexlab{a}},
  \apj, 696, 853

\bibitem[{{Laursen} {et~al.}(2009{\natexlab{b}}){Laursen}, {Sommer-Larsen}, \&
  {Andersen}}]{larsen09}
{Laursen}, P., {Sommer-Larsen}, J., \& {Andersen}, A.~C. 2009{\natexlab{b}},
  \apj, 704, 1640

\bibitem[{{Lequeux} {et~al.}(1995){Lequeux}, {Kunth}, {Mas-Hesse}, \&
  {Sargent}}]{lequeux95}
{Lequeux}, J., {Kunth}, D., {Mas-Hesse}, J.~M., \& {Sargent}, W.~L.~W. 1995,
  \aap, 301, 18

\bibitem[{{Malhotra} \& {Rhoads}(2002)}]{malhotra02}
{Malhotra}, S. \& {Rhoads}, J.~E. 2002, \apj, 565, L71

\bibitem[{{Malhotra} \& {Rhoads}(2004)}]{malhotra04}
{Malhotra}, S. \& {Rhoads}, J.~E. 2004, \apjl, 617, L5

\bibitem[{{Marscher} {et~al.}(1993){Marscher}, {Moore}, \&
  {Bania}}]{marscher93}
{Marscher}, A.~P., {Moore}, E.~M., \& {Bania}, T.~M. 1993, \aj, 419, 101

\bibitem[{{Mas-Hesse} {et~al.}(2003){Mas-Hesse}, {Kunth}, {Tenorio-Tagle},
  {Leitherer}, {Terlevich}, \& {Terlevich}}]{mashesse03}
{Mas-Hesse}, J.~M., {Kunth}, D., {Tenorio-Tagle}, G., {et~al.} 2003, \apj, 598,
  858

\bibitem[{{McKee} \& {Ostriker}(1977)}]{mckee77}
{McKee}, C.~F. \& {Ostriker}, J.~P. 1977, \apj, 218, 148

\bibitem[{{McLinden} {et~al.}(2011){McLinden}, {Finkelstein}, {Rhoads},
  {Malhotra}, {Hibon}, {Richardson}, {Cresci}, {Quirrenbach}, {Pasquali},
  {Bian}, {Fan}, \& {Woodward}}]{mclinden11}
{McLinden}, E.~M., {Finkelstein}, S.~L., {Rhoads}, J.~E., {et~al.} 2011, \apj,
  730, 136

\bibitem[{{Meier} \& {Terlevich}(1981)}]{meier81}
{Meier}, D.~L. \& {Terlevich}, R. 1981, \aj, 246, 109

\bibitem[{{Neufeld}(1990)}]{neufeld90}
{Neufeld}, D.~A. 1990, \apj, 350, 216

\bibitem[{{Neufeld}(1991)}]{neufeld91}
{Neufeld}, D.~A. 1991, \apjl, 370, L85

\bibitem[{{Nilsson} {et~al.}(2007){Nilsson}, {M{\o}ller}, {M\"oller}, {Fynbo},
  {Micha?owski}, {Watson}, \& {Ledoux}}]{nilsson07}
{Nilsson}, K.~K., {M{\o}ller}, P., {M\"oller}, O., {et~al.} 2007, \aap, 471, 71

\bibitem[{{{\"O}stlin} {et~al.}(2009){{\"O}stlin}, {Hayes}, {Kunth},
  {Mas-Hesse}, {Leitherer}, {Petrosian}, \& {Atek}}]{ostlin09}
{{\"O}stlin}, G., {Hayes}, M., {Kunth}, D., {et~al.} 2009, \apj, 138, 923

\bibitem[{{Ouchi} {et~al.}(2008){Ouchi}, {Shimasaku}, {Akiyama}, {Simpson},
  {Saito}, {Ueda}, {Furusawa}, {Sekiguchi}, {Yamada}, {Kodama}, {Kashikawa},
  {Okamura}, {Iye}, {Takata}, {Yoshida}, \& {Yoshida}}]{Ouchi08}
{Ouchi}, M., {Shimasaku}, K., {Akiyama}, M., {et~al.} 2008, \apjs, 176, 301

\bibitem[{{Ouchi} {et~al.}(2003){Ouchi}, {Shimasaku}, {Furusawa}, {Miyazaki},
  {Doi}, {Hamabe}, {Hayashino}, {Kimura}, {Kodaira}, {Komiyama}, {Matsuda},
  {Miyazaki}, {Nakata}, {Okamura}, {Sekiguchi}, {Shioya}, {Tamura},
  {Taniguchi}, {Yagi}, \& {Yasuda}}]{ouchi03}
{Ouchi}, M., {Shimasaku}, K., {Furusawa}, H., {et~al.} 2003, \apj, 582, 60

\bibitem[{{Ouchi} {et~al.}(2010){Ouchi}, {Shimasaku}, {Furusawa}, {Saito},
  {Yoshida}, {Akiyama}, {Ono}, {Yamada}, \& {Ota}}]{ouchi10}
{Ouchi}, M., {Shimasaku}, K., {Furusawa}, H., {et~al.} 2010, \apj, 723, 869

\bibitem[{{Partridge} \& {Peebles}(1967)}]{patridge67}
{Partridge}, R. \& {Peebles}, P. J.~E. 1967, \apj, 147, 868

\bibitem[{{Pierleoni} {et~al.}(2007){Pierleoni}, {Maselli}, \&
  {Ciardi}}]{pierleoni07}
{Pierleoni}, M., {Maselli}, A., \& {Ciardi}, B. 2007

\bibitem[{{Rhoads} {et~al.}(2000){Rhoads}, {Dey}, {Malhotra}, {Stern},
  {Spinrad}, \& {Jannuzi}}]{rhoads00}
{Rhoads}, J.~E., {Dey}, A., {Malhotra}, S., {et~al.} 2000, \apj, 545, 85

\bibitem[{{Rhoads} {et~al.}(2003){Rhoads}, {Dey}, {Malhotra}, {Stern},
  {Spinrad}, {Jannuzi}, {Dawson}, {Brown}, \& {Landes}}]{rhoads03}
{Rhoads}, J.~E., {Dey}, A., {Malhotra}, S., {et~al.} 2003, \aj, 125, 1006

\bibitem[{{Richling}(2003)}]{richling03}
{Richling}, S. 2003, \mnras, 344, 553

\bibitem[{Santos(2004)}]{santos04}
Santos. 2004, Galaxy formation near the epoch of reionization {{\TeX}book}
  (CaltechTHESIS)

\bibitem[{{Scarlata} {et~al.}(2009){Scarlata}, {Colbert}, {Teplitz}, {Panagia},
  {Hayes}, {Siana}, {Rau}, {Francis}, {Caon}, {Pizzella}, \&
  {Bridge}}]{scarlata09}
{Scarlata}, C., {Colbert}, J., {Teplitz}, H.~I., {et~al.} 2009, \apj, 704, 98

\bibitem[{{Schaerer}(2003)}]{schaerer03}
{Schaerer}, D. 2003, \aap, 397, 527

\bibitem[{{Schaerer} {et~al.}(2011){Schaerer}, {Hayes}, {Verhamme}, \&
  {Teyssier}}]{schaerer11}
{Schaerer}, D., {Hayes}, M., {Verhamme}, A., \& {Teyssier}, R. 2011, \aap, 531,
  A12

\bibitem[{{Schaerer} \& {Verhamme}(2008)}]{schaerer08}
{Schaerer}, D. \& {Verhamme}, A. 2008, \aap, 480, 369

\bibitem[{{Shimasaku} {et~al.}(2006){Shimasaku}, {Kashikawa}, {Doi}, {Ly},
  {Malkan}, {Matsuda}, {Ouchi}, {Hayashino}, {Iye}, {Motohara}, {Murayama},
  {Nagao}, {Ohta}, {Okamura}, {Sasaki}, {Shioya}, \& {Taniguchi}}]{shimasaku06}
{Shimasaku}, K., {Kashikawa}, N., {Doi}, M., {et~al.} 2006, \pasj, 58, 313

\bibitem[{{Steidel} {et~al.}(2011){Steidel}, {Bogosavljevic}, {Shapley},
  {Kollmeier}, {Reddy}, {Erb}, \& {Pettini}}]{steidel11}
{Steidel}, C.~C., {Bogosavljevic}, M., {Shapley}, A.~E., {et~al.} 2011, \apj,
  736, 160

\bibitem[{{Stutzki} \& {Guesten}(1990)}]{stutzki90}
{Stutzki}, J. \& {Guesten}, R. 1990, \aj, 256, 513

\bibitem[{{Swinbank} {et~al.}(2011){Swinbank}, {Papadopoulos}, {Cox}, {Krips},
  {Ivison}, {Smail}, {Thomson}, {Neri}, {Richard}, \& {Ebeling}}]{swinbank11}
{Swinbank}, A.~M., {Papadopoulos}, P.~P., {Cox}, P., {et~al.} 2011, \aj, 742,
  11

\bibitem[{{Taniguchi} {et~al.}(2003){Taniguchi}, {Ajiki}, {Murayama}, \&
  {Nagao}}]{taniguchi03}
{Taniguchi}, Y., {Ajiki}, M., {Murayama}, T., \& {Nagao}, T. 2003, \apj, 585,
  L97

\bibitem[{{Taniguchi} {et~al.}(2005){Taniguchi}, {Ajiki}, {Nagao}, {Shioya},
  {Murayama}, {Kashikawa}, {Kodaira}, {Kaifu}, {Ando}, {Karoji}, {Akiyama},
  {Aoki}, {Doi}, {Fujita}, {Furusawa}, {Hayashino}, {Iwamuro}, {Iye},
  {Kobayashi}, {Kodama}, {Komiyama}, {Matsuda}, {Miyazaki}, {Mizumoto},
  {Morokuma}, {Motohara}, {Nariai}, {Ohta}, {Ohyama}, {Okamura}, {Ouchi},
  {Sasaki}, {Sato}, {Sekiguchi}, {Shimasaku}, {Tamura}, {Umemura}, {Yamada},
  {Yasuda}, \& {Yoshida}}]{taniguchi05}
{Taniguchi}, Y., {Ajiki}, M., {Nagao}, T., {et~al.} 2005, PASJ, 57, 165

\bibitem[{{Tenorio-Tagle} {et~al.}(1999){Tenorio-Tagle}, {Silich}, {Kunth},
  {Terlevich}, \& {Terlevich}}]{tenorio99}
{Tenorio-Tagle}, G., {Silich}, S.~A., {Kunth}, D., {Terlevich}, E., \&
  {Terlevich}, R. 1999, \mnras, 309, 332

\bibitem[{{Teplitz} {et~al.}(2000){Teplitz}, {McLean}, {Becklin}, {Figer},
  {Gilbert}, {Graham}, {Larkin}, {Levenson}, \& {Wilcox}}]{teplitz00}
{Teplitz}, H.~I., {McLean}, I.~S., {Becklin}, E.~E., {et~al.} 2000, \apj, 533,
  L65

\bibitem[{{Thuan} {et~al.}(1997){Thuan}, {Izotov}, \& {Lipovetsky}}]{thuan97a}
{Thuan}, T.~X., {Izotov}, Y.~I., \& {Lipovetsky}, V.~A. 1997, \apj, 477, 661

\bibitem[{{Varosi} \& {Dwek}(1999)}]{varosi99}
{Varosi}, F. \& {Dwek}, E. 1999, arXiv:astro-ph/9905291

\bibitem[{{Verhamme} {et~al.}(2012){Verhamme}, {Dubois}, {Blaizot}, {Garel},
  {Bacon}, {Devriendt}, {Guiderdoni}, \& {Slyz}}]{verhamme12}
{Verhamme}, A., {Dubois}, Y., {Blaizot}, J., {et~al.} 2012, arXiv:1208.4781

\bibitem[{{Verhamme} {et~al.}(2008){Verhamme}, {Schaerer}, {Atek}, \&
  {Tapken}}]{verhamme08}
{Verhamme}, A., {Schaerer}, D., {Atek}, H., \& {Tapken}, C. 2008, \aap, 491, 89

\bibitem[{{Verhamme} {et~al.}(2006){Verhamme}, {Schaerer}, \&
  {Maselli}}]{Verhamme06}
{Verhamme}, A., {Schaerer}, D., \& {Maselli}, A. 2006, \aap, 460, 397

\bibitem[{{Vijh} {et~al.}(2003){Vijh}, {Witt}, \& {Gordon}}]{vijh03}
{Vijh}, U.~P., {Witt}, A.~N., \& {Gordon}, K.~D. 2003, \aj, 587, 533

\bibitem[{{Wang} {et~al.}(2004){Wang}, {Rhoads}, {Malhotra}, {Dawson}, {Stern},
  {Dey}, {Heckman}, {Norman}, \& {Spinrad}}]{wang04}
{Wang}, J.~X., {Rhoads}, J.~E., {Malhotra}, S., {et~al.} 2004, \aj, 608, 21

\bibitem[{{Witt} \& {Gordon}(1996)}]{witt96}
{Witt}, A.~N. \& {Gordon}, K.~D. 1996, \apj, 463, 681

\bibitem[{{Witt} \& {Gordon}(2000)}]{witt00}
{Witt}, A.~N. \& {Gordon}, K.~D. 2000, \apj, 528, 799

\end{thebibliography}

\end{document}